\documentclass[manuscript,screen]{acmart}
\usepackage{booktabs} 
\usepackage{grffile}
\usepackage[ruled]{algorithm2e} 
\usepackage{multirow}

\usepackage{epsfig}
\usepackage{amsmath}
\usepackage{color}
\usepackage{url,xspace,graphicx}
\usepackage{caption}
\usepackage{subcaption}
\usepackage{blindtext}
\usepackage{siunitx}
\usepackage[normalem]{ulem}
\usepackage{todonotes}

\newcommand{\SAIS}{\textsc{SAIS}\xspace}
\newcommand{\our}{\textsc{GCIS}\xspace}
\newcommand{\repair}{\textsc{RePair}\xspace}
\newcommand{\gzip}{\textsc{Gzip}\xspace}
\newcommand{\bzip}{\textsc{Bzip2}\xspace}
\newcommand{\szip}{\textsc{7-zip}\xspace}
\newcommand{\ppmdj}{\textsc{Ppmdj}\xspace}
\newcommand{\bigrepair}{\textsc{BigRePair}\xspace}
\newcommand{\relz}{\textsc{ReLZ}\xspace}

\newcommand{\lcp}{\texttt{lcp}\xspace}
\newcommand{\conc}{\cdot}
\newcommand{\etal}{{\it et al.}\xspace}
\newcommand{\gen}{\mathcal{G}}
\newcommand{\suff}{\mathrm{s}}

\newtheorem{definition}{Definition}

\newcommand{\SA}{\ensuremath{\mathsf{SA}}\xspace}
\newcommand{\LCP}{\ensuremath{\mathsf{LCP}}\xspace}

\begin{document}
\title{Grammar Compression by Induced Suffix Sorting}
\titlenote{A preliminary version of this work appeared in DCC 2018~\cite{NunesLGAN18}.}

\author{Daniel S. N. Nunes}
\affiliation{%
	\institution{Federal Institute of Bras\'ilia}
	\city{Taguatinga}
	\country{}}
\affiliation{
	\institution{Department of Computer Science, University of Bras\'ilia}
	\city{Bras\'ilia}
	\country{Brazil}
}
\email{daniel.nunes@ifb.edu.br}

\author{Felipe A. Louza}
\affiliation{
	\institution{Faculty of Electrical Engineering, Federal University of Uberl\^andia}
	\city{Uberl\^andia}
	\country{Brazil}
}
\email{louza@ufu.br}

\author{Simon Gog}
\affiliation{
	\institution{eBay Inc.}
	\city{San Jose}
	\country{USA}
}
\email{sgog@ebay.com}

\author{Mauricio~Ayala-Rinc\'on}
\affiliation{
	\institution{Departments of Computer Science and Mathematics, University of Bras\'ilia}
	\city{Bras\'ilia}
	\country{Brazil}
}
\email{ayala@unb.br}

\author{Gonzalo Navarro}
\affiliation{
	\institution{Center for Biotechnology and Bioengineering (CeBiB)}
}
\affiliation{
	\institution{Department of Computer Science, University of Chile}
	\city{Santiago}
	\country{Chile}
}
\email{gnavarro@dcc.uchile.cl}

\renewcommand\shortauthors{Nunes, D. S. N. \etal}

\begin{abstract}
	A grammar compression algorithm, called GCIS, is introduced in this work. GCIS is based on the induced suffix sorting algorithm SAIS, presented by Nong \etal in 2009. The proposed solution builds on the factorization performed by SAIS during suffix sorting.
	A context-free grammar is used to replace factors by non-terminals. The algorithm is then recursively applied on the shorter sequence of non-terminals.
	The resulting grammar is encoded by exploiting some redundancies, such as common prefixes between right-hands of rules,  sorted according to SAIS. GCIS excels for its low space and time required for compression while obtaining competitive compression ratios. Our experiments on regular and repetitive, moderate and very large texts, show that GCIS stands as a very convenient choice compared to well-known compressors such as Gzip, 7-Zip, and RePair, the gold standard in grammar compression.
	In exchange, GCIS is slow at decompressing. Yet, grammar compressors are more convenient than Lempel-Ziv compressors in that one can access text substrings directly in compressed form, without ever decompressing the text. We demonstrate that GCIS is an excellent candidate for this scenario because it shows to be competitive among its RePair based alternatives. We also show, how GCIS relation with SAIS makes it a good intermediate structure to build the suffix array and the LCP array during decompression of the text.
\end{abstract}

%
%


%
%


\maketitle

\section{Introduction}

Text compression is a fundamental task in Computer Science that consists in transforming an input string into another string whose bit sequence representation is smaller.
The suffix array~\cite{Manber1993,Gonnet1992} is a key data structure used to compute lossless compression transforms ~\cite{Ohlebusch2011,Karkkainen2013,Goto2014}, such as the Burrows-Wheeler transform (BWT)~\cite{Burrows1994} and the Lempel-Ziv factorization
(LZ77) \cite{Ziv1977,DBLP:journals/spe/LiuNCW16}, which are at the heart of the popular data compression tools like \bzip, \szip and \gzip.

The suffix array construction (or suffix sorting) may be performed in linear time (see~\citet{Puglisi2007} and \citet{Dhaliwal2012} for good reviews).
\citet{Nong2009a} introduced a remarkable suffix array construction algorithm called \SAIS, which runs in linear time and is fast in practice. Subsequently, \SAIS was adapted to directly compute the BWT~\cite{Okanohara2009}, the $\Phi$-array~\cite{Karkkainen2009,Goto2014}, the LCP array~\cite{Fischer2011}, and the suffix array for string collections~\cite{Louza2017c}.

Grammar compression \cite{KY00} is a compression technique based on finding a small context-free grammar that generates (only) the text. Finding the smallest such grammar is NP-hard \cite{CLLPPSS05}, but heuristics like \repair \cite{Larsson1999b} work very well in practice. Like Lempel-Ziv, grammar compression performs particularly well on repetitive text collections. An advantage of grammar compression is that text substrings can be extracted from the compressed representation without the need to decompress the text \cite{BLRSRW15}. One can then aim at never decompressing the text but work on it directly in compressed form.

This article introduces \our, a new grammar-based compression algorithm
that builds on  \SAIS framework.
\our  constructs  a context-free grammar recursively, based on the string factorization performed by \SAIS.
The rules are encoded exploiting the common prefixes between the right-hands of consecutive rules, which are sorted lexicographically by \SAIS. \our is the first grammar-compressor algorithm, as far as we know, based on induced suffix sorting.

Our experiments show that, on repetitive texts and compared to the best grammar-compressor  (\repair~\cite{Larsson1999b}) and the best Lempel-Ziv compressor (\szip~\cite{pavlov}), \our is an interesting alternative in practice because it displays the fastest compression speed and low memory usage, while reaching a
compression ratio close to that of \repair. We also show that \our is competitive with the relevant alternatives when processing regular (not highly repetitive) and very large texts.

In exchange, \our is slower than most alternatives for decompressing the text. As explained, however, we may aim at extracting any desired substring while never decompressing the whole text.
When compared to variants of \repair that allow extraction, \our turns out to be very efficient while using less space.
Further, it is possible to build the suffix and LCP arrays as a byproduct of text decompression by \our in  competitive time.

This work differs from its early version \cite{NunesLGAN18} in that we support efficient extraction of substrings and computation of suffix and LCP arrays directly from decompression. We also present more thorough descriptions and experimental results.



\section{Background}\label{s:background}

\subsection{Suffix array}

Let $T$ be a string of length $|T|=n$, over an ordered alphabet $\Sigma$. We assume that our alphabet $\Sigma$ has an integer size, but limited to $n$, that is, $1\leq |\Sigma| \leq n$. The concatenation of strings or symbols is denoted by the dot operator ($\conc$). The symbol $<$ is used for the lexicographic order relation between strings.

Let $T[i]$ be the $i$-th symbol of $T$. The substring (factor) from $T[i]$ to $T[j]$, both included, is denoted by $T[i,j]$, for $1 \le i \le j \le n$. A prefix of $T$ is a substring of the form $T[1,i]$ and a suffix is a substring of the form $T[i,n]$,  also denoted by $T_i$. For convenience, it is assumed that $T$ always ends with a special symbol $T[n]=\$$, which is not present elsewhere in $T$ and lexicographically precedes every symbol in $T[1,n-1]$.

	The suffix array (\SA)~\cite{Manber1993,Gonnet1992} of a string $T[1,n]$ is an array of integers in the range $[1, n]$ that gives the lexicographic order of all suffixes of $T$, such that $T_{\SA[1]} < T_{\SA[2]} < \ldots < T_{\SA[n]}$.
	The length of the longest common prefix (\LCP) of two strings $X$ and $Y$ in $\Sigma^*$ is denoted  $\lcp(X,Y)$.
	The \LCP array of $T[1,n]$ is an array of integers that stores the \lcp value between consecutive suffixes in \SA, such that $\LCP[i]=\lcp(T_{\SA[i-1]},T_{\SA[i]})$, for $1<i\leq n$, and we define $LCP[1]=0$.
	The suffixes starting with the same symbol $c \in \Sigma$ form a $c$-bucket in \SA. The head and the tail of a $c$-bucket refer to the first and the last position of the $c$-bucket in \SA.

	\subsection{Grammar compression}

	Let $G = (\Sigma, \Gamma, P, X_S)$ be a reduced context-free grammar (i.e,  with no unreachable non-terminals) such that $\Sigma$ is the terminal alphabet of $G$; $\Gamma$ is the set of non-terminal symbols (disjoint from $\Sigma$);  $P \subseteq \Gamma \times (\Sigma \cup \Gamma)^*$ is the set of production rules;  and $X_S \in \Gamma$ is the start symbol.

	A production rule $(X_i, \alpha_i)$ is also denoted $X_i \rightarrow \alpha_i$. In this case, it is said that $\alpha_i$ is derived from $X_i$. For strings $S,R \in (\Sigma \cup \Gamma)^*$,  if $R$ is obtained from $S$ by production rules in $P$, then $R$ is derived from $S$. When $R$ is obtained by a (possibly empty) sequence of derivations from $S$, then $R$ is generated from $S$.

	Given a string $T$, the grammar compression problem is to find a grammar $G$ that generates only $T$, such that $G$ can be represented in less space than the original $T$.
	Given that $G$ grammar-compresses $T$, for $(X_i,\alpha_i)\in P$, $\gen(X_i)=S$ is defined  as the single string $S\in\Sigma^*$ that is generated from $X_i$.
	The language generated by $G$ contains the unique string $\gen(X_S) = T$. This notion can be extended further for a string of terminals and non-terminals $S$, such that:

$$ \gen(S) = \{W_1 \cdot W_2 \cdot \ldots \cdot W_{|S|} \; \lvert \; W_k \in \gen(S[k]), 1\leq k \leq |S| \}. $$

	When each $S[i]$, $1\leq i \leq |S|$, generates a single sequence, the previous definition can be replaced by the concatenation of the strings generated by $S[i]$, $1\leq i \leq |S|$:

$$\gen(S) = \gen(S[1]) \cdot \gen(S[2]) \cdot \ldots \cdot \gen(S[|S|])  $$

\subsection{Integer Encoding}

We cover various techniques to encode sequences of integers, when most of them are expected to be small. Some of the techniques allow us to directly access any integer in the sequence.

\paragraph{Simple8b}

The {\tt Simple8b} scheme, proposed by \citet{AnhM10}, encodes a sequence of small integers in a $64$-bit word using the number of bits required by the largest integer. Basically, it identifies a word with a $4$-bit tag called \textit{selector}, which specifies the number of integers encoded in the rest of the word and the width of such integers. {\tt Simple8b} also has specific selectors for a run consisting of zeroes. If a run of $240$ or $120$ zeros is encountered, it can be represented with a single $64$-bit word.

\paragraph{Directly Addressable Codes}

The Directly Addressable Codes (DAC) proposed by \citet{Brisaboa2013} allow efficient retrieval of any given value $A[i]$ from an array of integers $A[1,n]$ while encoding such integers compactly. Let $l_i$ be the length (number of bits) of $A[i]$, then this encoding splits each $A[i]$ into $\lceil l_i/b\rceil$ blocks $v_{i,1}$, $v_{i,2}$,\ldots, $v_{i,k}$ of $b$ bits each. A bit $b_{i,j} = 1$ is associated with a block $v_{i,j}$ if $j<k$, that is, $v_{i,j}$ is not the last block of $A[i]$. Otherwise, $b_{i,j} = 0$. Then a layered data structure is constructed in such a way that the $k$-th layer contains two bitmaps: the first bitmap is the concatenation of blocks $v_{i,k}$, for every $1\leq i \leq n$, whereas the second bitmap is the concatenation of the bits $b_{i,k}$ associated with each block $v_{i,k}$.

To retrieve any given value $A[i]$, one must first recover $b_{i,1}$ and check if its value is zero. If so $A[i]$ equals $v_{i,1}$, otherwise, it is necessary to proceed recursively to the $j$-th entry of the next layer, where $j=\sum_{m=1}^{i} b_{m,1}$, and append the result of the recursive call to $v_{i,1}$.  The prefix sum value $j$ can be computed in constant time by using auxiliary data structures on top of the bitmaps.

\paragraph{Elias-Fano Encoding}

This format permits the encoding of a monotonically increasing sequence of $n$ integers over the interval $[0,m-1]$ within $2n+n\lceil \lg \frac{m}{n}\rceil$ bits and allows the retrieval of any integer of such sequence in constant time \cite{DBLP:conf/wsdm/Vigna13}.  Each integer $a_i$ is divided into two parts: $u_i$, the $\lceil \lg n \rceil$ most significant bits of $a_i$ and $l_i$, the $\lceil \lg \frac{m}{n}\rceil$ remaining bits of $a_i$. The $l_i$ values are concatenated in a single array of $n\lceil \lg \frac{m}{n}\rceil$ bits and each value $a_i$ is classified in one of the total of $n$ possible buckets. Then, the number of elements of each bucket is represented in a negated unary representation and such representations are concatenated in a bitmap $B$ of $2n$ bits, $n$ bits for each possible bucket and further $n$ bits for every element $a_i$. 

To retrieve the $i$-th value of the sequence of integers, one just needs to search for the position $k$ of the $i$-th one bit on $B$, and append $l_i$ to the binary representation of  $k-i$.  The position $k$ can be retrieved in constant time using auxiliary data structures on top of $B$.

\section{SAIS: Induced Suffix Sorting}\label{s:related}

\SAIS~\cite{Nong2009a} builds on the induced suffix sorting technique introduced by previous algorithms~\cite{Itoh1999,Ko2003}. Induced suffix sorting consists in deducing the order of unsorted suffixes from a (smaller) set of already ordered suffixes.

The next definition classifies suffixes and symbols of  strings.

\begin{definition}[L-type and S-type]
	For any string $T$, $T_n=\$$ has type S. A suffix $T_i$ is an S-suffix if $T_i < T_{i+1}$, otherwise $T_i$ is an L-suffix. Each symbol $T[i]$ has the type of $T_i$.
\end{definition}

The suffixes can be classified in linear time by scanning $T$ once from right to left, so that the type of each suffix is stored in a bitmap of size $n$. 

Note that, within a $c$-bucket, the L-suffixes precede the S-suffixes.

Further, the classification of suffixes is refined as follows:

\begin{definition}[LMS-type]
	Let $T$ be a string. Then $T_i$ is an LMS-suffix if $T_i$ is an S-suffix and $T_{i-1}$ is an L-suffix.
\end{definition}

\citet{Nong2009a} showed that the order of the LMS-suffixes is enough to induce the order of all suffixes.
This is the basis of the \SAIS algorithm.

\subsection{SAIS framework}

\begin{enumerate}

	\item Sort the LMS-suffixes. This step is explained later.

	\item Insert the LMS-suffixes into the tail of their respective $c$-buckets in $\SA[1,n]$, without changing
	      their order. Now $\SA$ contains the LMS-suffixes positions, in sorted order, on the end of each $c$-bucket. The remaining values of \SA are initialized with a sentinel $\bot$ value.

	\item Induce L-suffixes by scanning $\SA[1,n]$ from left to right: for each suffix
	      $\SA[i]\neq \bot$, if $T[\SA[i]-1]$ is L-type, insert $\SA[i]-1$ into the
	      head of its $c$-bucket. 

	\item Induce S-suffixes by scanning $\SA[1,n]$ from right to left: for each suffix
	      $\SA[i]\neq \bot$, if $T[\SA[i]-1]$ is S-type, insert $\SA[i]-1$ into the
	      tail of its $c$-bucket. 

\end{enumerate}

Whenever a value is inserted in the head (or tail) of a $c$-bucket, the pointer to the head (or tail) is increased (or decreased) by one.



In order to sort the LMS-suffixes in Step 1, $T[1,n]$ is divided (factorized)
into LMS-substrings.

\begin{definition}\label{def:lms-type}
	$T[i,j]$ is an LMS-substring if both $T_i$ and $T_j$ are LMS-suffixes,
	but no suffix between $i$ and $j$ has LMS-type.
	The last suffix $T_n$ is an LMS-substring.
\end{definition}

Let $r^1_1, r^1_2, \dots, r^1_{n^1}$ be the $n^1$ LMS-substrings of $T$ read from left to right.
A modified version of \SAIS is applied to sort the LMS-substrings.
Starting from Step 2, $T[1,n]$ is scanned (right-to-left) and each new LMS-suffix starting with $c$ is
inserted (bucket-sorted) at the tail of its $c$-bucket.
Steps 3 and 4 work exactly the same.
At the end, the  LMS-substrings are sorted and the beginning positions of each LMS-substring are stored in their corresponding
$c$-buckets in \SA.

\subsection{Naming}

A {\it name} $v^1_i$ is assigned to each LMS-substring $r^1_i$ according to its
lexicographical rank in $[1, \sigma^1]$, such that $v^1_i \le v^1_j$ iff $r^1_i \le r^1_j$, where $\sigma^1$ is the number of different
LMS-substrings in $T$.
In order to compute the names, 
each consecutive LMS-substrings in \SA, say $r^1_i$ and $r^1_{i+1}$, are compared
to determine if either $r^1_i = r^1_{i+1}$ or $r^1_i < r^1_{i+1}$.
In the former case $v^1_{i+1}$ is set to $v^1_i$, whereas in the latter case $v^1_{i+1}$ is set to $v^1_i+1$.
This procedure may be sped up by comparing the LMS-substrings first by symbol and then by
type, with L-type symbols being smaller than S-type symbols in case of
ties~\cite{Nong2011}.


\subsection{Recursive call}

A new (reduced) string $T^1 = v^1_1 \conc v^1_1 \cdots v^1_{n^1}$ is created, whose
length ${n^1}$ is at most $n/2$, and the alphabet size $\sigma^1$ is integer.
If every $v^1_i\ne v^1_j$ then all LMS-suffixes are already sorted.
Otherwise, \SAIS is recursively applied to sort all the suffixes of $T^1$.
Nong \etal~\cite{Nong2009a} showed that the relative order of the LMS-suffixes
in $T$ and the order of the respective suffixes in $T^1$ are the same.
Therefore, the order of all LMS-suffixes can be determined by the result of the
recursive algorithm.

%

\section{\our: Grammar Compression by Induced Suffix Sorting}\label{s:algorithm}

This section introduces the Grammar Compression algorithm by Induced Sorting
(\our).

\subsection{Compressing}
\label{subsec:compressing}


First, a context-free grammar $G = (\Sigma, \Gamma, P, X_S)$ that generates only $T[1,n]$ is computed. To do this \SAIS is modified as follows.

Consider the $j$-th recursion level. In Step 1, after the input string $T^j[1,n^{j}]$ is divided into the LMS-substrings $r^j_1, r^j_2, \dots, r^j_{n^{j+1}}$ and named $v^j_1, v^j_2, \dots, v^j_{n^{j+1}}$,  a new rule $v^j_i \rightarrow r^j_i$ is created for each different LMS-substring $r^j_i$. Moreover, an additional rule $v_0^j \rightarrow v_0^{j-1}T^j[1,j_1 -1]$ if $j>0$ or $v_0^j \rightarrow T[1,j_1-1]$ if $j=0$, with $j_1$ standing for the index of the leftmost LMS-type suffix of $T^j$, is created for the prefix of $T^j$ that is not included in any LMS-substring. In this context, when $j=0$, $n^0 = n$ and $T^0=T$.

The algorithm is then called recursively with the reduced string $T^{j+1} = v^j_1~\conc~v^j_2~\cdots~v^j_{n^{j+1}}$ as input as long as $\sigma^{j+1}<n^{j+1}$, that is, the LMS-substrings are not pairwise distinct. At the end, when $\sigma^\ell=n^\ell$, the last recursion level $j=\ell$ is reached, and the start symbol of $X_S$ of $G$ is created so that the initial production $X_S \rightarrow v^\ell_0 \conc v^\ell_1 \conc v^\ell_2 \cdots v^\ell_{n^\ell}$ generates the original string $T[1,n]$.

The algorithm stops after computing $X_S$, since we are not interested in constructing the suffix array; Steps 2, 3 and 4 of \SAIS are not executed.  The recursive calls return to the top level and a grammar $G$
that generates only $T[1,n]$ has been computed.

\begin{figure}[ht]
	\centering
	\includegraphics[width=\textwidth]{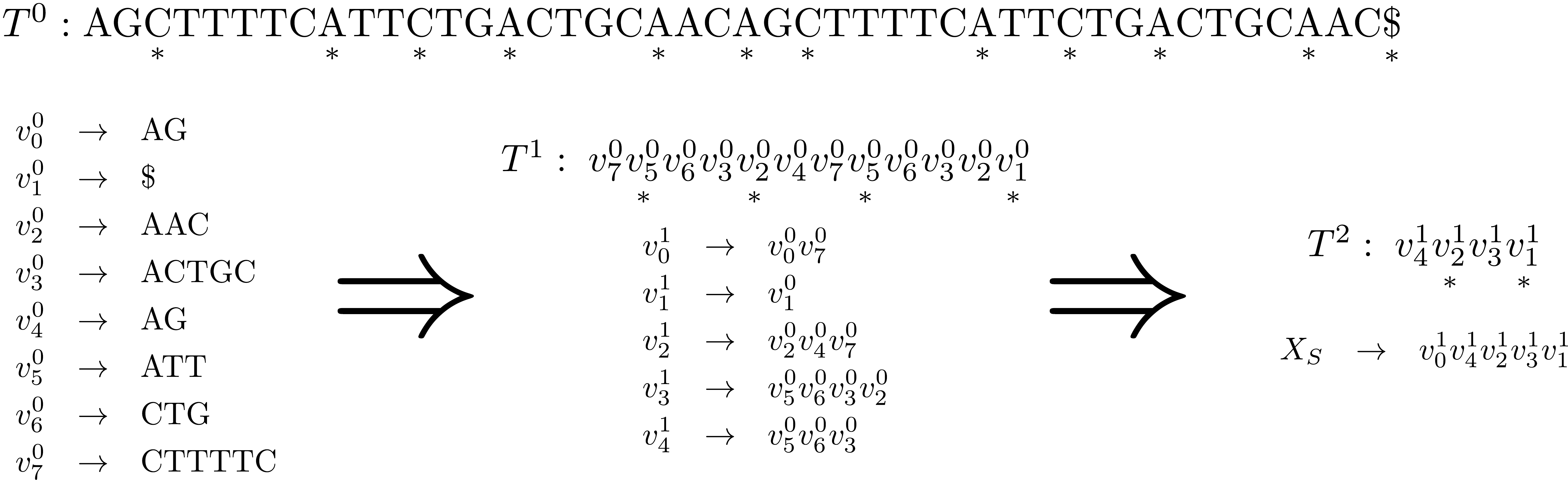}
	\caption{Grammar construction during \our. All LMS-substrings (those starting with a `$*$' symbol), are sorted according to \SAIS framework, and then rules $v^0_0\rightarrow \mathrm{AG}, v^0_1\rightarrow \mathrm{\$}, \ldots, v^0_7 \rightarrow CTTTTC$ are created. Next, $T^1$ is obtained by replacing every LMS-substring by the left-hand side of its rule. The procedure is applied recursively to $T^1$. When $T^2$ is created, the alphabet size is equal to $|T^2|=n^2$, and thus the starting rule $X_S$ that generates $T^0$ is obtained.}
	\label{fig:gcis-1}
\end{figure}

Since for each LMS-substring there is a unique $v^j_i$, there are no cycles in any generation. Further, there is only one path of derivations that from a string $S$ generates a string $S'$. The consequence of this deterministic choice, for every derivation, is that $\gen(X_i)$, for  $X_i \in \Gamma$, is a fixed string of terminals. Figure \ref{fig:gcis-1} shows the grammar construction on \our.

\paragraph{Grammar compression}

Consecutive entries in the set of productions $P$ are likely to share a common
prefix, since the LMS-substrings are given lexicographically ordered by \SAIS.
Therefore, each rule $X_i \rightarrow \alpha_i \in P$ is encoded using two
values $(l_i, \suff(\alpha_i))$, such that $l_i$ encodes \lcp$(\alpha_{i-1},\alpha_{i})$, and the
remaining symbols of $\alpha_{i}$ are given by
$\suff(\alpha_i)=\alpha_i[l_i+1,|\alpha_i|]$.
For each starting rule $v^j_0$, we define $l_i=0$.
This technique is known as Front-coding~\cite{Witten1999}.

The computation of $(l_i, \suff(\alpha_i))$ is performed with no additional
cost with a slight modification in the naming procedure of \SAIS.
Each consecutive LMS-substring in \SA, say $r^j_{i-1}$ and $r^j_{i}$, are compared first
by symbol until a mismatch is found, and then compared by type, to check if either $r^j_{i-1} = r^j_{i}$ or $r^j_{i-1} < r_{i}$.
The symbol-wise comparison reveals $\lcp(r^j_{i-1},r^j_{i})$ as well, so the resulting complexity is the same with a small slowdown in the running time.

\paragraph{Time complexity}
\our runs in $O(n)$ time, since each step of the modified \SAIS runs in linear time and
the length of the reduced string $T^j$ is at most $|T^{j-1}|/2$.

\paragraph{Implementation details}\label{s:implementation}

%

Each non-terminal $X_i$ is represented by a pair $\alpha_i = (l_i,\suff(\alpha_i))$, as explained.
The $l_i$ values tend to be small
and, considering the $j$-th recursion value, the sum of such values cannot be
greater than $n^j$, since no two LMS-substrings overlap by more than one symbol.

One can encode all $l_i$ values by using the {\tt Simple8b} encoding in an integer array $W$.
All strings $\suff(\alpha_{i})$ are encoded in a single fixed-width integer array $Y$,
of cell width $\lfloor\lg(\sigma^j)\rfloor+1$ bits. The length of
each $\suff(\alpha_{i})$ is also encoded using {\tt Simple8b} into a word
array $Z$. The same observation of the $\lcp$ sum can be done here: the sum
of all $|\suff(\alpha_{i})|$ on the $j$-th recursion level is no larger than $n^j$.

%
%

\subsection{Decompressing}\label{s:decompression}

The decoding process is done level-wise, starting from the last recursion
level $j=\ell$, by decoding the right side of each rule.
At the end, $T$ is decoded from $T^1$.

In the $j$-th recursion level, the values $(w,y,z)$
from $W$, $Y$ and $Z$, the data structures mentioned in the implementation details of Section \ref{subsec:compressing}, are decoded sequentially.
In order to obtain the right-hand side of the production rules $\alpha_{k+1}$ from $\alpha_k$, the first $w$ symbols of
$\alpha_k$ are copied to $\alpha_{k+1}$ and the $z$ symbols from $Y$, which
correspond to the string $y$, are appended to $\alpha_{k+1}$. After this process the plain representation of each rule is stored, in a single array of cells with fixed width
$\lfloor \lg(\sigma^j)\rfloor+1$ bits. An additional array of pointers $D$ is also created to find the starting position of each rule in this fixed-width array.

With the fixed-width array and the array of pointers $D$,  $T^{j-1}$ now can be decoded from $T^j$.  First, the
right side of $v_0^j$ is copied into $T^{j-1}$. Then, $T^j$ is scanned in a
left-to-right fashion and for each $T^j[i]$ the algorithm appends to $T^{j-1}$ the right-hand side of the non-terminal $T^j[i]$, which can be easily found with support of array $D$ in constant time.

\paragraph{Time complexity} The whole decompression process takes $O(n)$ time.


\subsection{Extracting substrings}\label{s:extract}

In order to support extraction of substrings from the compressed text, it is necessary to augment the dictionary with two additional data structures: $P_S$, a partial-sum on the lengths of the symbols in the reduced string $T^\ell$ of the last recursion level, and $L$, a data structure that for each non-terminal $X_i$ stores $|\gen(X_i)|$. Formally, those data structures are defined as:

\begin{equation*}
	P_S(i) ~~=~~ \sum_{j=1}^{i-1} |\gen(X_{S_i})|,\quad X_S \rightarrow X_{S_1},\ldots X_{S_k} \text{ and } \quad 1\leq i \leq k +1
\end{equation*}

\begin{equation*}
	L(X) ~~=~~ |\gen(X)|, \quad X \in \Sigma \cup \Gamma
\end{equation*}

The data structure $L$ can also be defined recursively as:

\begin{equation*}
	L(X) = \left\{
	\begin{array}{l}
		1, \quad X \in \Sigma \\
		\displaystyle{\sum_{i=1}^{|S|} L(S[i])},\quad  X \rightarrow S
	\end{array}
	\right.
\end{equation*}

To obtain a substring $T[l,r]$, we then proceed as follows:

\begin{enumerate}
	\item
	      With a binary search, locate \textit{indices} $a$ and $b$ from $P_S$ such that:
	      \begin{eqnarray*}
		      & a = \max\{ 1\leq k \leq |T^\ell|\; \lvert \; P_S(k) \leq l \}\\
		      & b = \min \{ 1 \leq k \leq |T^\ell|  + 1\;\lvert \; P_S(k)> r\} -1
	      \end{eqnarray*}

	\item
	      Let $\ell$ be the number of levels in \our grammar and $S$ the string derived from $X_S$. Then define $E^\ell = S[a,b]$ and follow the next steps for $i=\ell$ to $i=1$.

	\item
	      Apply a derivation step to each non-terminal $X\in E^i$ to obtain a new string $E^{i-1}$. Note that $\gen(E^{i-1}) = T[l',r']$ is a superstring of $T[l,r]$.

	\item
	      Trim $E^{i-1}$ from the left and right as much as possible as long as it generates a superstring of $T[l,r]$. This can be done efficiently because we know the lenght of $\gen(X)$, for every $X\in \Sigma \cup \Gamma$.

	      \begin{enumerate}
		      \item
		            If $i=1$, then $E^{0}$ contains only terminal symbols and generates a superstring $T[l',r'] = E^{0}$ of $T[l,r]$. Thus, one simply  extracts the symbols $E^{0}[l-l'+1,r-l'+1]$ to obtain $T[l,r]$.

		      \item
		            If $i>1$, then $E^{i-1}$ contains only non-terminal symbols and generates a superstring $T[l',r'] =  \gen(E^{i-1}) $ of $T[l,r]$. We then trim $E^{i-1}$ by using $L$ and finding, with a linear search, two indices $a$ and $b$ of $E^{i-1}$ such that:

		            \begin{eqnarray*}
			            a = \max\left\{1\leq k \leq |E^{i}| \; \left \lvert  \; l' + \sum_{j=1}^{k-1} L(E^{i}[j])  \leq l \right.\right\}\\[2mm]
			            b = \max\left\{1\leq k \leq |E^{i}| \; \left \lvert  \; r' - \sum_{j=k+1}^{|E^{i}|} L(E^{i}[j])  \geq r  \right.\right\} \\
		            \end{eqnarray*}

		            $E^{i-1}$ is then trimmed to $E^{i-1}[a,b]$ before proceeding.
	      \end{enumerate}
\end{enumerate}

\begin{figure}[ht]
	\centering
	\includegraphics[scale=.5]{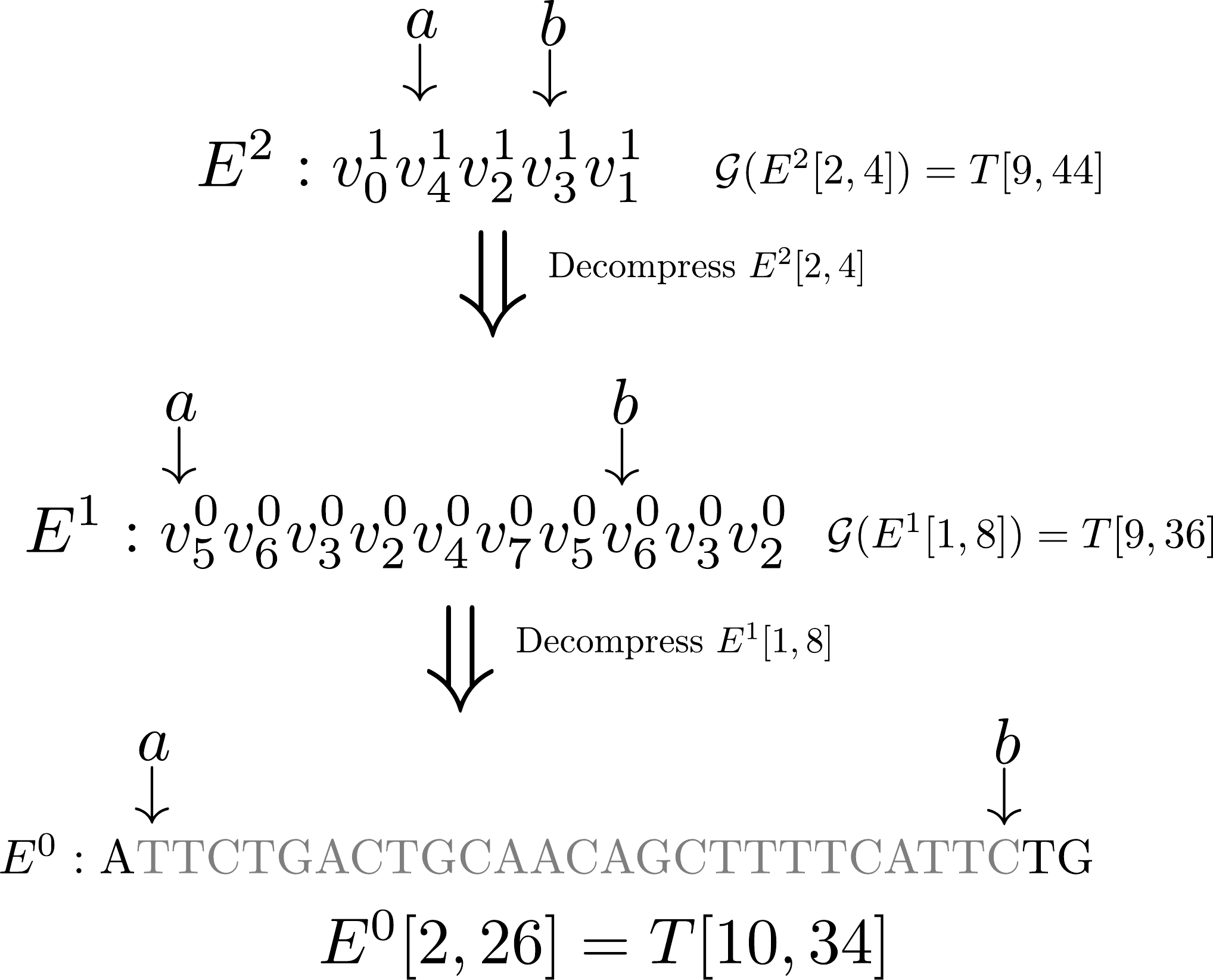}
	\caption{Extraction of the substring $T[10,34]$ of the text of Figure \ref{fig:gcis-1}. Initially, a binary search is performed on $P_S$ to identify the substring of $E^2$ that shall be decompressed: $E^2[2,4]$, which generates $T[9,43]$, is decompressed to obtain $E^1$. A linear scan is performed in both ends considering the length of the terminals generated by each rule of $E^1$ to find the indexes $a=2$ and $b=8$. $E^1[2,8]$ is then decompressed and $E^0 = T[9,36]$ is obtained, which makes possible to extract $T[10,34]$ by simply ignoring both ends.}
	\label{fig:extraction-example}
\end{figure}

Figure \ref{fig:extraction-example} shows an example for extracting a text using the aforementioned procedure.

\paragraph{Implementation details}\label{parag:implementation-extraction}

Since the length of the string $T^{\ell}$ is much shorter than the original text in practice, the verbatim representation of $P_S$ as an array of integers is affordable.

The array $L$ is represented using DACs. This representation allows efficient access while representing the data in a compact way.

To support fast extraction, we need to efficiently decompress a single rule. {\tt Simple8b} encoding works very well when the objective is compressing or decompressing since all the rules are first expanded sequentially in the decompressing stage. However, when the aim is to extract symbols, we need to expand individual rules. Thus, instead of encoding all the \lcp values with the {\tt simple8b} scheme, Elias-Fano encoding is employed, allowing us to retrieve a random \lcp value of a rule efficiently and hence the decoding of a random rule. The length of each $\suff(\alpha_{i})$ is also encoded using Elias-Fano and the $\suff(\alpha_i)$ values are encoded in a fixed-width integer array. Since the \lcp values are front-encoded, we force that every $k$-th \lcp value is set to $0$, with $k\in O(1)$. This setting does not have a significant impact on compression and ensures that we have to backtrack a constant number of rules to extract an individual rule prefix.

\subsection{Suffix array construction}\label{s:saca}

The suffix array (\SA) construction boils down to sorting all suffixes of $T$.
Although \our compression does not sort suffixes, it executes Step~1
of \SAIS and the production rules created correspond to the LMS-substrings already
sorted, which is used by \SAIS for sorting all suffixes. We show next how
to modify our decompression algorithm for building \SA as a
byproduct with, asymptotically,  no additional overhead.

First, when $j=\ell$, $X_S \rightarrow T^\ell$ is taken, and its suffix array $\SA^{j}$ is built directly as $\SA^j[T^j[i]]=i$.  \citet{Nong2009a} observed that $\SA^j$ also gives the order of all
LMS-suffixes of string $T^{j-1}$.
Then, $T^{j-1}$ is decoded (Section~\ref{s:decompression}), and Steps 2, 3 and
4 of \SAIS (Section~\ref{s:related}) are executed to obtain $\SA^{j-1}$, and so
on.
The algorithm proceeds for $j=\ell-1, \dots, 1$, obtaining the reduced string
$T^{j-1}$ together with $\SA^{j-1}$ at each iteration.
At the end, the original string $T$ is decoded from $T^1$ and \SA is induced
from $\SA^{1}$.

\paragraph{Time Complexity}
\SA is built in $O(n)$ time, since each step of \SAIS is linear and the length
of all reduced strings is $O(n)$.

\subsection{LCP array construction}\label{s:saca-lcp}

When $|\Sigma| \in O(1)$, the longest common prefix (\LCP) array can also be computed in linear time within the induced
suffix sorting framework~\cite{Fischer2011, Louza2017a}.
We show below how to modify our decompression algorithm to compute \SA and \LCP together with, asymptotically,  no additional cost.

When $j=1$, the original string $T$ is decoded from $T^1$, and
$\SA^{1}$ stores the order of all LMS-suffixes of $T$.
Then, in linear time, we compute the \LCP array of the LMS-suffixes using a sparse variant of $\Phi$-algorithm by~\citet{Karkkainen2009}, which avoids storing auxiliary arrays by reusing the space of $\SA[n/2,n]$ and $\LCP[n/2,n]$. The \lcp-values between the LMS-suffixes are used to induce the \lcp-values between the L-suffixes during Step 3 (Section~\ref{s:related}), and these are used to induce the S-suffixes during Step 4 (see~\citet{Louza2017a} for details). Given an additional stack of $O(\sigma \log n)$ bits~\cite{Gog2011b}, each \lcp-value induction is done in $O(\sigma)$ time, which we assume to be constant at the top recursion level.

Therefore, at the end  we have computed \SA and \LCP as a byproduct of \our decompression.

\paragraph{Time complexity}
\SA and \LCP are built in $O(n)$ time, since each step of \SAIS is linear and the \LCP-values induction can be done in constant time.

\section{Experiments}\label{s:experiments}


To confirm the practical value of \our, we conducted experiments using several corpora. We measured compression and decompression speed, compression ratio, and memory usage during compression and decompression of \our against classical and grammar-based compressors (Section \ref{subsec:codec}), evaluated the extraction of symbols (Section \ref{subsec:extract}) and showed the efficiency of suffix array construction of \our during the decompression (Section \ref{subsec:saca}).

In the following subsections we describe and discuss the experimental setup and the results.

\subsection{Texts}
\label{subsec:text}

Regular texts were taken from the corpora {\tt large-corpus} \cite{large-corpus}, {\tt enwiki} \cite{enwiki-corpus}, {\tt manzini} \cite{manzini-corpus}, {\tt pizza-chili} \cite{pizza-chili-corpus} and {\tt silesia} \cite{silesia-corpus}.
Repetitive texts were chosen from {\tt pizza-chili-repetitive} corpus \cite{pizza-chili-repetitive-corpus}.
Very large inputs were built by repeating and mutating strings such as {\tt chr19} \cite{chr19} , {\tt sars-cov} \cite{covid19} and {\tt salmonella} \cite{salmonela} with a mutation rate of $0.1\%$, thus these texts are highly repetitive as well; each filename has an integer suffix that represents the number of repetitions. In addition, a $20$GB prefix from November 2019 Wikipedia dump was taken \cite{enwiki-dump}. Tables \ref{tab:regular-texts}, \ref{tab:repetitive-texts} and \ref{tab:huge-texts}   summarize the chosen texts and their size, grouping in boxes texts from the same corpus.

\begin{minipage}[t]{0.25 \textwidth}
	\begin{table}[H]
		\caption{Regular texts.}
		\renewrobustcmd{\bfseries}{\fontseries{b}\selectfont}
		\sisetup{detect-weight,mode=text}
		\begin{tabular}{|l|S|}
			\hline
			{Regular text}    & {Size (MB)} \\ \hline \hline
			archive           & 27.07       \\
			{emacs}           & 47.46       \\
			{linux}           & 47.60       \\
			{samba}           & 41.58       \\
			{spamfile}        & 84.22       \\\hline \hline
			{enwiki8}         & 100.00      \\
			{enwiki9}         & 1000.00     \\\hline \hline
			{chr22}           & 34.55       \\
			{etext99}         & 105.28      \\
			{gcc-3.0.tar}     & 86.83       \\
			{howto}           & 39.42       \\
			{jdk13c}          & 69.73       \\
			{linux-2.4.5.tar} & 116.25      \\
			{rctail96}        & 114.71      \\
			{rfc}             & 116.42      \\
			{sprot34.dat}     & 109.62      \\
			{w3c2}            & 104.20      \\\hline \hline
			{dblp.xml}        & 296.14      \\
			{dna}             & 403.93      \\
			{english}         & 2210.40     \\
			{pitches}         & 55.83       \\
			{sources}         & 210.87      \\\hline \hline
			{dickens}         & 10.19       \\
			{mozilla}         & 51.22       \\
			{mr}              & 9.97        \\
			{nci}             & 33.55       \\
			{oofice}          & 6.15        \\
			{osdb}            & 10.09       \\
			{reymont}         & 6.63        \\
			{samba}           & 21.61       \\
			{sao}             & 7.25        \\
			{webster}         & 41.46       \\
			{xray}            & 8.457        \\
			{xml}             & 5.35        \\\hline
		\end{tabular}
		\label{tab:regular-texts}
	\end{table}
\end{minipage}\hspace{1mm}
\begin{minipage}[t]{0.28 \textwidth}
	\begin{table}[H]
		\caption{Repetitive texts.}
		\renewrobustcmd{\bfseries}{\fontseries{b}\selectfont}
		\sisetup{detect-weight,mode=text}
		\begin{tabular}{|l|S|}
			\hline
			{Repetitive text} & {Size (MB)} \\\hline
			cere              & 461.29      \\
			coreutils         & 205.28      \\
			dblp.xml.00001.1  & 104.86      \\
			dblp.xml.00001.2  & 104.86      \\
			dblp.xml.0001.1   & 104.86      \\
			dblp.xml.0001.2   & 104.86      \\
			dna.001.1         & 104.86      \\
			einstein.de.txt   & 92.76       \\
			einstein.en.txt   & 467.63      \\
			english.001.2     & 104.86      \\
			Escherichia\_Coli & 112.69      \\
			influenza         & 154.81      \\
			kernel            & 257.96      \\
			para              & 429.27      \\
			proteins.001.1    & 104.86      \\
			sources.001.2     & 104.86      \\
			world\_leaders    & 46.97       \\\hline
		\end{tabular}
		\label{tab:repetitive-texts}
	\end{table}
\end{minipage} \hspace{-12mm}
\begin{minipage}[t]{0.5\textwidth}
	\begin{table}[H]
		\caption{Very large texts.}
		\renewrobustcmd{\bfseries}{\fontseries{b}\selectfont}
		\sisetup{detect-weight,mode=text}
		\begin{tabular}{|l|S|}
			\hline
			{Very large text}   & {Size (MB)} \\\hline
			c050                & 2956.45     \\
			c100                & 5912.90     \\
			c150                & 8869.35     \\
			c200                & 11825.80    \\
			c250                & 14782.25    \\
			c300                & 17738.69    \\
			c350                & 20695.14    \\\hline \hline
			sars-cov100000      & 2990.30     \\
			sars-cov200000      & 5980.60     \\
			sars-cov300000      & 8970.90     \\
			sars-cov400000      & 11961.20    \\
			sars-cov500000      & 14951.50    \\
			sars-cov600000      & 17941.80    \\
			sars-cov700000      & 20932.10    \\\hline	\hline
			enwiki-20191120-20G & 20000.00    \\\hline \hline
			salmonella1000      & 4928.40     \\
			salmonella2000      & 9856.80     \\
			salmonella3000      & 14785.20    \\
			salmonella4000      & 19713.60    \\\hline
		\end{tabular}
		\label{tab:huge-texts}
	\end{table}
\end{minipage}

\subsection{Compressors and Extractors}
\label{subsec:compressors}

To evaluate \our in compression speed, decompression speed and compression ratio, we chose the well-known compressors \gzip \cite{Gailly2003}, \bzip \cite{Seward1997}, \szip \cite{pavlov}; the statistical compressor \ppmdj \cite{ppmdj}; the grammar compressor \repair \cite{WanRepair}, the Lempel-Ziv approximation for very large texts \relz \cite{VKNP20}, and the \repair approximation for very large texts \bigrepair \cite{GIMNST19}.


Regarding extraction of symbols, we compared \our with different encodings of \repair grammars that allow fast extraction. These encodings can be represented in a more straight-forward way, storing $\gen(X)$, for $X\in \Sigma \cup \Gamma$, or in a more elaborated way, creating succinct tree data structures that replace the original grammar encoding while allowing one obtain the right-hand side of any rule, as described by \citet{DBLP:conf/dcc/MaruyamaT14}. The implementation of such data structures was based on the work of \citet{GagieSLP} and can be found in \cite{TomohiroI}. We used the following encodings:

\begin{itemize}
	\item {\tt PlainSlp\_32Fblc}: uses $32$-bit integers for the array representations.
	\item {\tt PlainSlp\_FblcFblc}: employs the minimum bit length required to represent the maximum value of a given integer array.
	\item {\tt PlainSlp\_IblcFblc}: uses roughly $\lceil \lg i \rceil$ bits to represent the $i$-th rule exploiting that the $i$-th rule is less than $i$. For representing $\gen(X)$, for $X\in \Sigma \cup \Gamma$, it uses the same strategy of {\tt PlainSlp\_FblcFblc}.
	\item {\tt PoSlp\_Iblc}: employs the approach POSLP of \citet{DBLP:conf/dcc/MaruyamaT14} to represent the parse tree and encodes the leaves using roughly $\lceil \lg i \rceil$ bits for the $i$-th rule.
	\item {\tt PoSlp\_Sd}: applies the POSLP approach of \citet{DBLP:conf/dcc/MaruyamaT14} to represent the parse tree and encodes the leaves with Elias-Fano.
\end{itemize}

In order to assess the computation of suffix and LCP arrays directly from decompression, \our was compared with efficient suffix and \LCP construction algorithms implemented by {\tt sais-lite} \cite{sais-lite,sais-lite-lcp} and {\tt divsufsort} \cite{divsufsort,divsufsort-lcp}.

\our source code and a detailed description of the processed data are  available at \url{https://github.com/danielsaad/gcis}.

\subsection{Environment Setup}
\label{subsec:environment}

Due to memory capacity and availability, we conducted the experiments in two machines, one for the regular and repetitive corpora and another for the very large datasets. Their specifications follow:

Machine \#1, used for regular and repetitive texts:
\begin{itemize}
	\item CPU: {\tt 2x Intel(R) Xeon(R) CPU E5-2640 v3 @ 2.60GHz} CPUs;
	\item RAM Memory: $64$GB;
	\item Operating System: {\tt Centos7}, kernel version {\tt 3.10}.
\end{itemize}

Machine \#2, used for very large datasets:
\begin{itemize}
	\item CPU: {\tt 2x Intel(R) Xeon(R) E5-2630 v3 @ 2.40GHz};
	\item RAM Memory: $386$GB;
	\item Operating System: {\tt Debian GNU/Linux 8}, kernel version {\tt 3.16}.
\end{itemize}

We compiled \our, \repair (and its extractors), \bigrepair, \ppmdj and \relz under {\tt gcc} with {\tt -O3 -NDEBUG} flags. The default command line parameters of \gzip, \bzip, \ppmdj, \relz were used on the experiments. A dictionary size of $1$ GB was used in \szip. \bigrepair RAM usage was limited to $10$ GB.

\our was implemented in {\tt C++11} using the Succinct Data Structure Library (SDSL) version 2.0 \cite{gbmp2014sea}.

\subsection{Compression and decompression}
\label{subsec:codec}

We evaluated all compressors in terms of compression ratio, compression and decompression speed. We also considered their peak memory usage during compression and decompression. \bigrepair could not compress some texts, so its corresponding data in the graphs are missing. Decompression in \relz is not implemented, nonetheless, \relz serves as a compression benchmark since it approximates the Lempel-Ziv parse.

It is important to remark that \bigrepair does not produce a compact representation of rules, since it represents the right-hand side of its rules with $2$ integers (all the rules are of length two).  However, we optimized it by  representing each rule with at most $\lceil \log_2 r \rceil$ bits, $r$ being the number of rules, and integrating the non-terminals that  occur only once in their corresponding  right-hand side. This saves $\lceil \log_2 r\rceil$ bits for each eliminated non-terminal.

For very large texts, only \szip, \our, \relz and \bigrepair were evaluated, since they are the best choices for repetitive data. \repair was not considered because its data structures do not fit in main memory on such texts.

\paragraph{Compression ratio}
\label{par:compression-ratio}

It stands for the ratio between the compressed and the original text size, and it is given as percentage.

\begin{figure}[t!]
	\centering
	\includegraphics[width=1.25\textwidth,angle=-90]{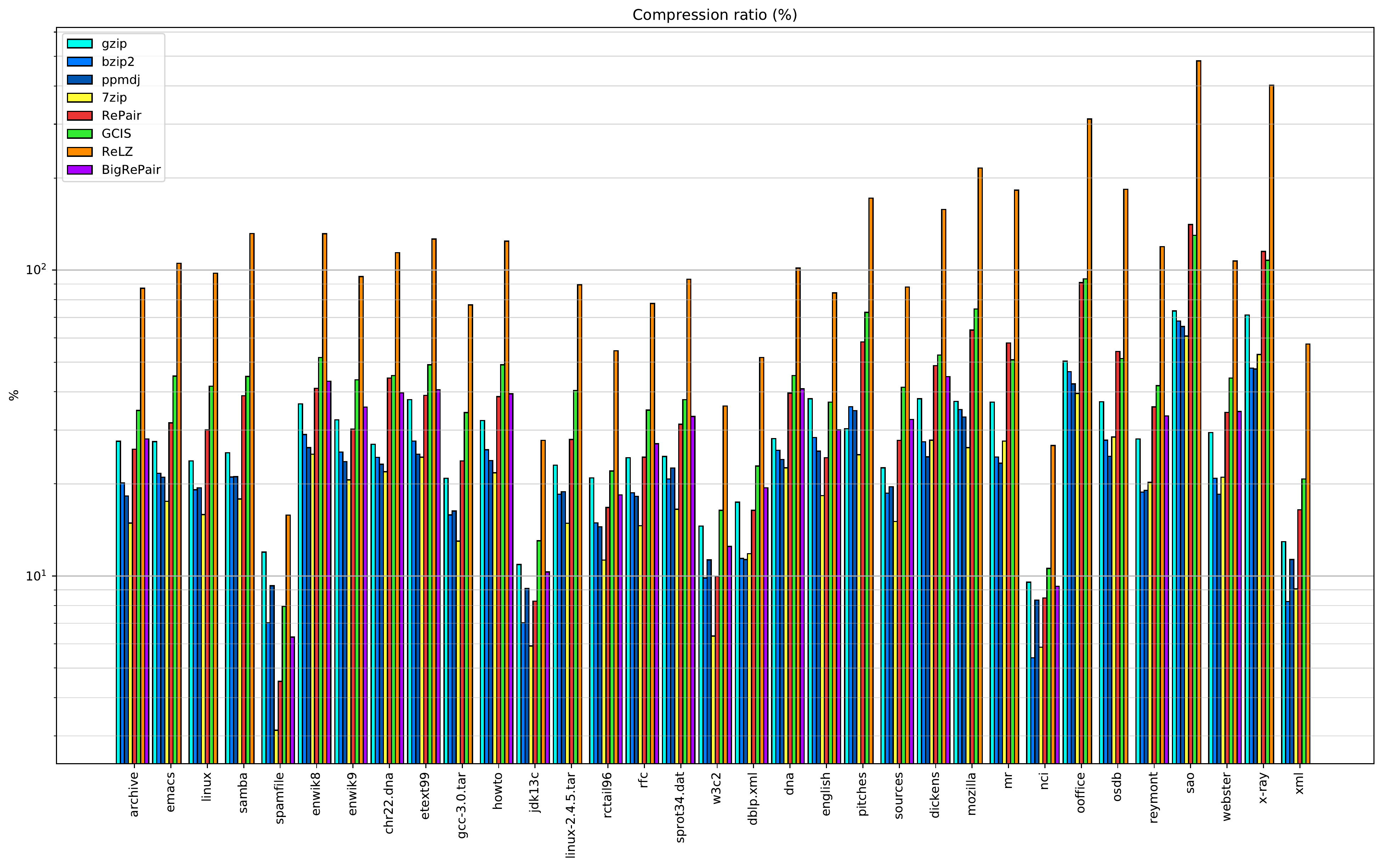}
	\caption{Compression ratio on regular texts.}
	\label{fig:compression-ratio-regular-texts}
\end{figure}

Figure \ref{fig:compression-ratio-regular-texts} shows that \repair outperforms \our and the approximations designed for very large texts, and it is competitive with a basic Lempel-Ziv compressor such as \gzip. However, it is clearly outperformed by \bzip, \ppmdj and \szip. The latter displays the best compression ratio overall, being \ppmdj a close competitor in some cases.

\begin{figure}[t!]
	\centering
	\includegraphics[width=0.95\textwidth]{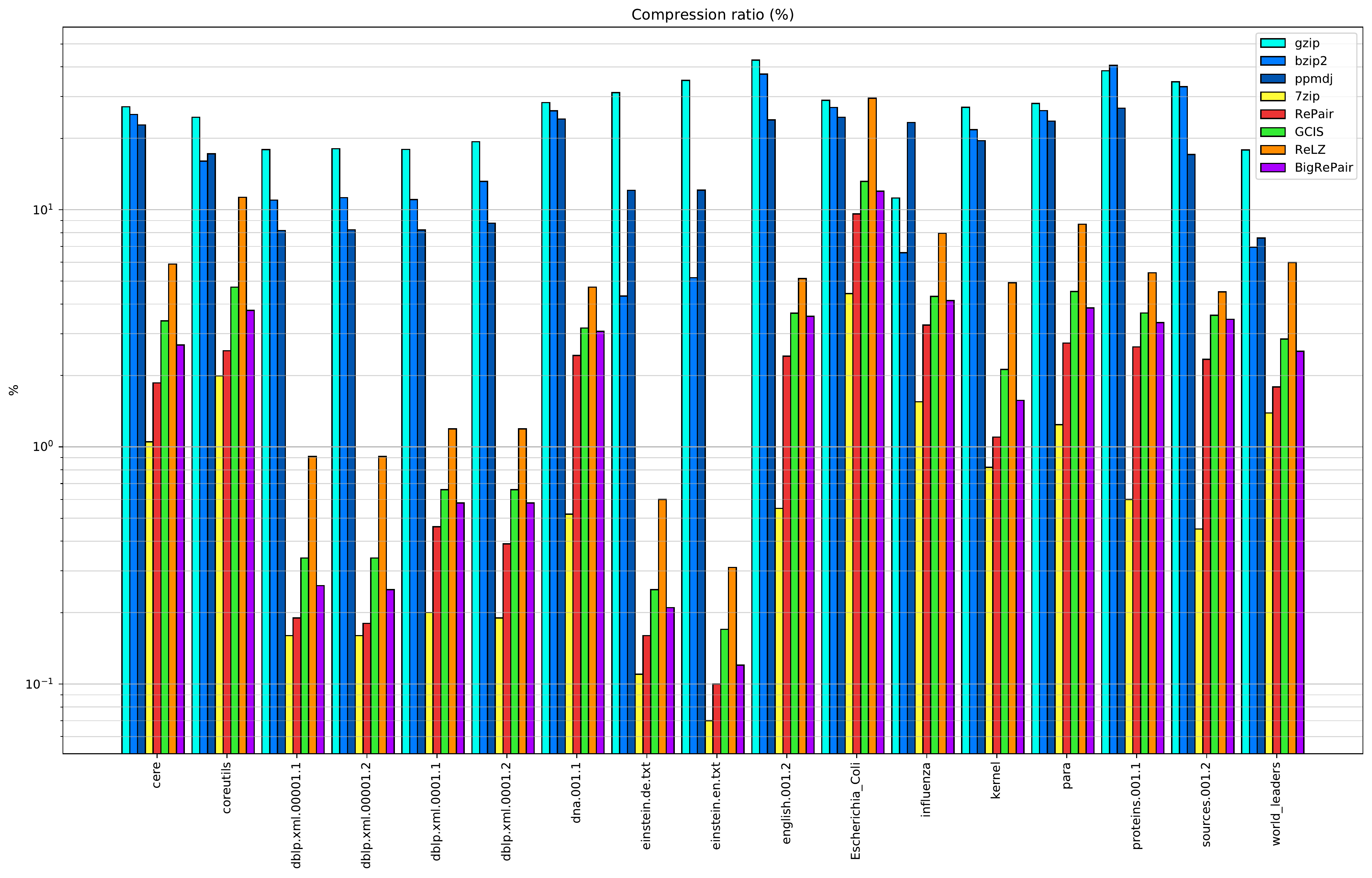}

	\vspace*{-5mm}
	\caption{Compression ratio on repetitive texts.}
	\label{fig:compression-ratio-repetitive-texts}
\end{figure}

Figure \ref{fig:compression-ratio-repetitive-texts}  shows the compression ratios for the repetitive corpora. The compressors that exploit repetitiveness obtain much better compression ratios this time, whereas \gzip, \bzip and \ppmdj obtain similar compression ratios as on the regular texts. In particular, \szip obtains the best compression ratio in all cases, closely followed by \repair. The compression ratio of \our is about twice that of \repair in most cases, but it is still very good in absolute terms and outperforms \relz.
\bigrepair obtains a compression ratio between those of \repair and \our.

\begin{figure}[t!]
	\centering
	\includegraphics[width=0.95\textwidth]{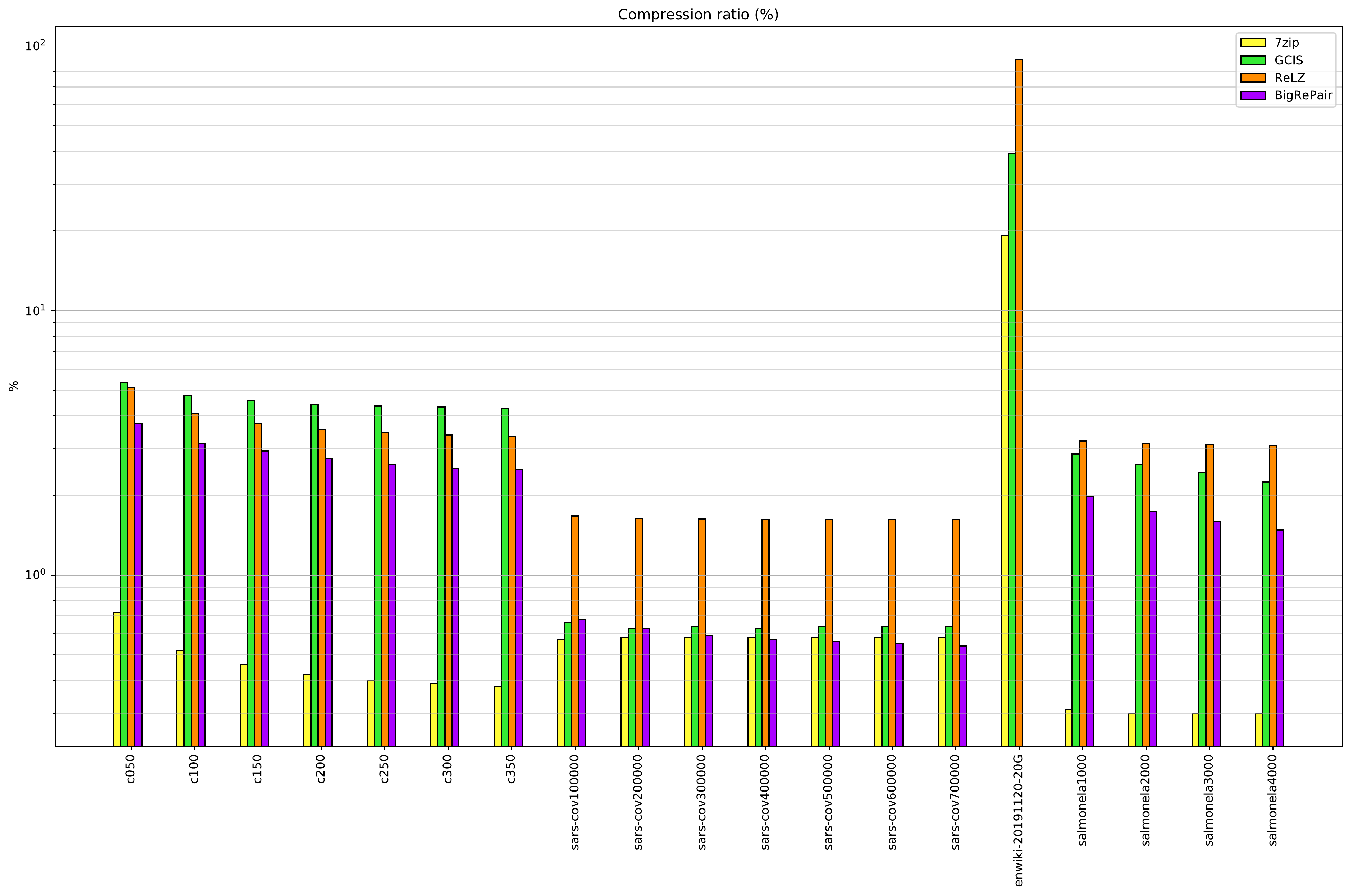}

	\vspace*{-5mm}
	\caption{Compression ratio on very large texts.}
	\label{fig:compression-ratio-huge-texts}
\end{figure}

The results for the very large texts are depicted in Figure \ref{fig:compression-ratio-huge-texts}. \szip is better for the text {\tt enwiki-20191120-20G}. The situation stays as in the smaller repetitive files: \bigrepair compresses more than \our, and \our compresses better than \relz, except for the the chromosome 19 based texts.

\paragraph{Compression speed}
\label{par:compression-speed}

\begin{figure}[t!]
	\centering
	\includegraphics[width=1.25\textwidth,angle=-90]{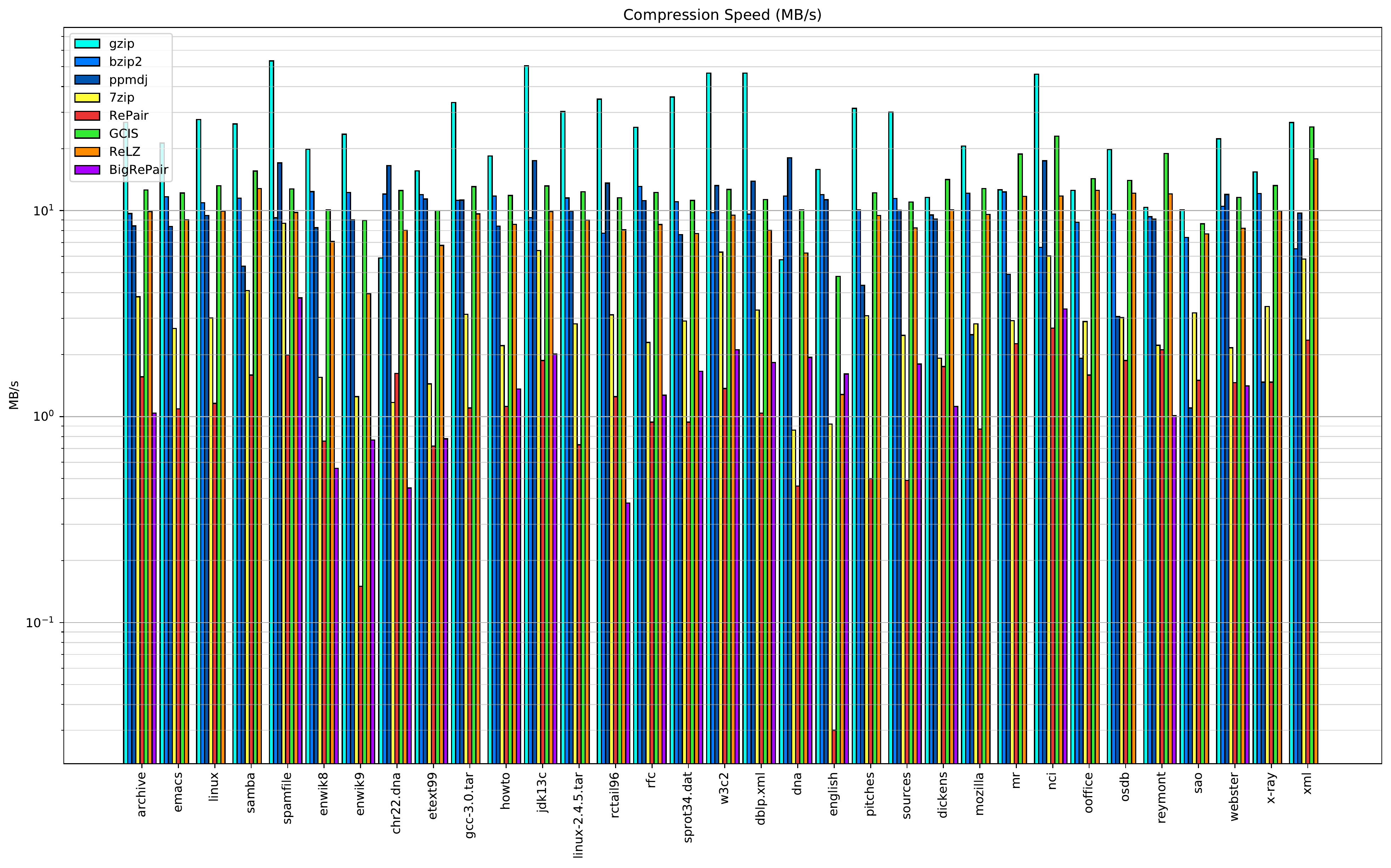}
	\caption{Compression speed on regular texts.}
	\label{fig:compression-speed-regular-texts}
\end{figure}

\begin{figure}[t!]
	\centering
	\includegraphics[width=0.95\textwidth]{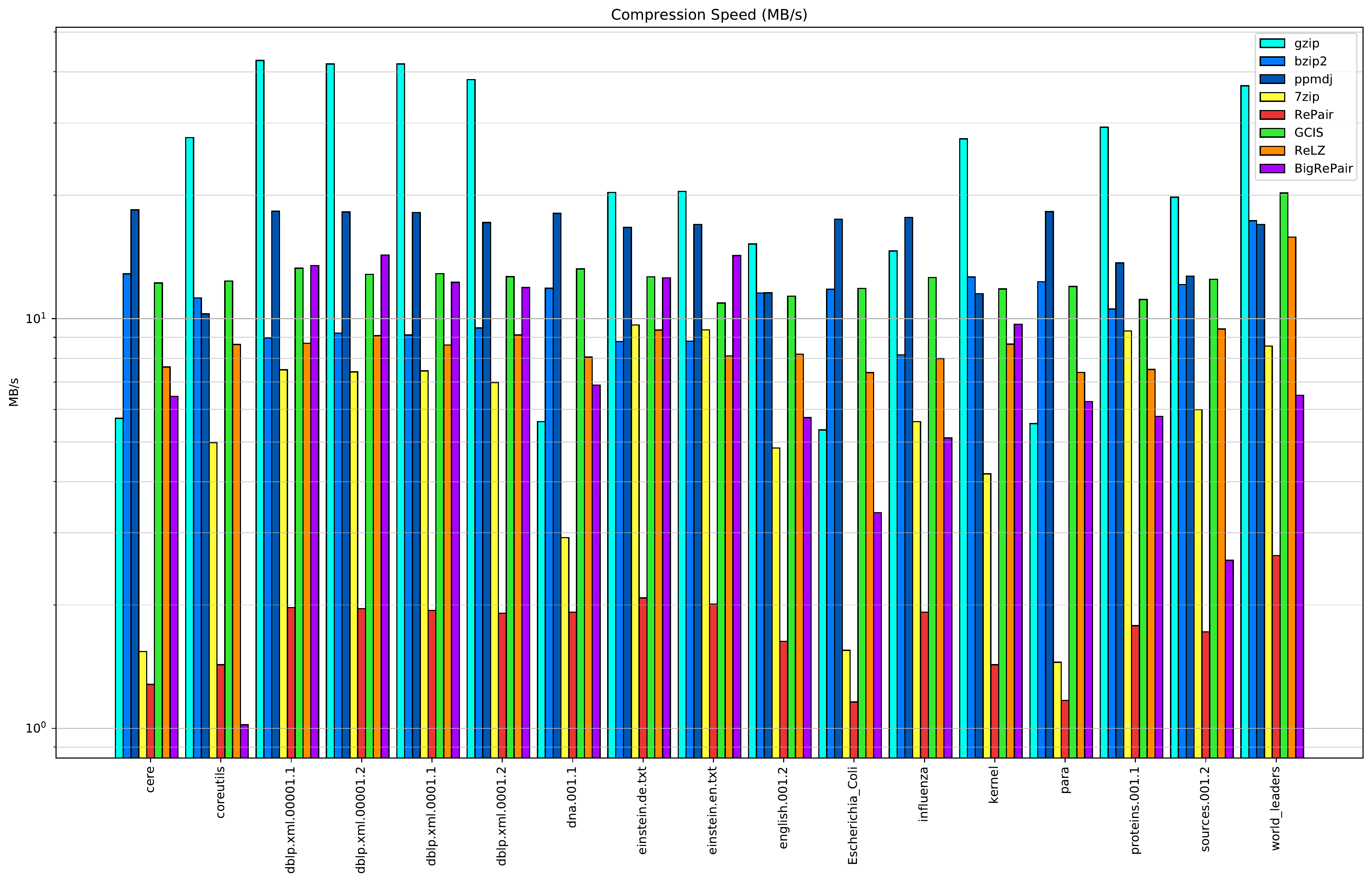}

	\vspace*{-5mm}
	\caption{Compression speed on repetitive texts.}
	\label{fig:compression-speed-repetitive-texts}
\end{figure}

\begin{figure}[t!]
	\centering
	\includegraphics[width=0.95\textwidth]{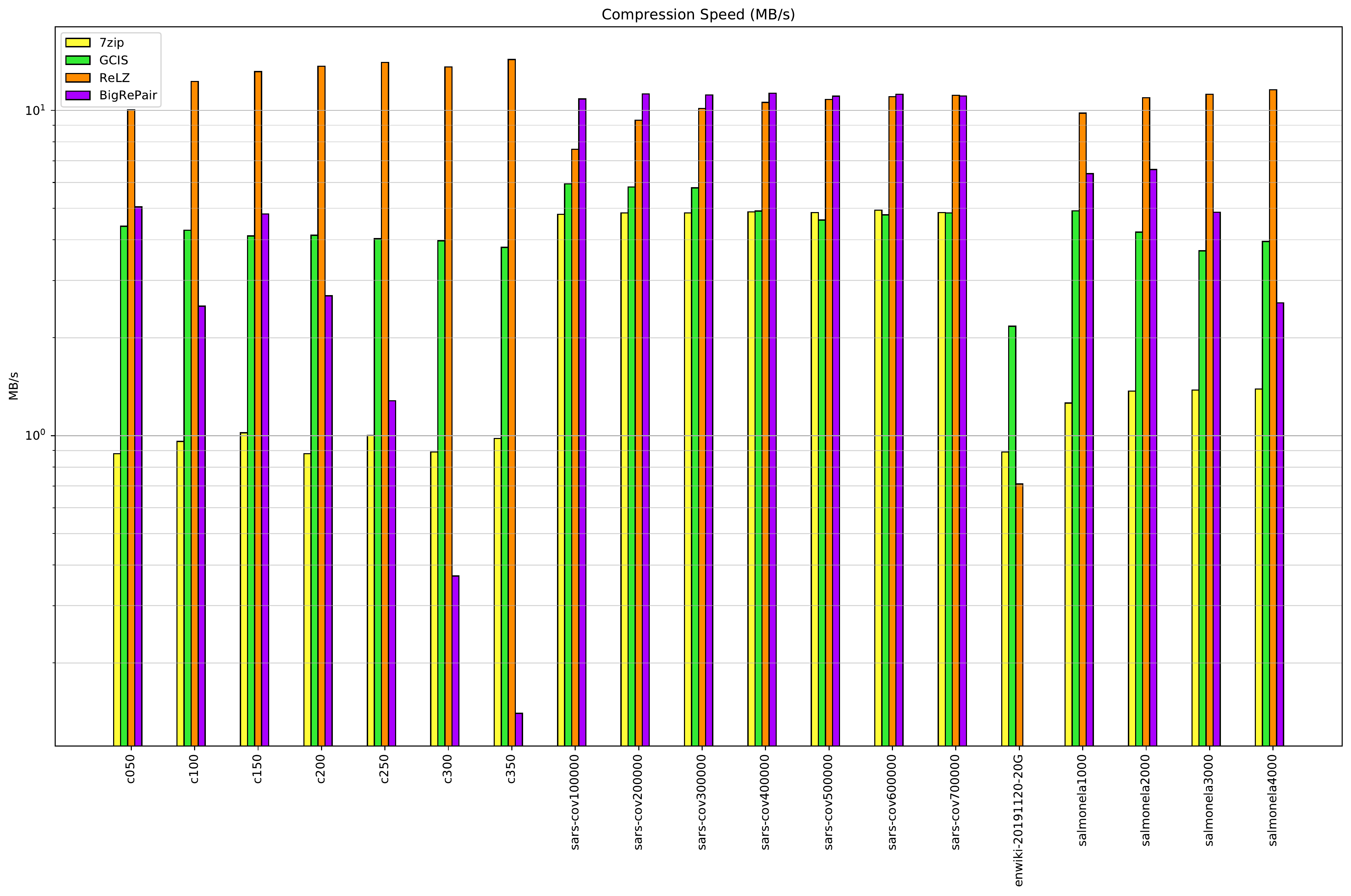}

	\vspace*{-5mm}
	\caption{Compression speed on very large texts.}
	\label{fig:compression-speed-huge-texts}
\end{figure}

Figures \ref{fig:compression-speed-regular-texts},\ref{fig:compression-speed-repetitive-texts} and \ref{fig:compression-speed-huge-texts} show the compression speed, in MB/s, of the compressors for each text type.

\gzip is the fastest compressor in most regular texts. \our is the second-fastest compressor, followed by \ppmdj. \our outperforms \bzip and \relz and is faster than the others by a wide margin. In particular, \our is typically an order of magnitude faster than the other grammar compressors (\repair and \bigrepair), which are its direct competitors.

Considering repetitive texts, \our is still faster than \relz and \szip (and often faster than \bzip); it is also orders of magnitude faster than \repair.
\gzip is still generally the fastest, but its compression ratio is unacceptable for repetitive data. \bigrepair is also slower than \our in most cases and, in the exceptions where \bigrepair is faster, it outperforms \our by a small margin.

For very large texts, \our is much faster than \szip, which becomes the slowest of the considered compressors. However, \relz and \bigrepair become much faster than \our,  as expected from being designed for this scenario. \relz is clearly the fastest compressor, though its compression ratio is the worst. A problem for \our on these very large files is that, once the text exceeds 2 GiB, it needs to use 64-bit integers, which doubles the memory requirements. \bigrepair and \relz do not suffer from this problem and require a small amount of main memory during compression.

\paragraph{Decompression speed}

\begin{figure}[t!]
	\centering
	\includegraphics[width=1.25\textwidth,angle=-90]{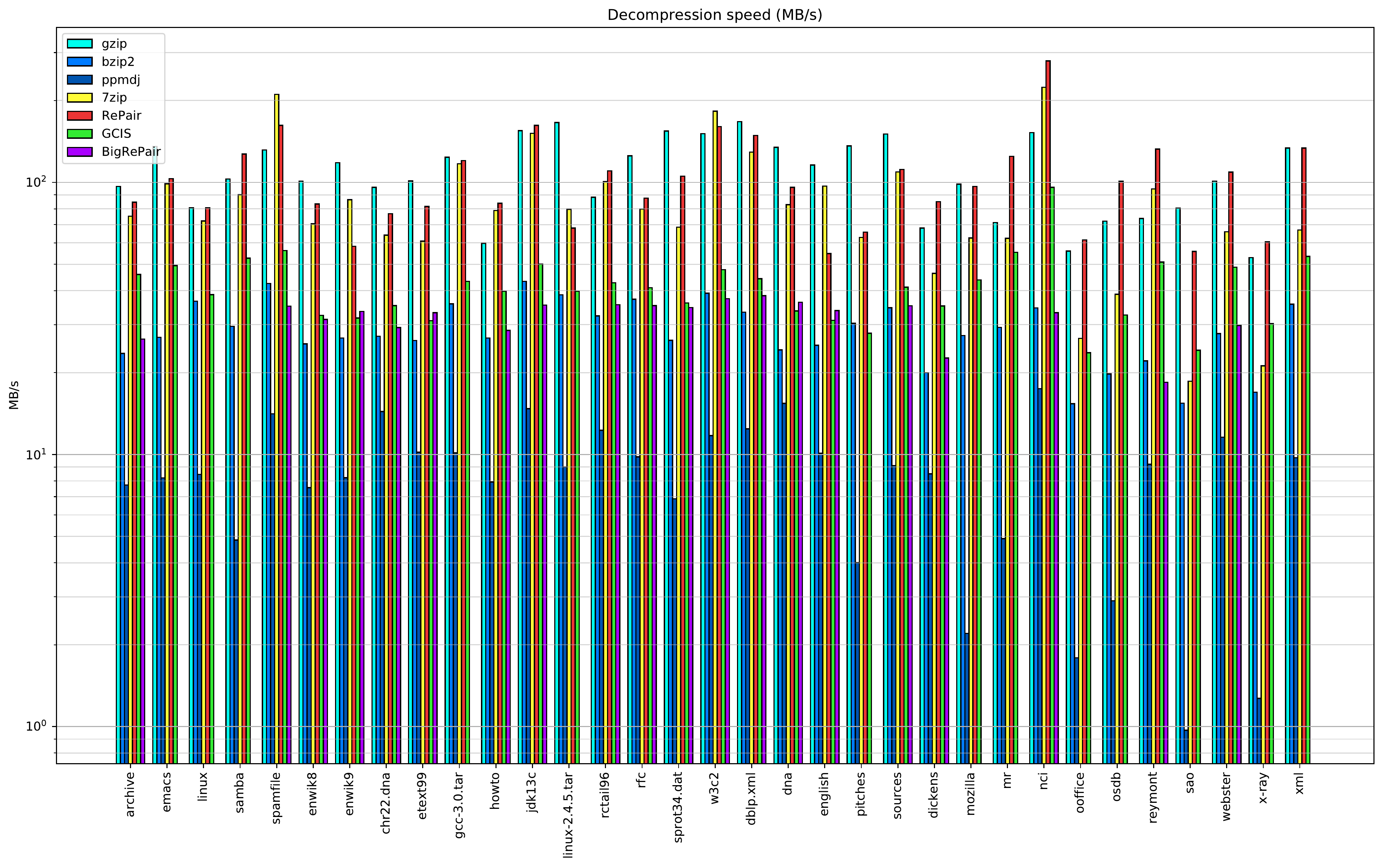}
	\caption{Decompression speed on regular texts.}
	\label{fig:decompression-speed-regular-texts}
\end{figure}

Figure \ref{fig:decompression-speed-regular-texts} depicts the results for regular texts. \gzip and \repair are the fastest at decompressing, followed by \szip and \our. \bzip and \bigrepair are the slowest decompressors. 

\begin{figure}[t!]
	\centering
	\includegraphics[width=0.95\textwidth]{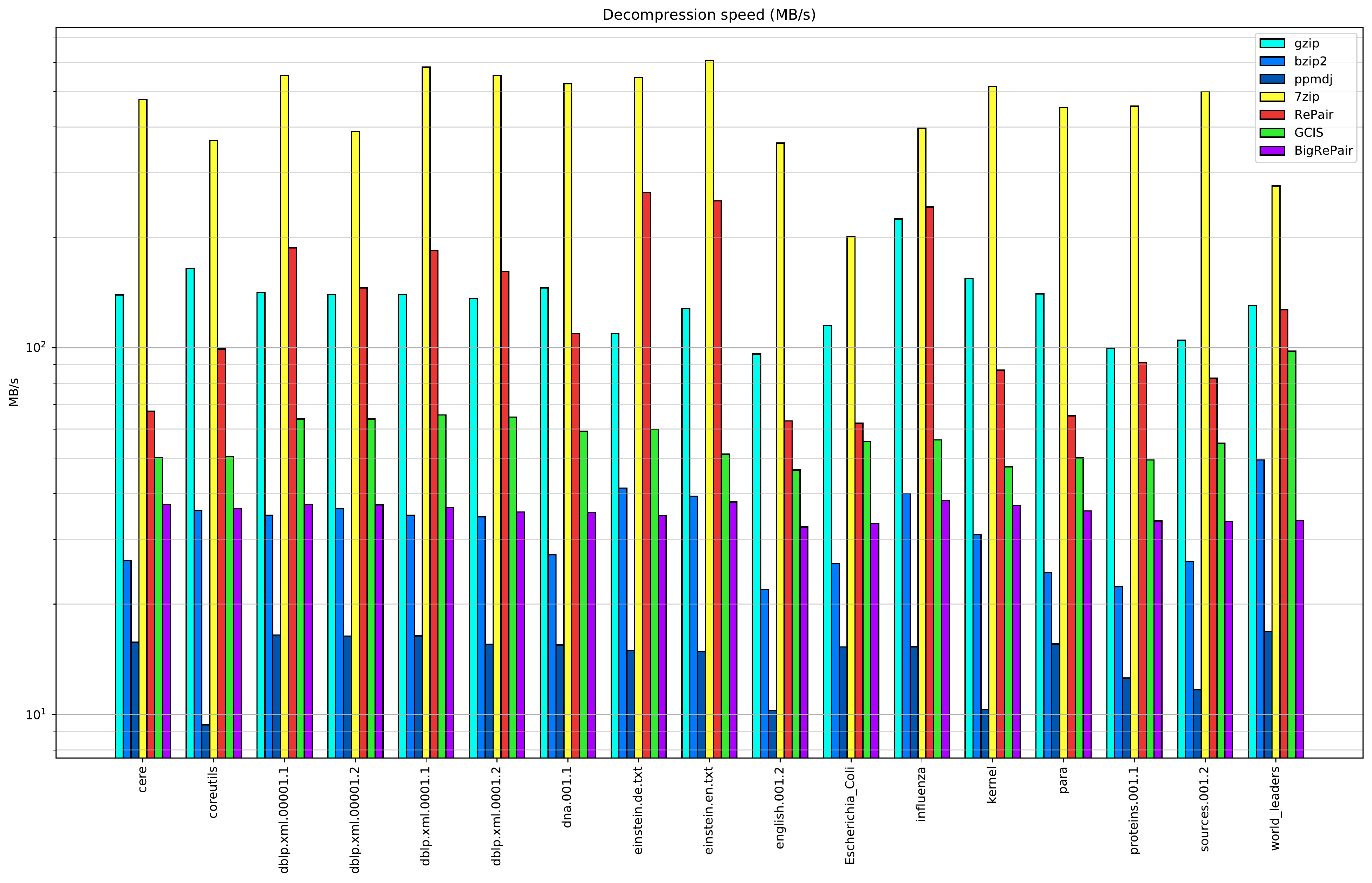}

	\vspace*{-5mm}
	\caption{Decompression speed on repetitive texts.}
	\label{fig:decompression-speed-repetitive-texts}
\end{figure}

Figure \ref{fig:decompression-speed-repetitive-texts} shows that the situation is similar on repetitive texts, except that  \szip becomes way faster than the others in almost all cases.
Despite the relative differences, in absolute terms \our is still fast, decompressing the files in around 5 MB/s.

\begin{figure}[t!]
	\centering
	\includegraphics[width=0.95\textwidth]{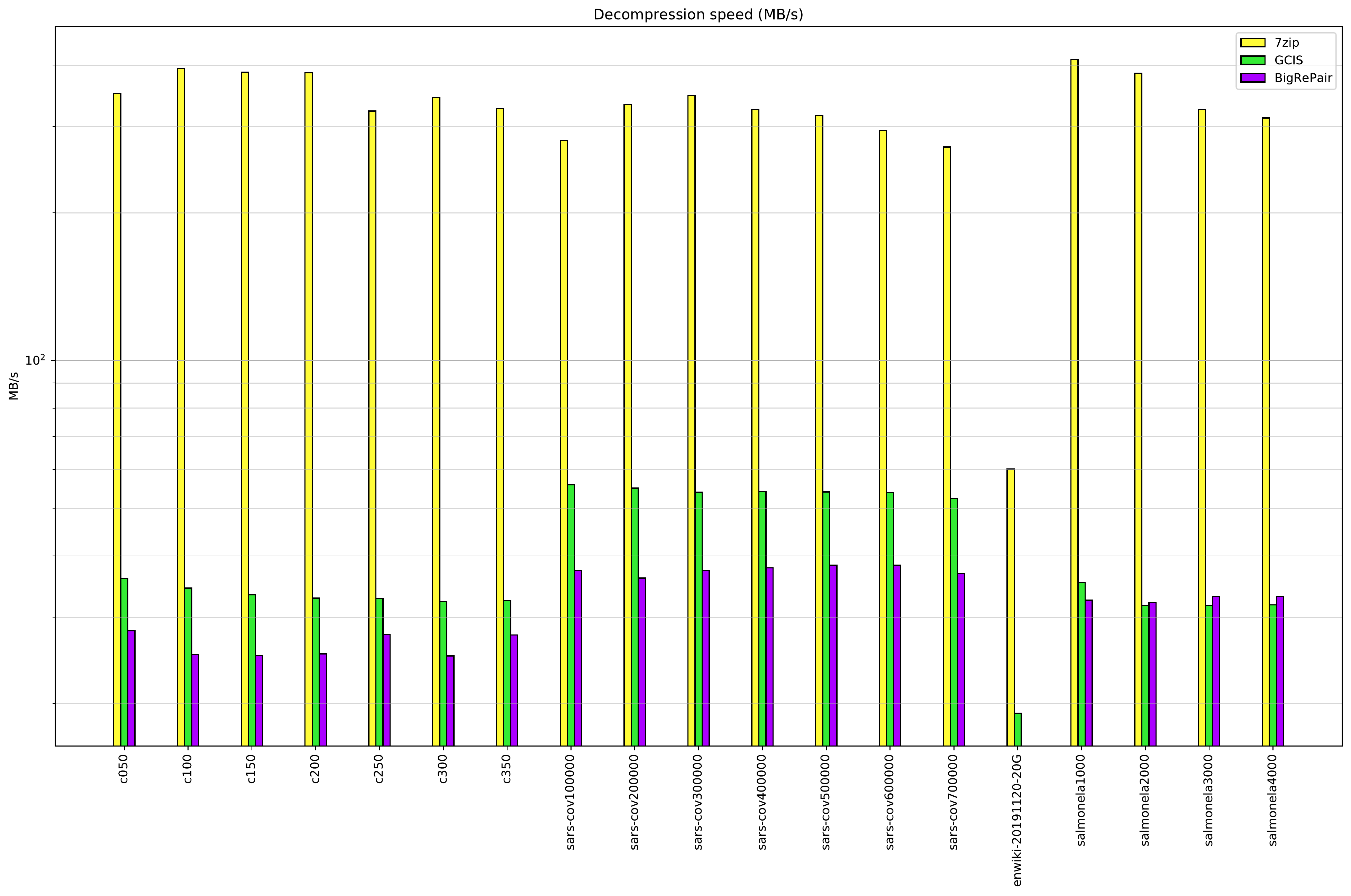}

	\vspace*{-5mm}
	\caption{Decompression speed on very large texts.}
	\label{fig:decompression-speed-huge-texts}
\end{figure}

The situation stays the same on the compressors that run on very large texts: as shown in Figure \ref{fig:decompression-speed-huge-texts}, \szip is the fastest, followed by \our and then by \bigrepair. 

\paragraph{Peak memory}
\label{sec:peak-memory}

\begin{figure}[t!]
	\centering
	\includegraphics[width=1.25\textwidth,angle=-90]{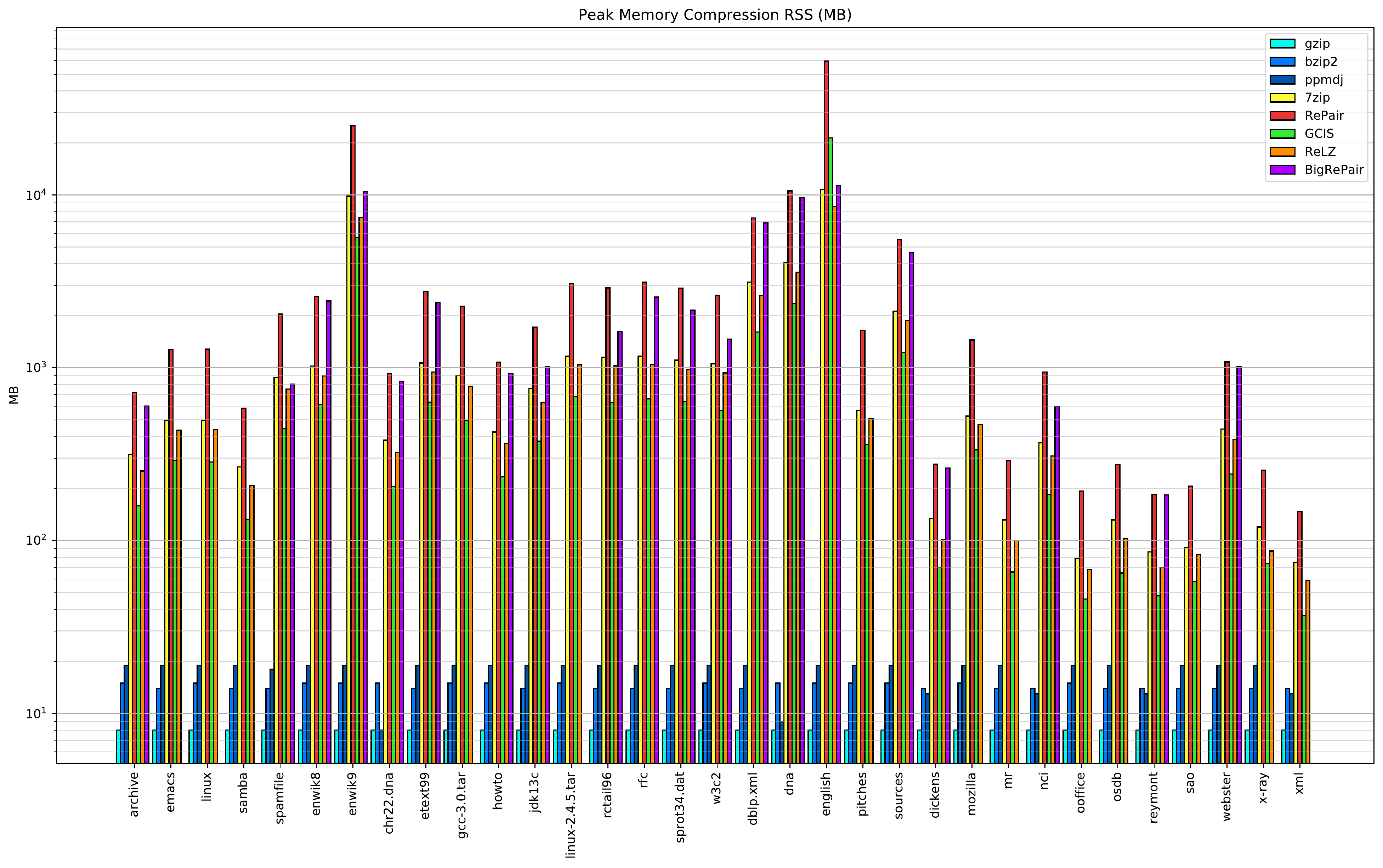}
	\caption{Peak memory (in MB) used by the compressors during compression for regular texts.}
	\label{fig:mprof-compress-regular-texts}
\end{figure}

\begin{figure}[t!]
	\centering
	\includegraphics[width=0.95\textwidth]{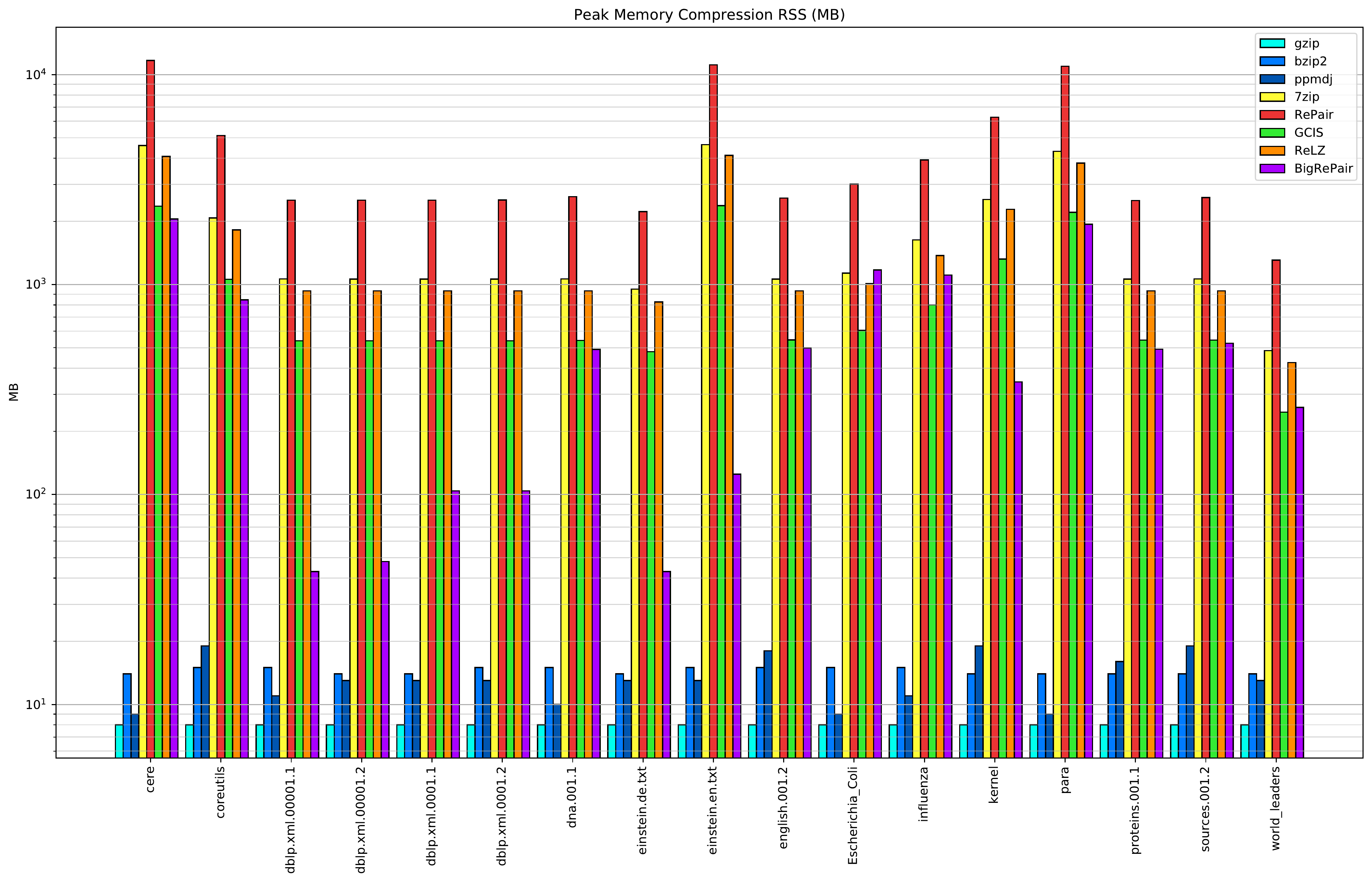}

	\vspace*{-5mm}
	\caption{Peak memory (in MB) used by the compressors during compression for repetitive texts.}
	\label{fig:mprof-compress-repetitive-texts}
\end{figure}

\begin{figure}[t!]
	\centering
	\includegraphics[width=0.95\textwidth]{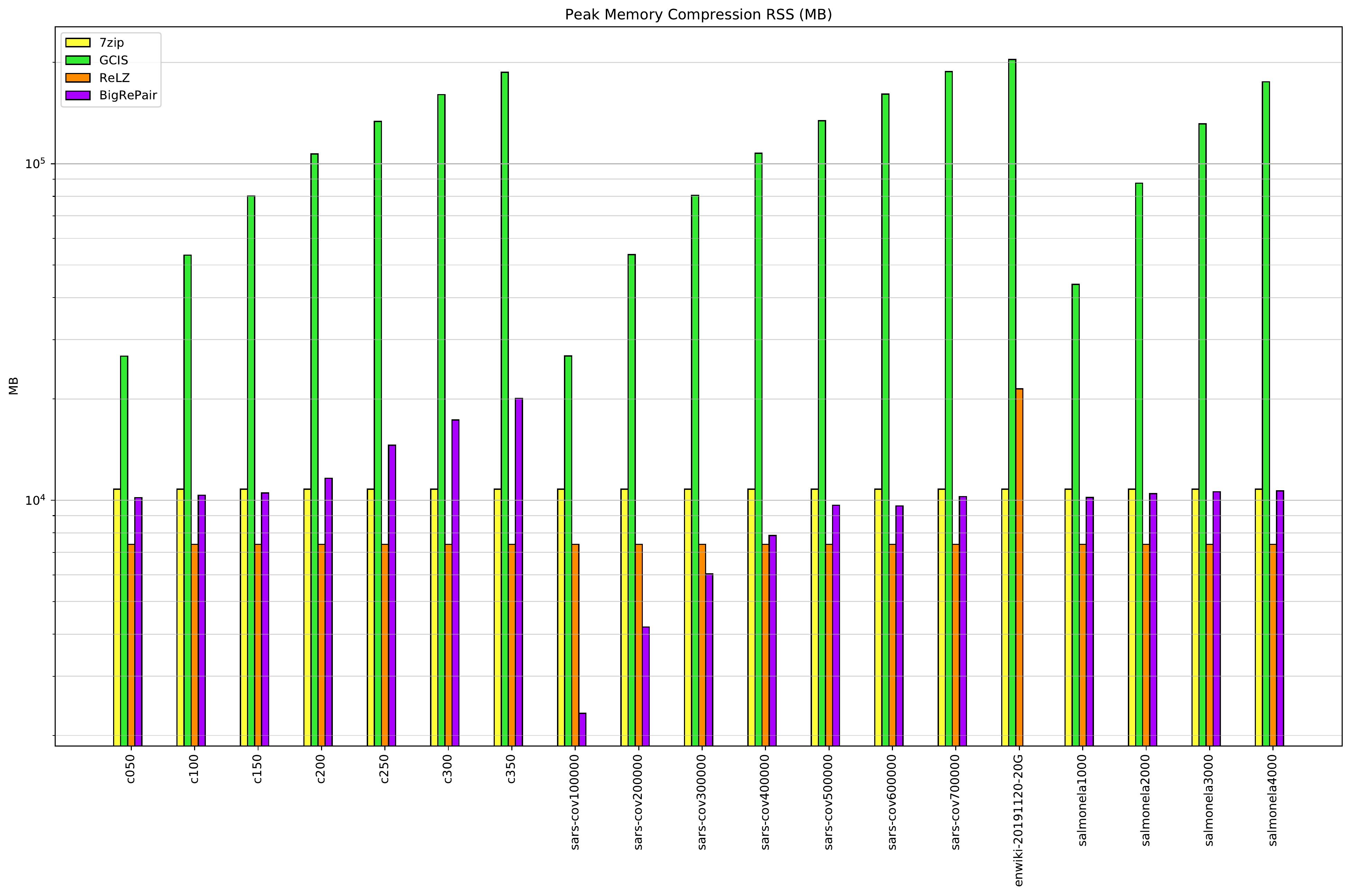}

	\vspace*{-5mm}
	\caption{Peak memory (in MB) used by the compressors during compression for very large texts.}
	\label{fig:mprof-compress-huge-texts}
\end{figure}

\begin{figure}[t!]
	\centering
	\includegraphics[width=1.25\textwidth,angle=-90]{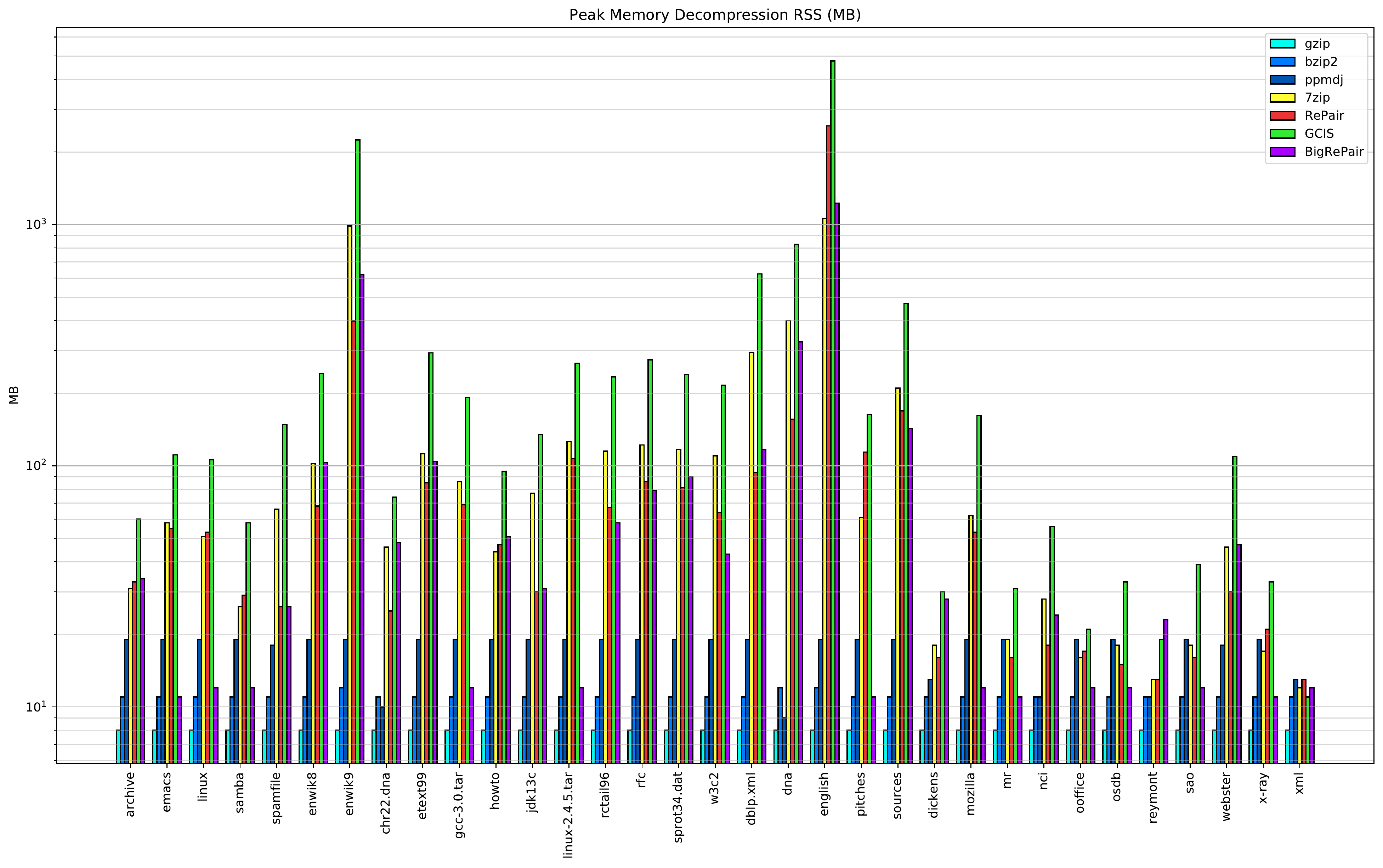}
	\caption{Peak memory (in MB) used by the compressors during decompression for regular texts.}
	\label{fig:mprof-decompress-regular-texts}
\end{figure}

\begin{figure}[t!]
	\centering
	\includegraphics[width=0.95\textwidth]{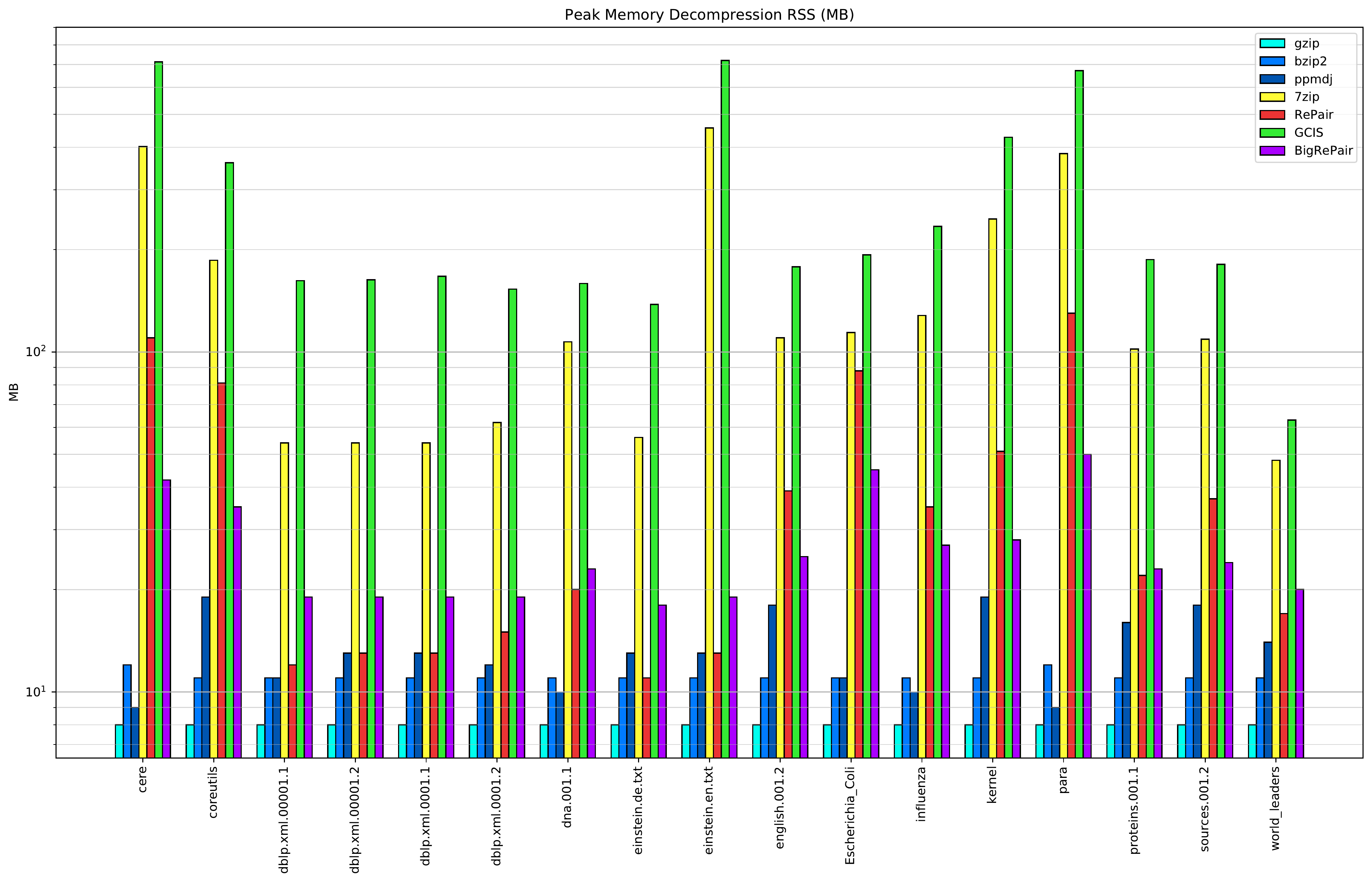}

	\vspace*{-5mm}
	\caption{Peak memory (in MB) used by the compressors during decompression for repetitive texts.}
	\label{fig:mprof-decompress-repetitive-texts}
\end{figure}

\begin{figure}[t!]
	\centering
	\includegraphics[width=0.95\textwidth]{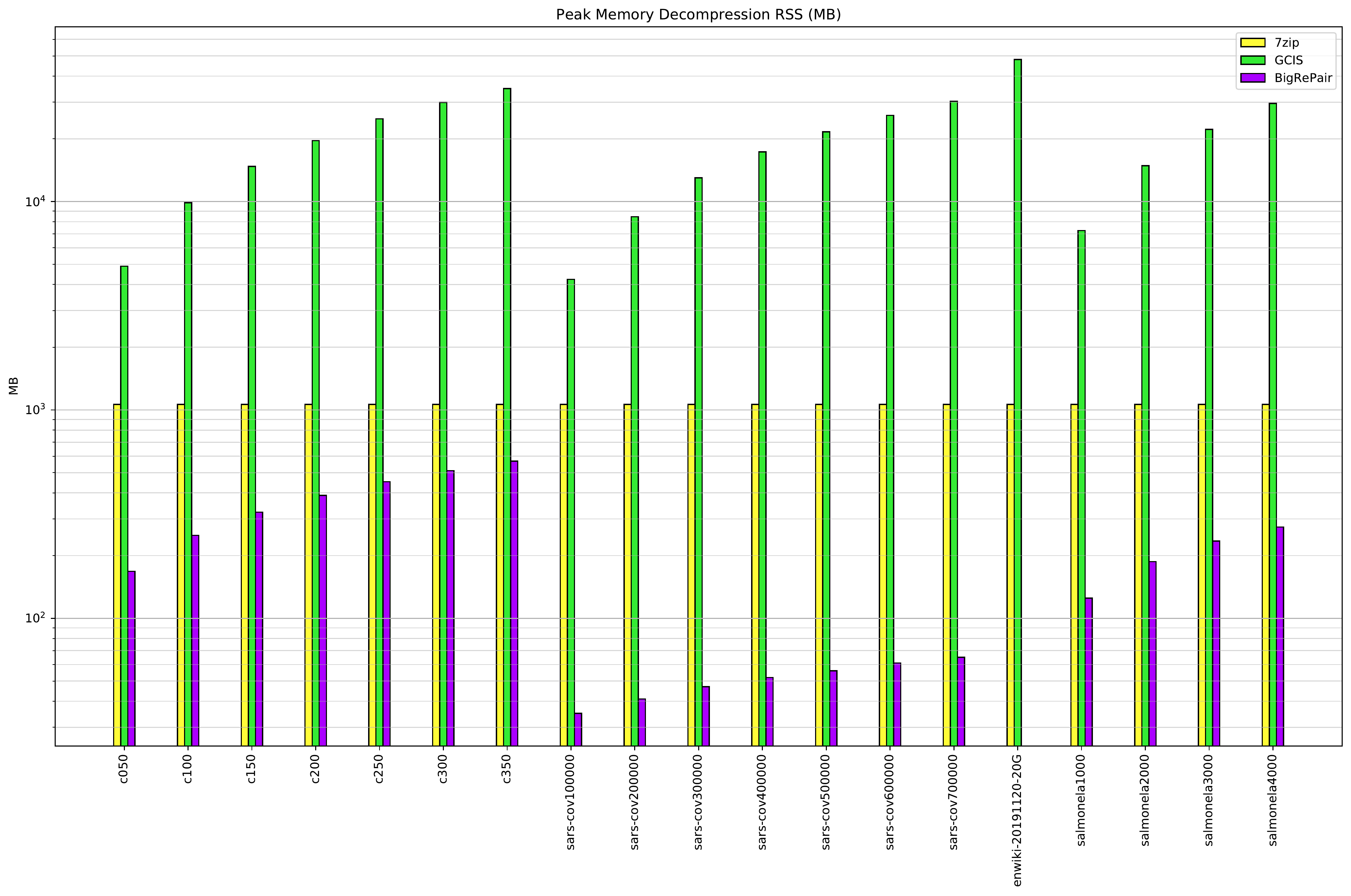}

	\vspace*{-5mm}
	\caption{Peak memory (in MB) used by the compressors during decompression for very large texts.}
	\label{fig:mprof-decompress-huge-texts}
\end{figure}

We evaluated the peak memory consumption (resident size) of all compressors during compression and decompression for each type of text; the results are shown in Figures \ref{fig:mprof-compress-regular-texts} to \ref{fig:mprof-decompress-huge-texts}. Overall, \gzip, \bzip and \ppmdj require negligible space to compress or decompress. On regular texts, considering compression, \our is followed by \relz, \szip, \bigrepair and \repair, the last being behind by a large margin. The situation is reversed in decompression: \bigrepair is followed by  \repair, \szip, and \our.  On repetitive texts, the scenario is similar for compression, but \our escorts \bigrepair. On very large texts, during compression \relz is the most space-efficient, followed by \bigrepair, \szip and lastly by \our. Considering decompression the order stays the same, except for \relz, which does not decompress.

\subsection{Overview}

\begin{figure}[p]
	\includegraphics[width=\linewidth]{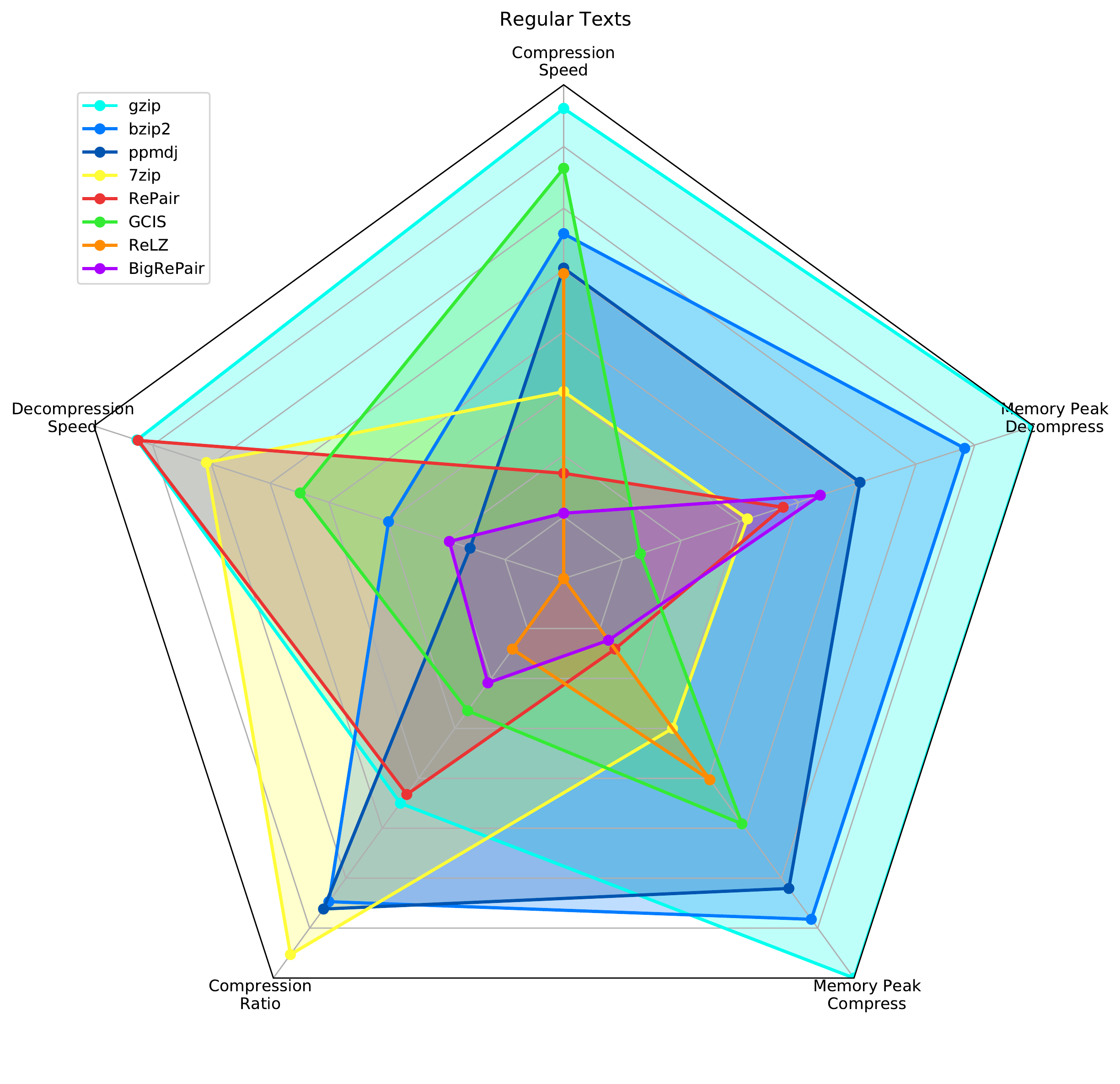}
	\caption{Compressor comparison on regular texts.}
	\label{fig:compressor-comparisons-regular-texts}
\end{figure}

\begin{figure}[p]
	\includegraphics[width=\linewidth]{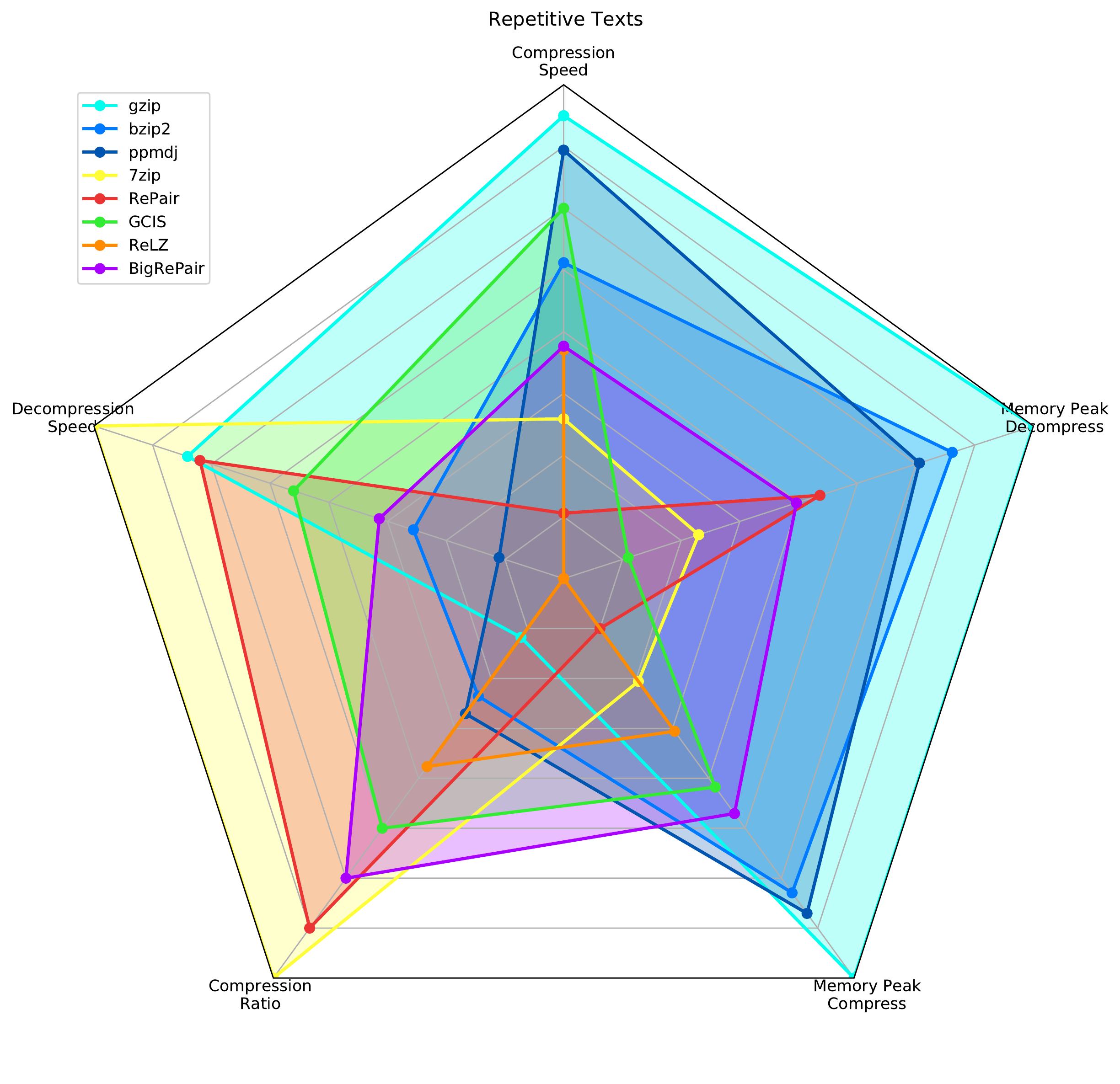}
	\caption{Compressor comparison on repetitive texts.}
	\label{fig:compressor-comparisons-repetitive-texts}
\end{figure}

\begin{figure}[p]
	\includegraphics[width=\linewidth]{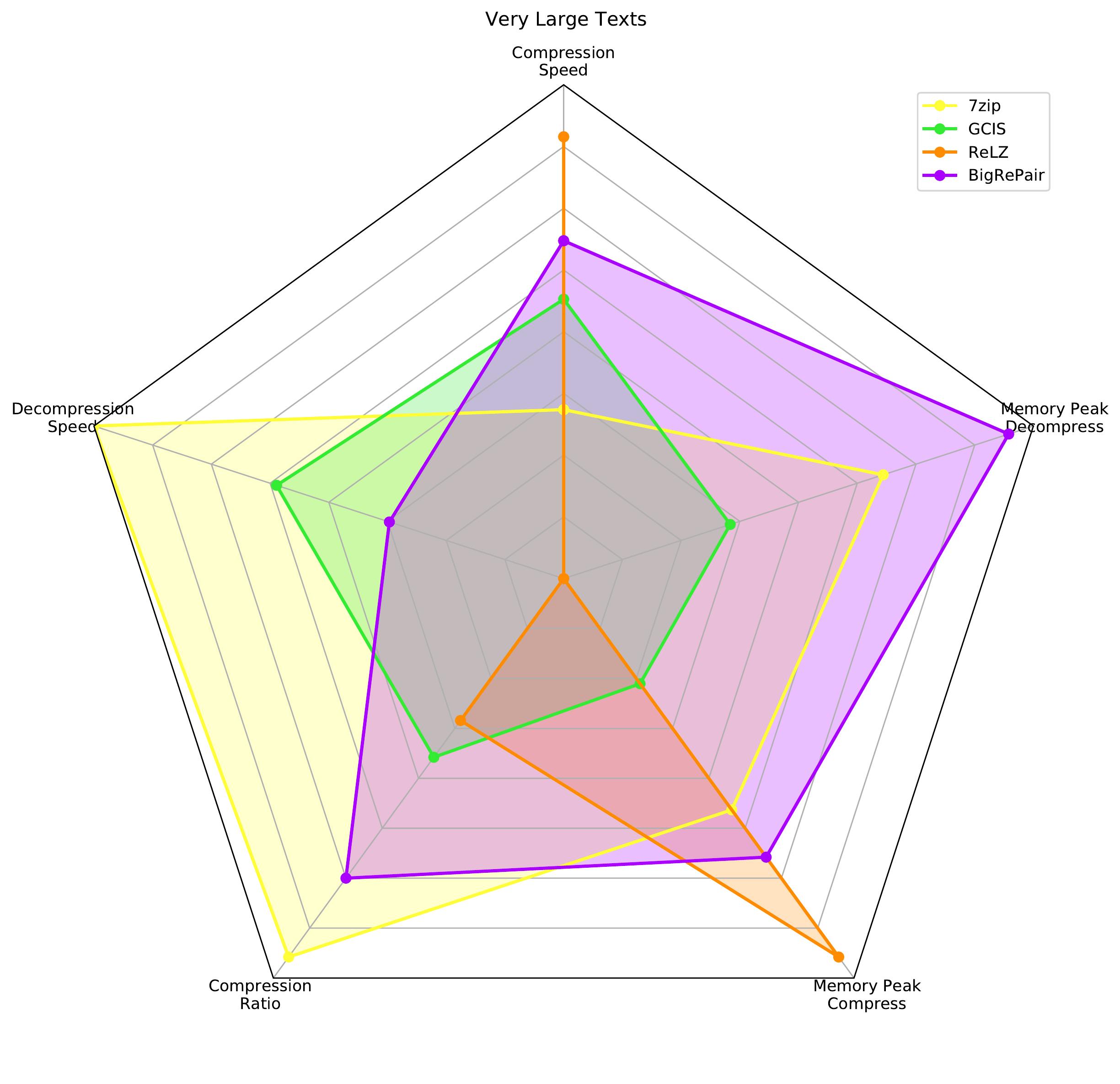}
	\caption{Compressor comparison on very large texts.}
	\label{fig:compressor-comparisons-huge-texts}
\end{figure}

Figures \ref{fig:compressor-comparisons-regular-texts} to \ref{fig:compressor-comparisons-huge-texts} present conceptual radar charts that summarize, for each text type, the performance of all compressors in each rated aspect. The closer the values are to the pentagon borders, the better the compressor performed on the corresponding aspect.



\subsection{Extract operation}
\label{subsec:extract}

\begin{figure}[t!]
	\centering
	\begin{subfigure}[t]{.457\textwidth}
		\centering
		\includegraphics[width=\textwidth]{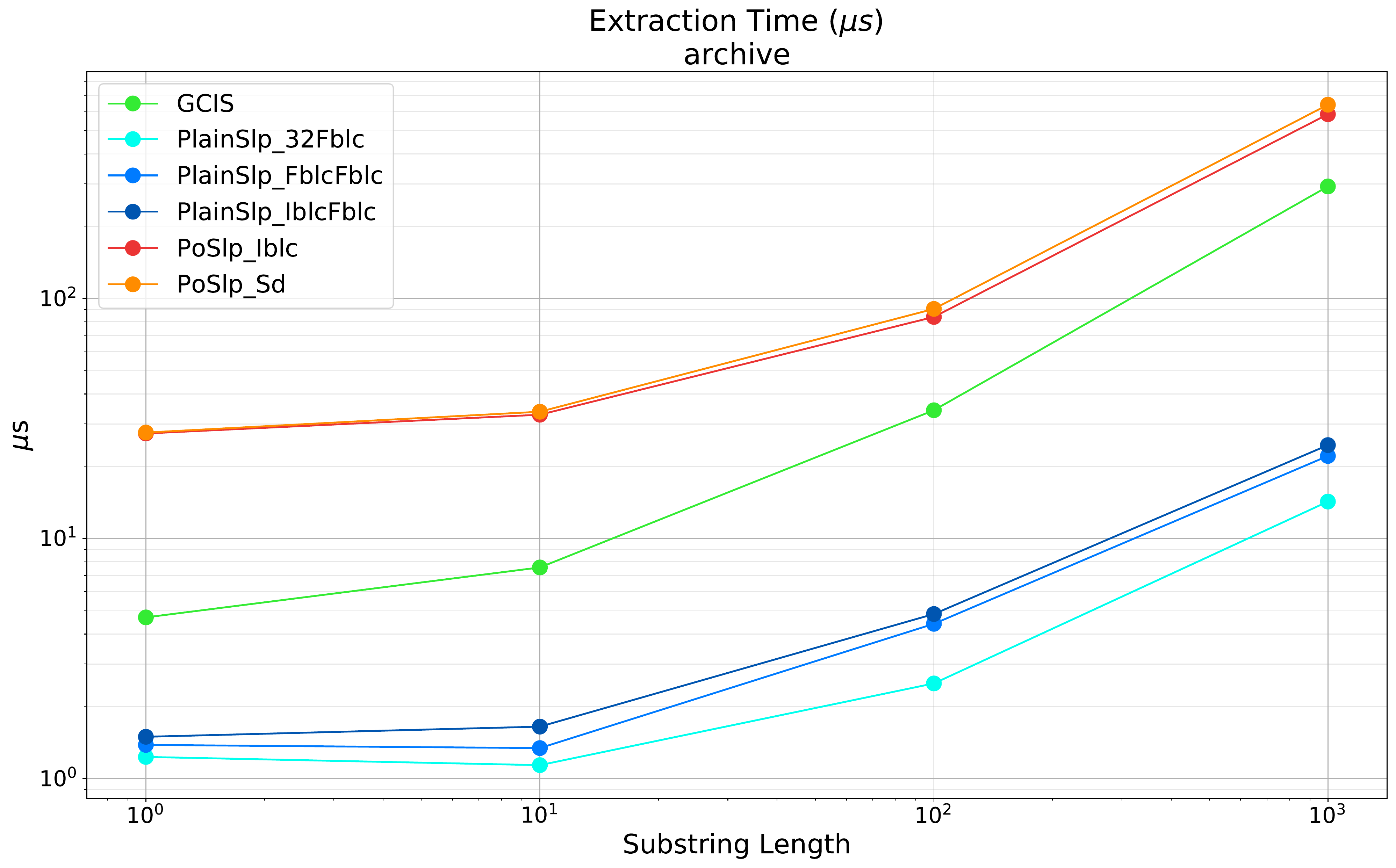}
		\label{fig:extract-archive}
	\end{subfigure}
	\begin{subfigure}[t]{.457\textwidth}
		\centering
		\includegraphics[width=\textwidth]{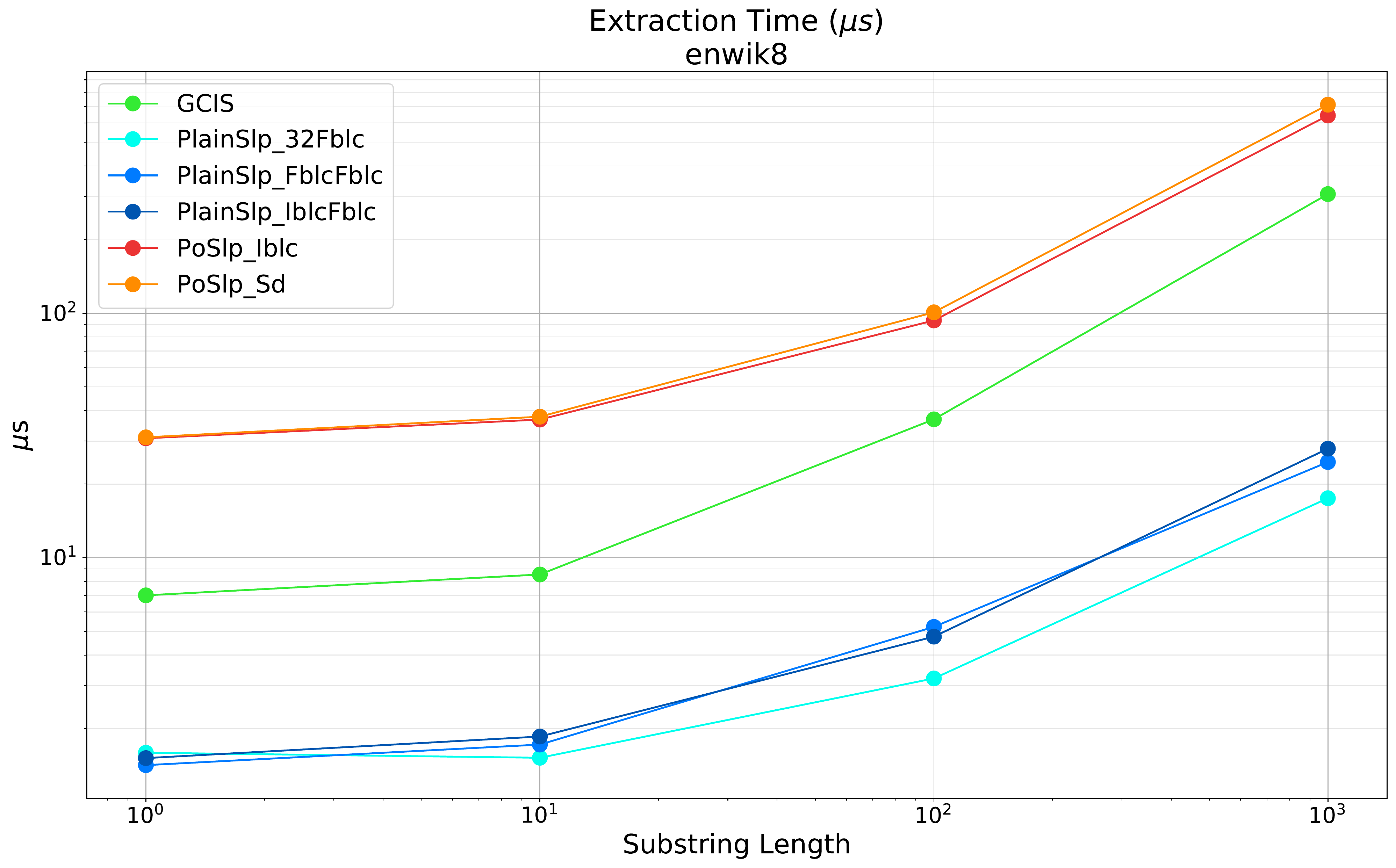}
		\label{fig:extract-enwik8}
	\end{subfigure}
	\begin{subfigure}[t]{.457\textwidth}
		\centering
		\includegraphics[width=\textwidth]{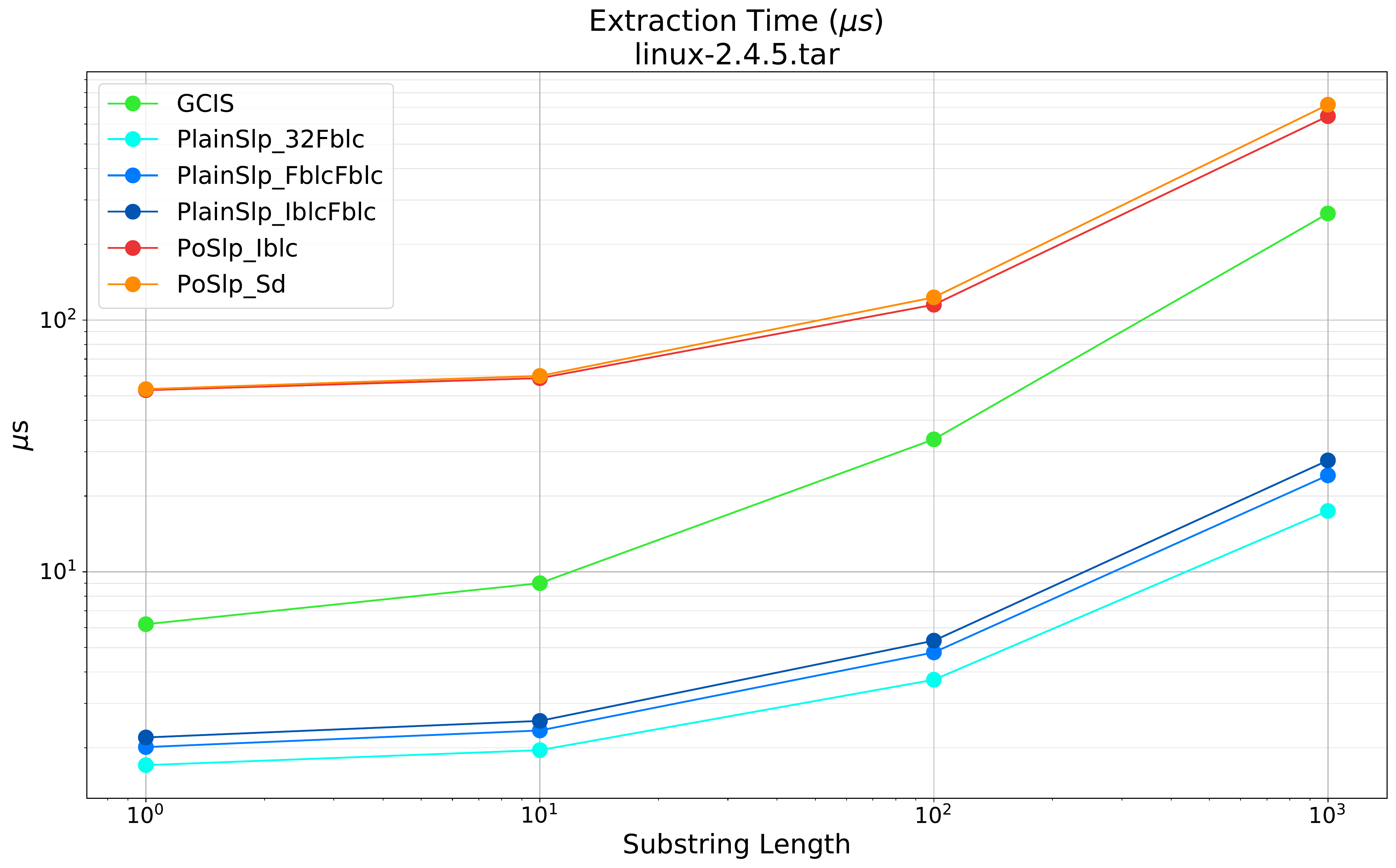}
		\label{fig:extract-linux}
	\end{subfigure}
	\begin{subfigure}[t]{.457\textwidth}
		\centering
		\includegraphics[width=\textwidth]{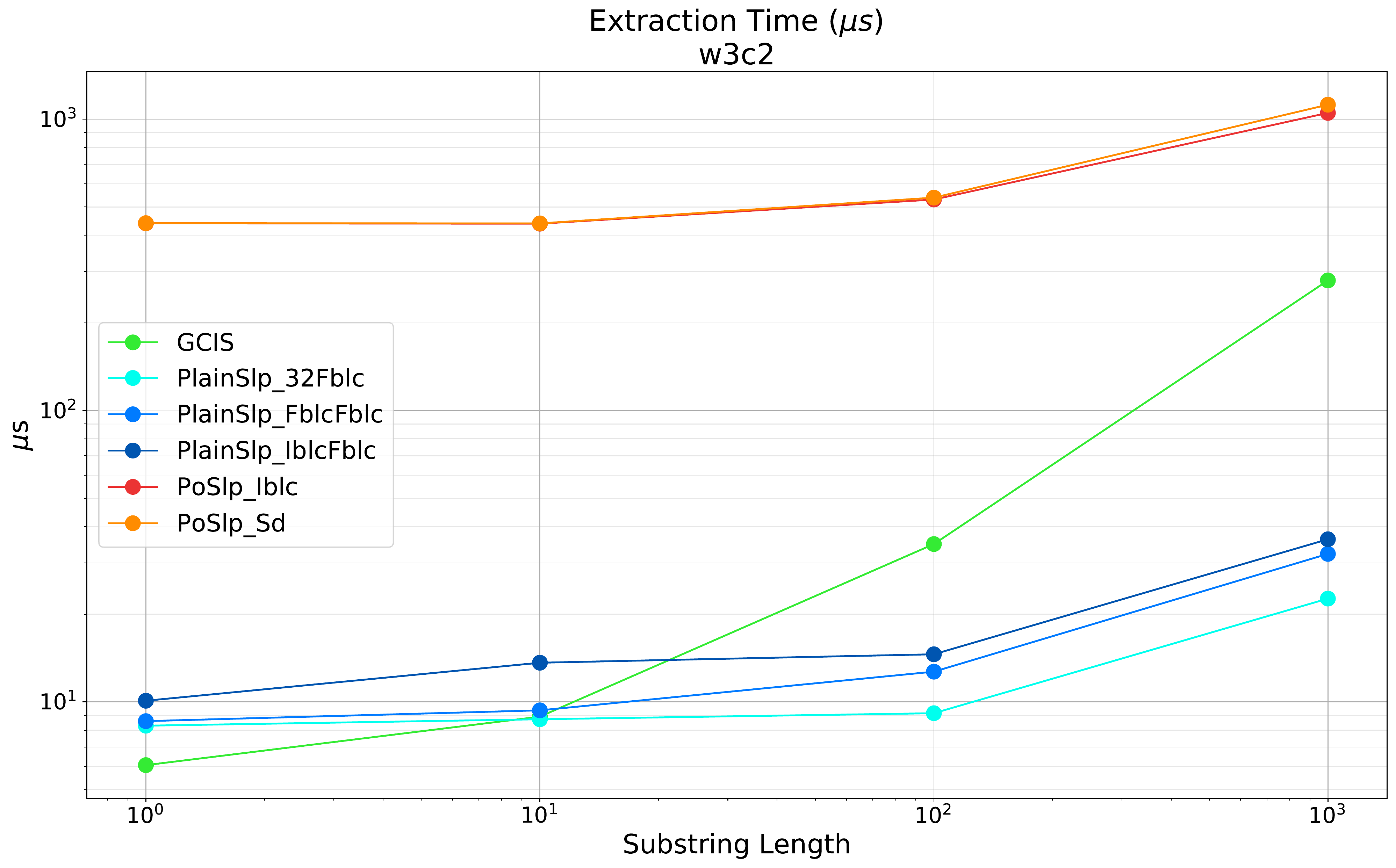}
		\label{fig:extract-w3c2}
	\end{subfigure}
	\begin{subfigure}[t]{.457\textwidth}
		\centering
		\includegraphics[width=\textwidth]{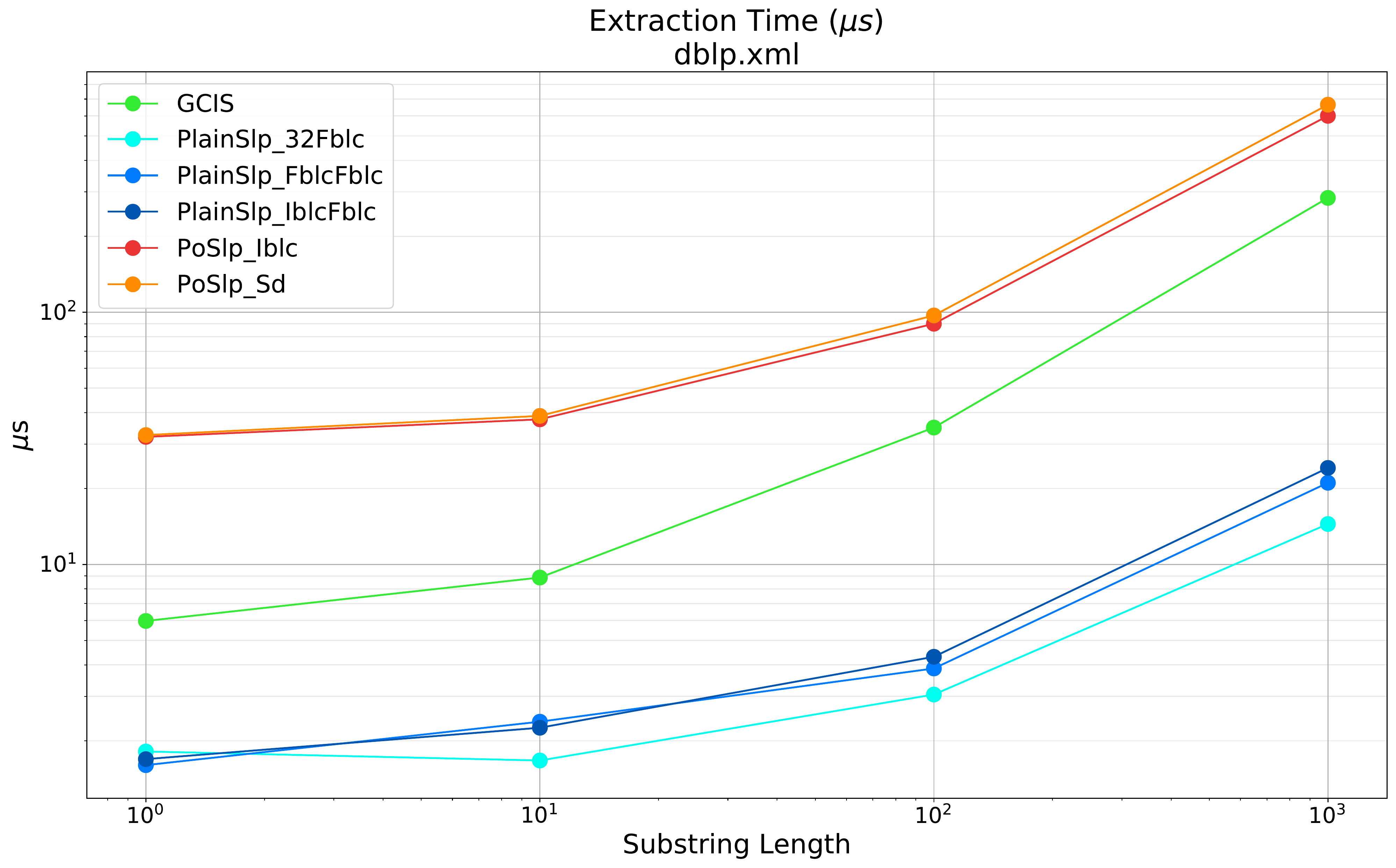}
	\end{subfigure}
	\begin{subfigure}[t]{.457\textwidth}
		\centering
		\includegraphics[width=\textwidth]{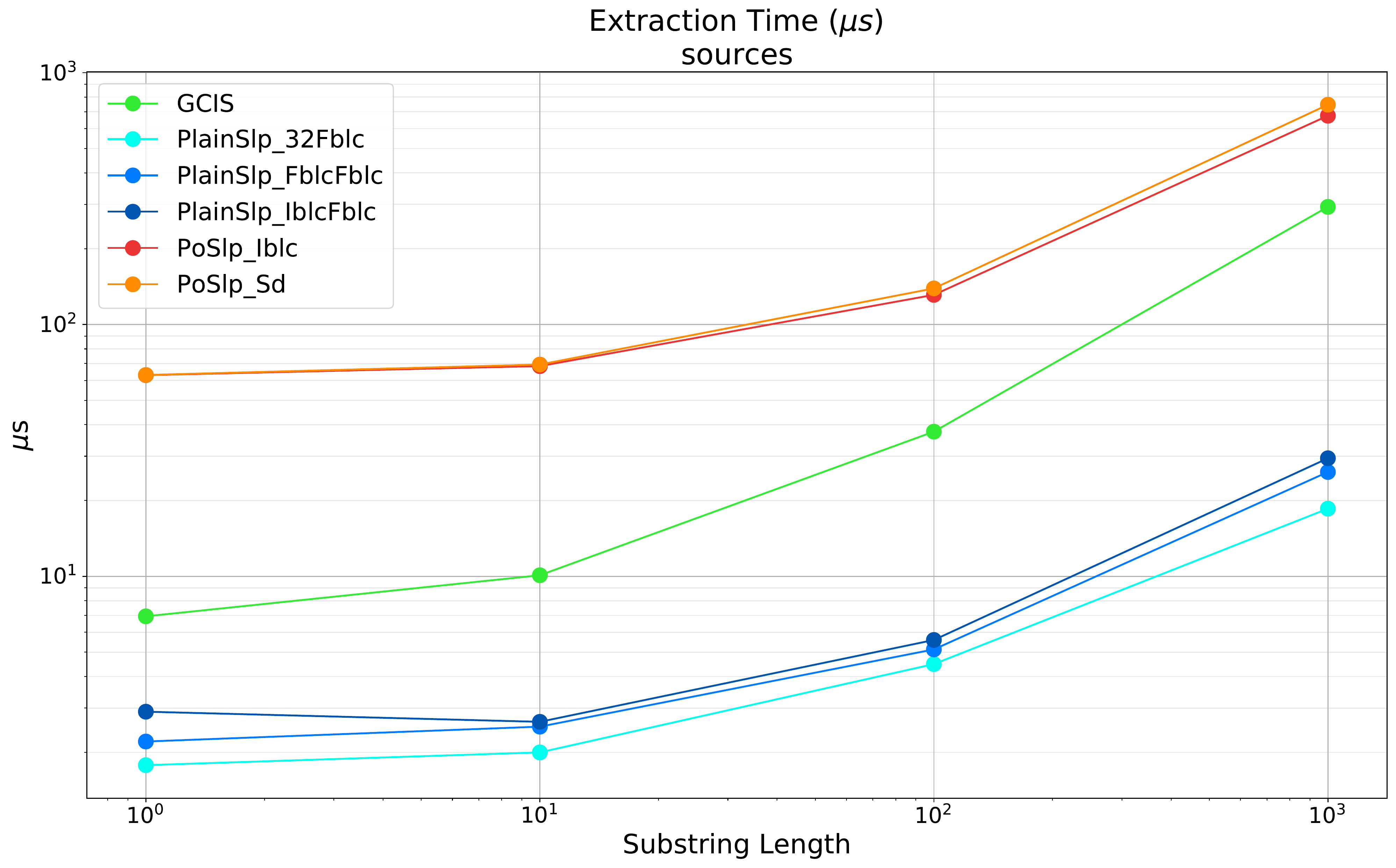}
	\end{subfigure}
		
	\vspace{4mm} 
	
	\begin{subfigure}[t]{.457\textwidth}
		\centering
		\includegraphics[width=\textwidth]{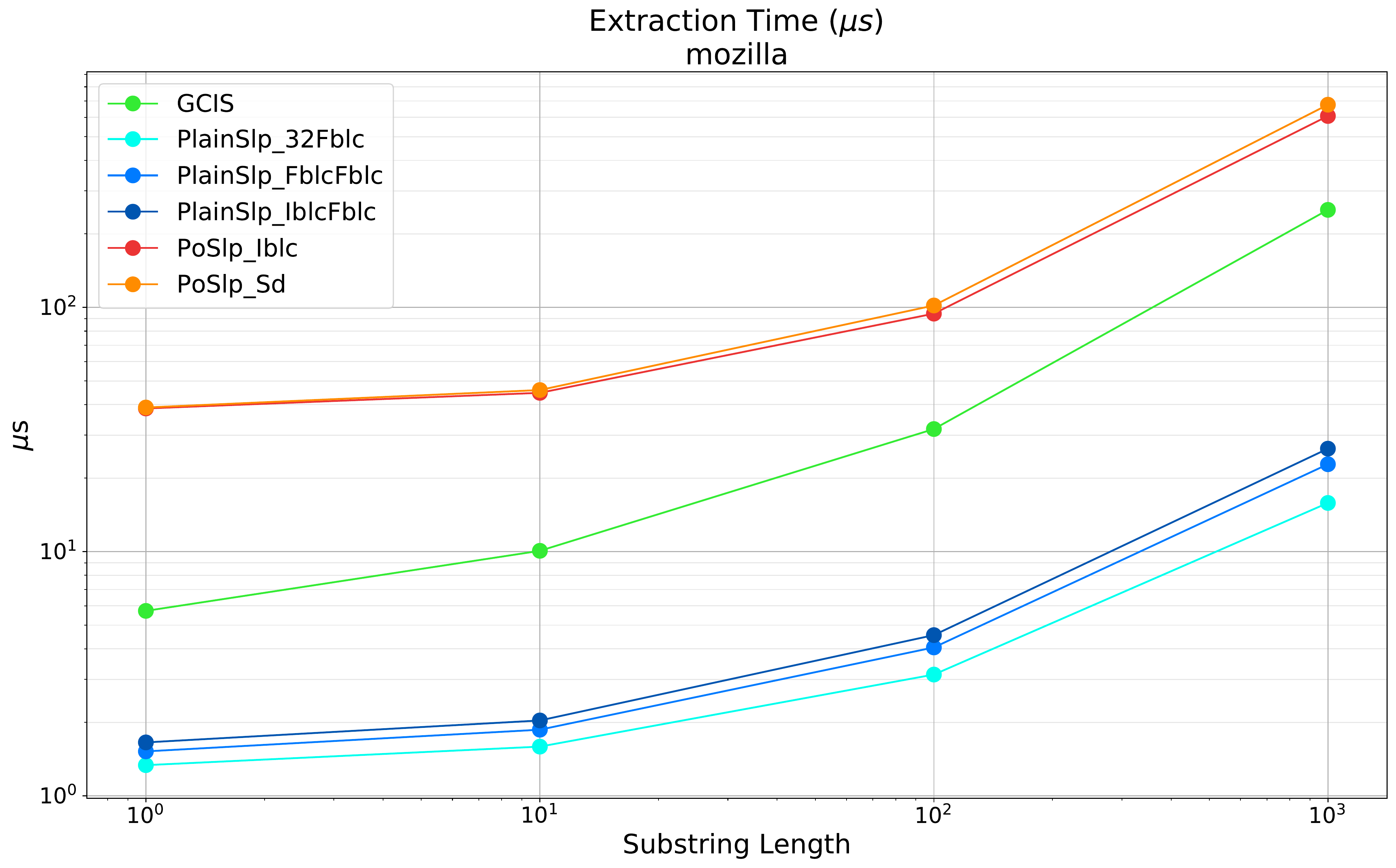}
	\end{subfigure}
	\begin{subfigure}[t]{.457\textwidth}
		\centering
		\includegraphics[width=\textwidth]{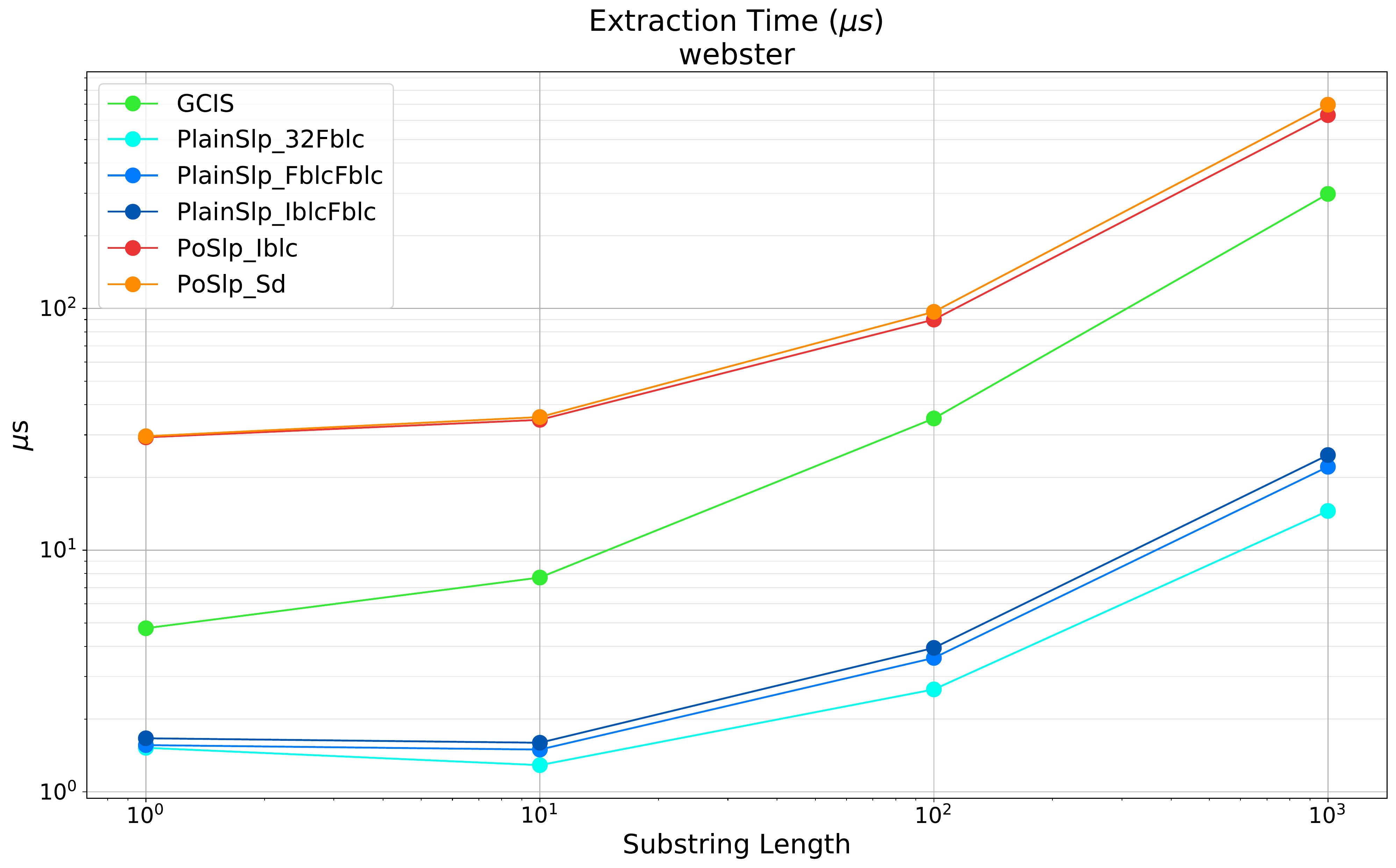}
	\end{subfigure}
	\caption{Substring length \emph{vs.} extraction time (microseconds) on regular texts.}
	\label{fig:extract-regular-texts}
\end{figure}

\begin{figure}[ht]
	\centering
	\begin{subfigure}[t]{.457\textwidth}
		\centering
		\includegraphics[width=\textwidth]{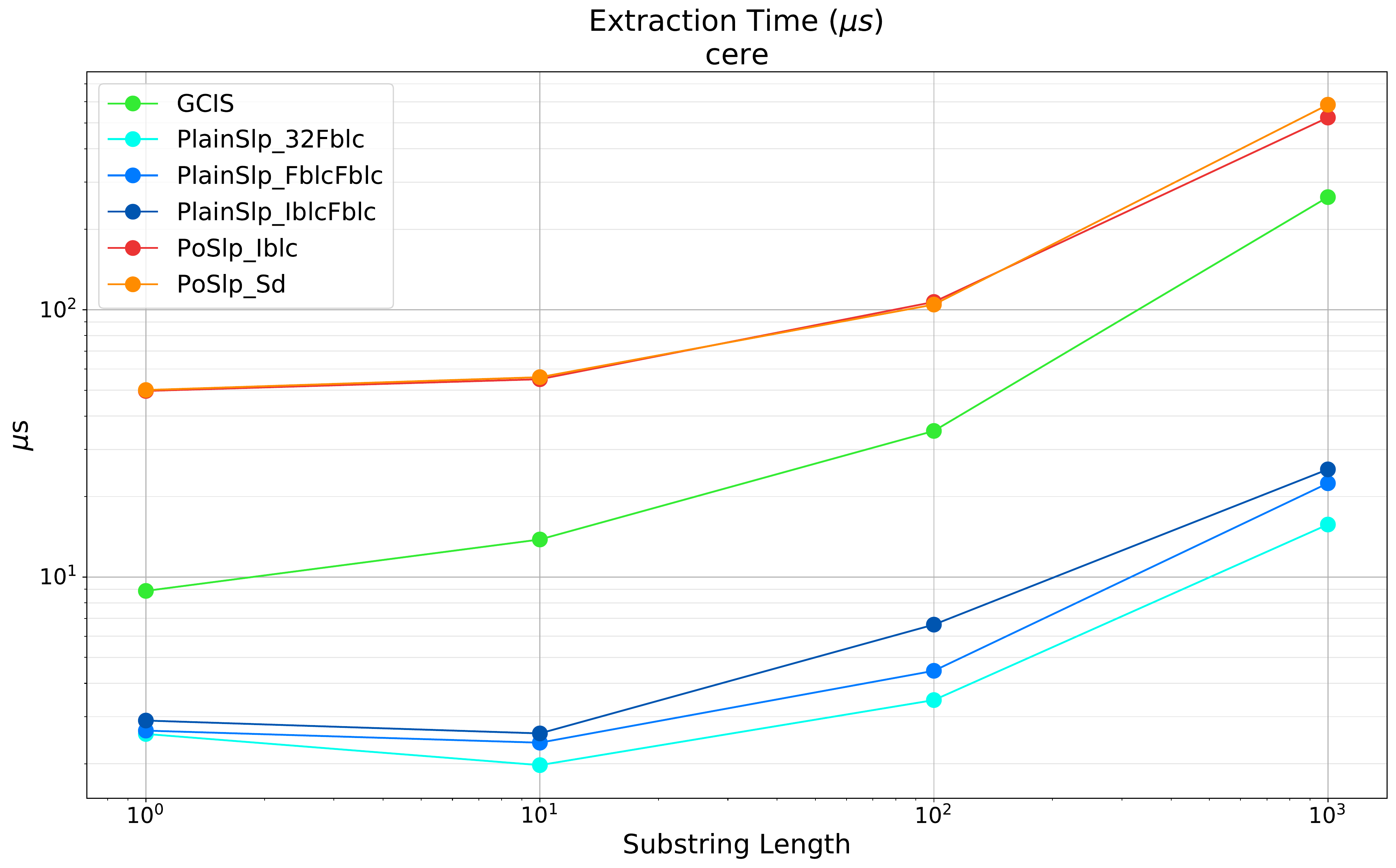}

		\label{fig:extract-cere}
	\end{subfigure}
	\begin{subfigure}[t]{.457\textwidth}
		\centering
		\includegraphics[width=\textwidth]{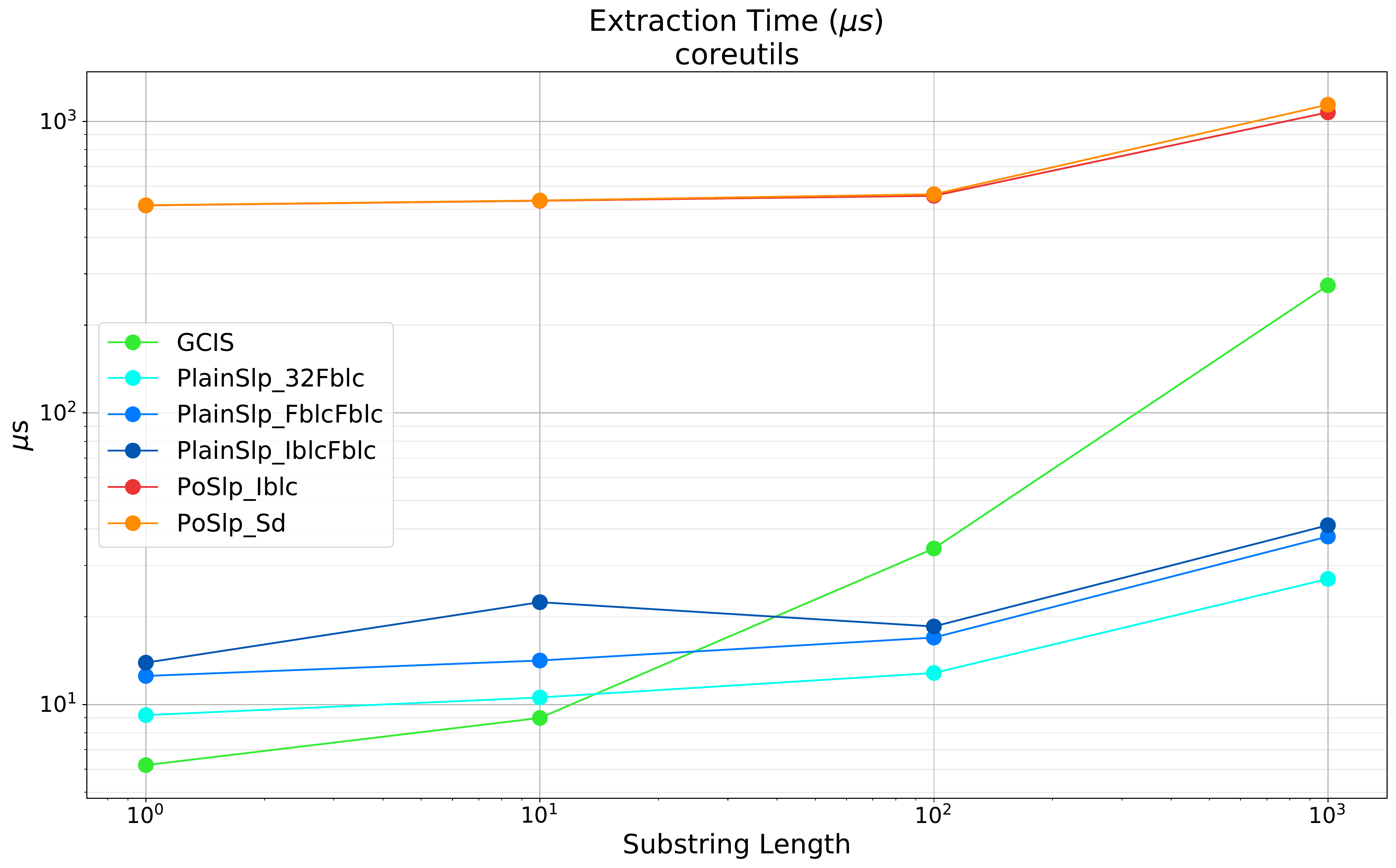}

		\label{fig:extract-coreutils}
	\end{subfigure}
	\begin{subfigure}[t]{.457\textwidth}
		\centering
		\includegraphics[width=\textwidth]{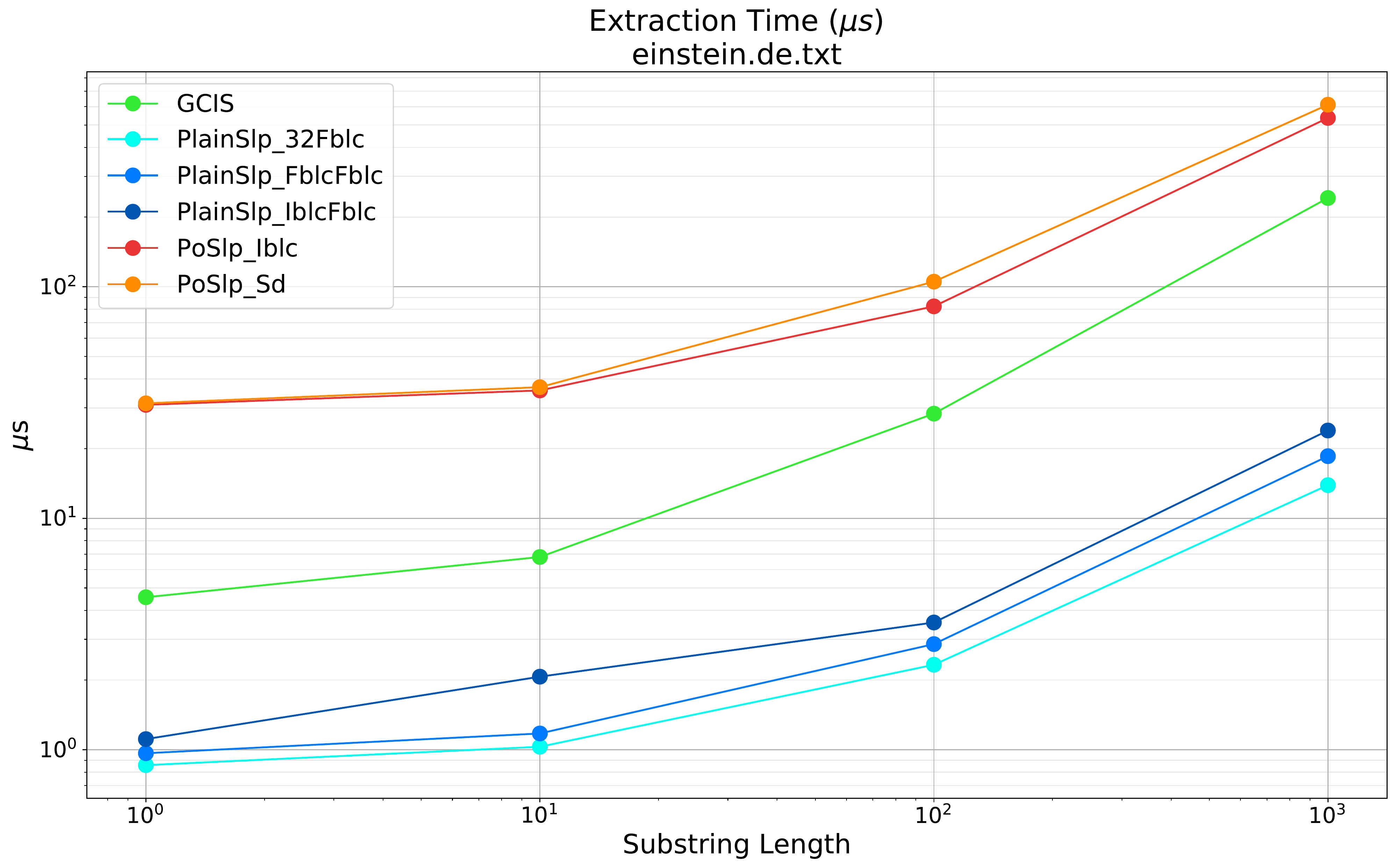}
		\label{fig:extract-einstein-de}
	\end{subfigure}
	\begin{subfigure}[t]{.457\textwidth}
		\centering
		\includegraphics[width=\textwidth]{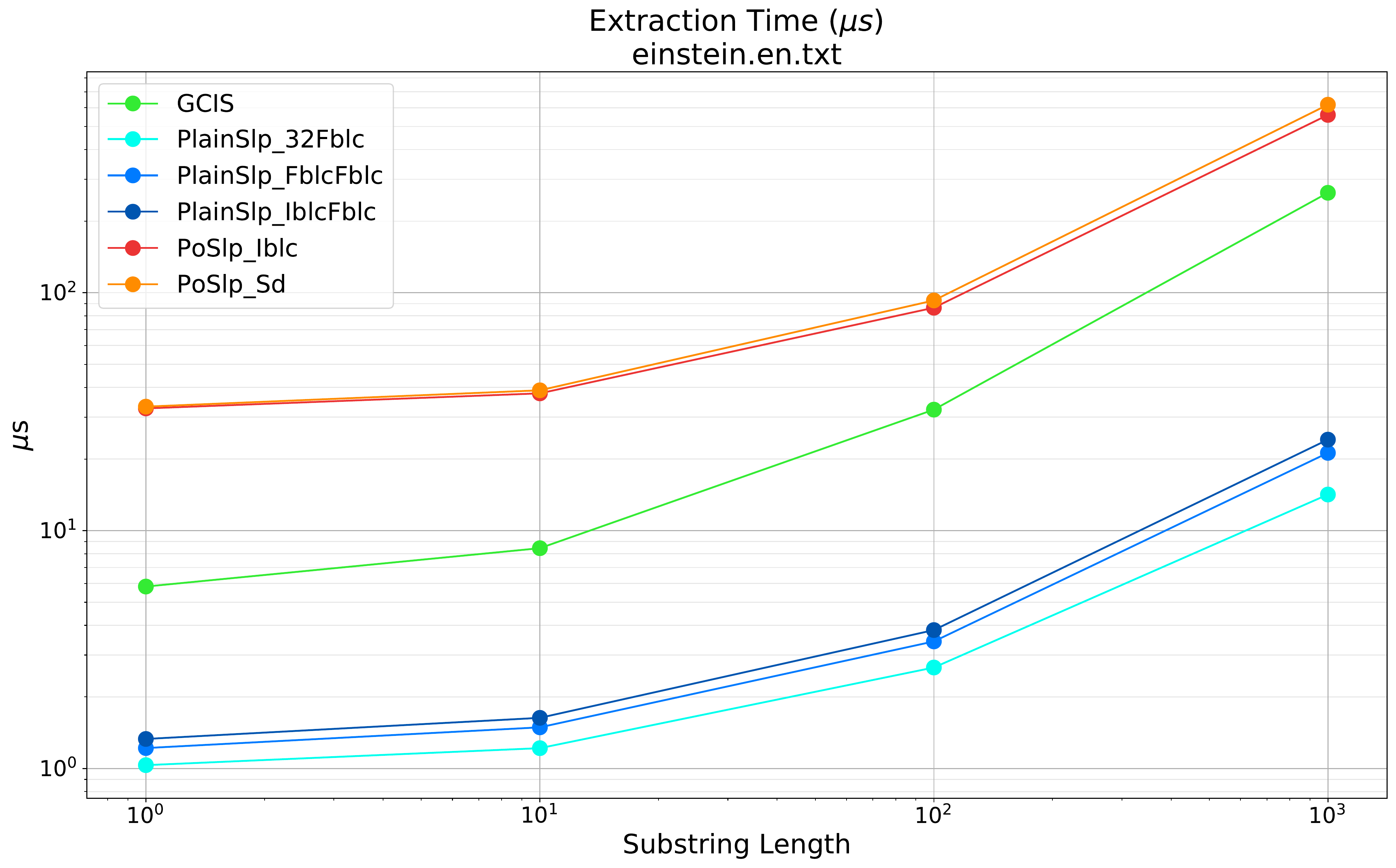}
		\label{fig:extract-einstein-en}
	\end{subfigure}
	\begin{subfigure}[t]{.457\textwidth}
		\centering
		\includegraphics[width=\textwidth]{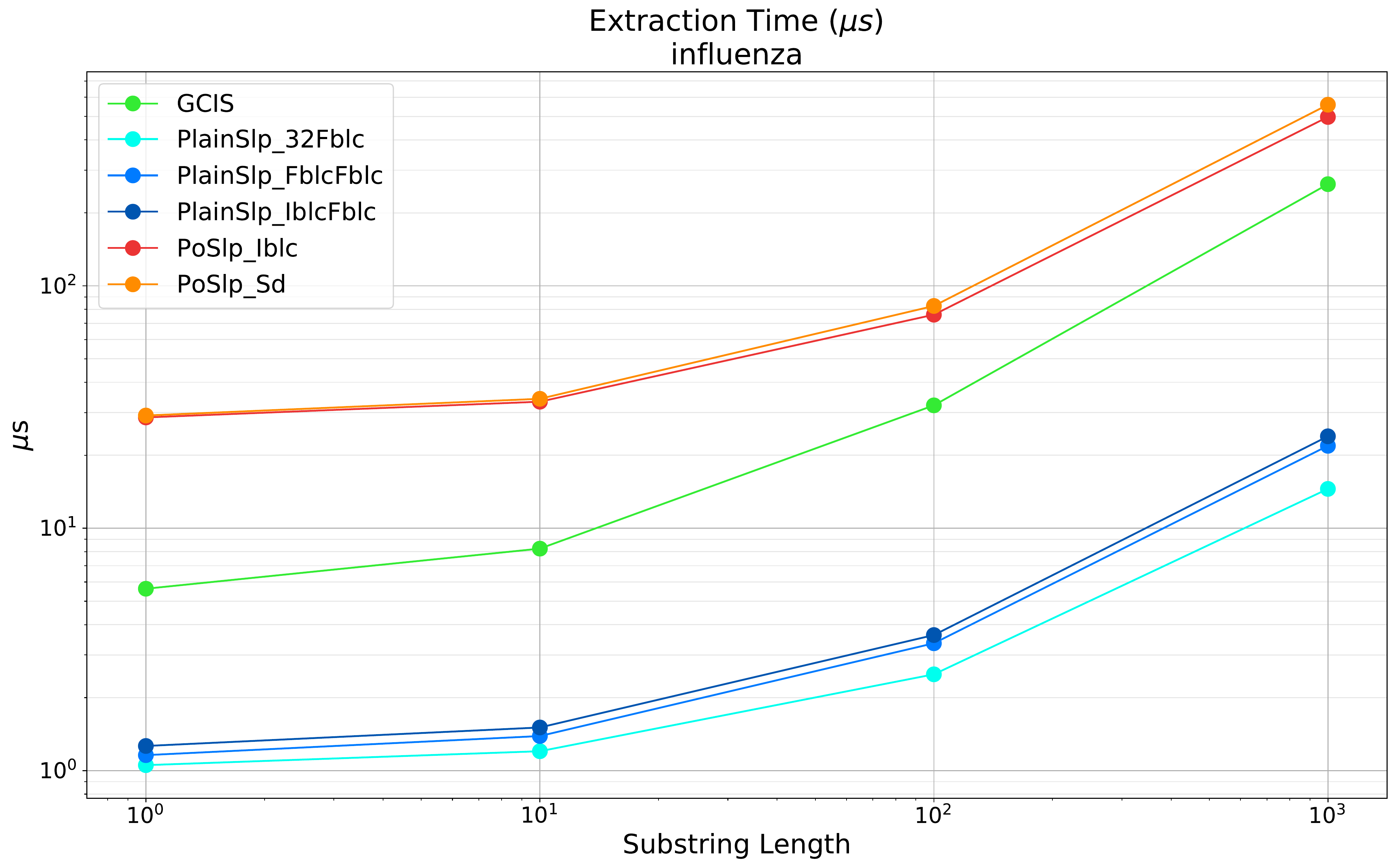}
		\label{fig:extract-influenza}
	\end{subfigure}
	\begin{subfigure}[t]{.457\textwidth}
		\centering
		\includegraphics[width=\textwidth]{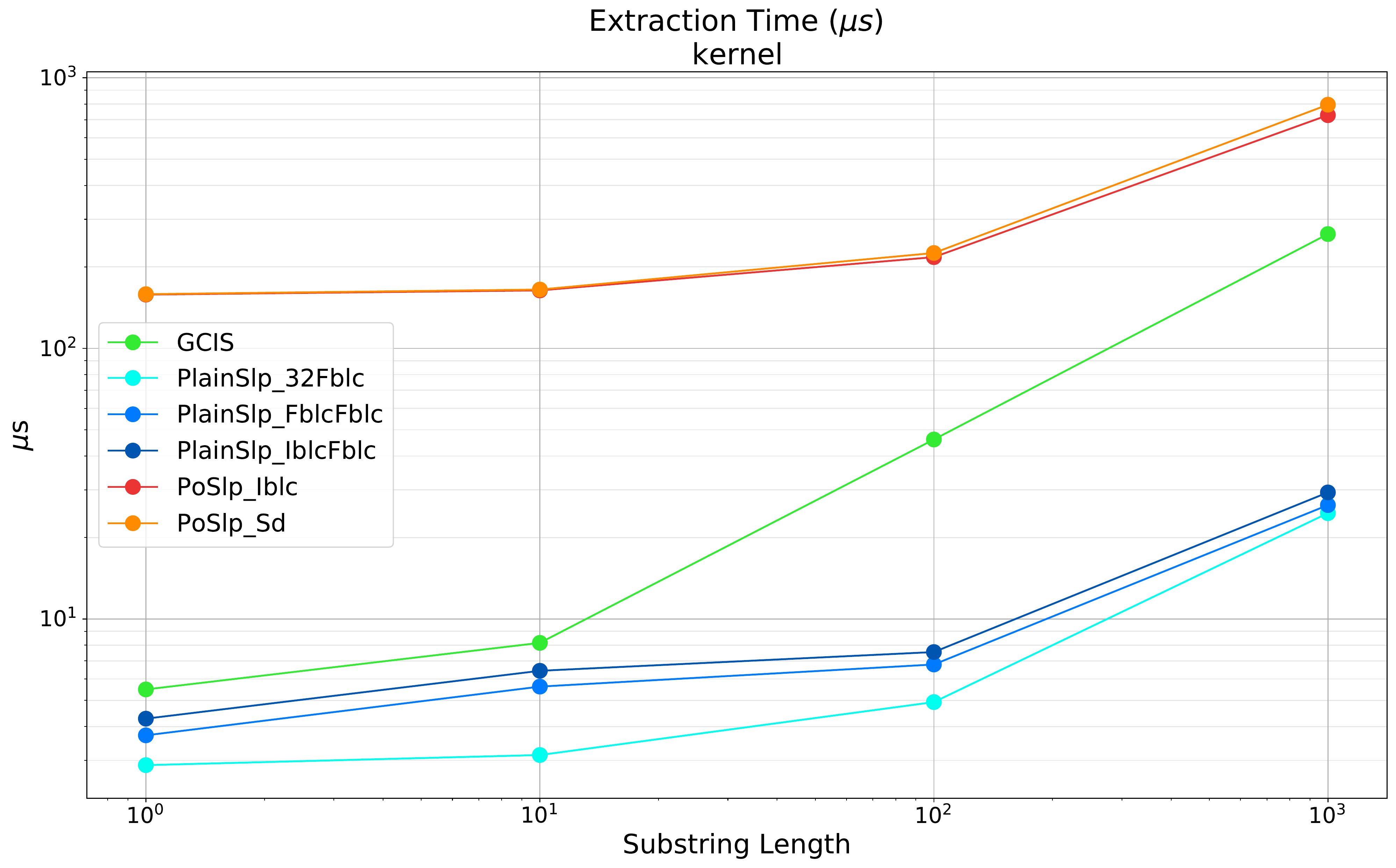}
		\label{fig:extract-kernel}
	\end{subfigure}
	\vspace{4mm}
	\begin{subfigure}[t]{.45\textwidth}
		\centering
		\includegraphics[width=\textwidth]{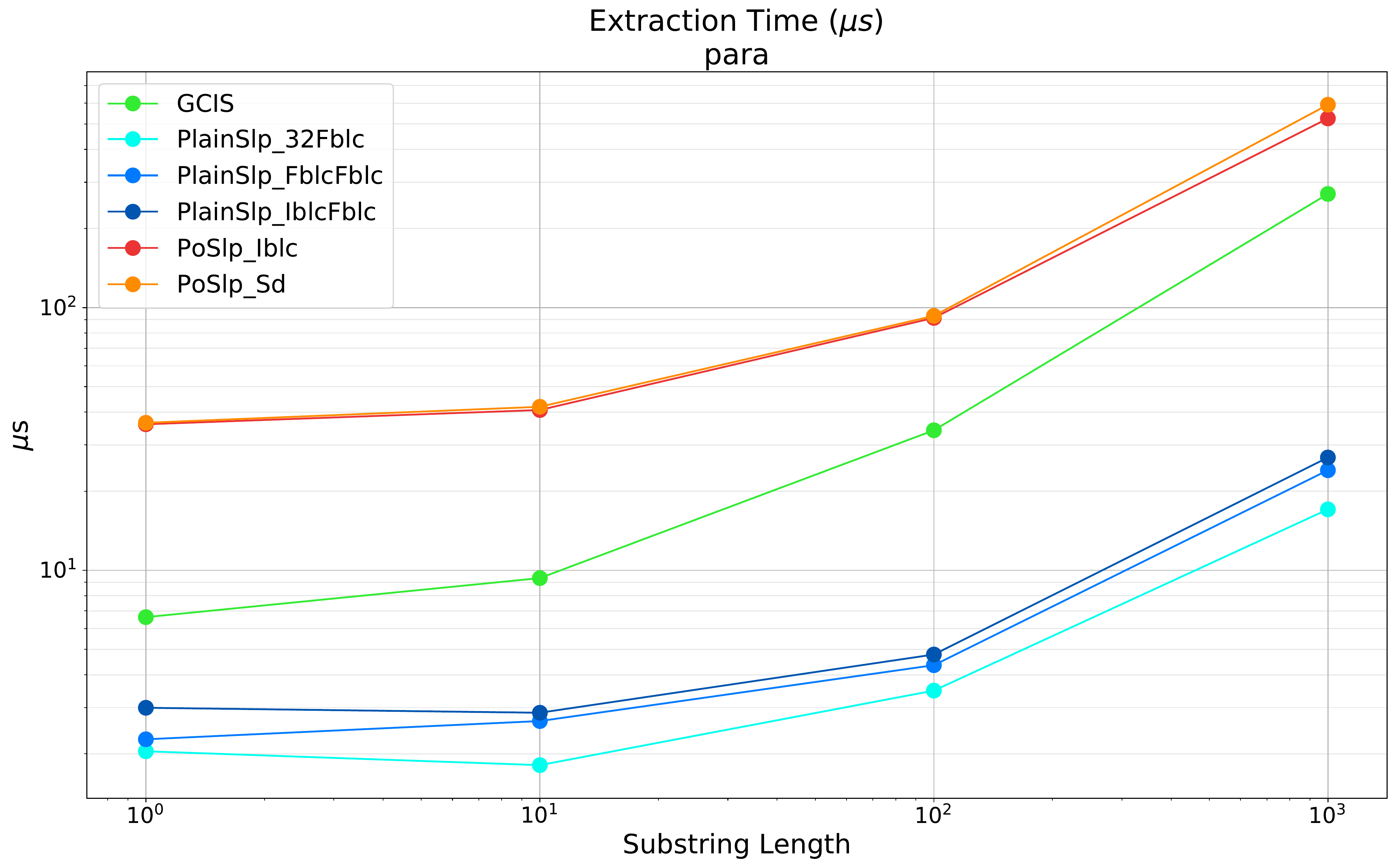}
		\label{fig:extract-para}
	\end{subfigure}
	\begin{subfigure}[t]{.45\textwidth}
		\centering
		\includegraphics[width=\textwidth]{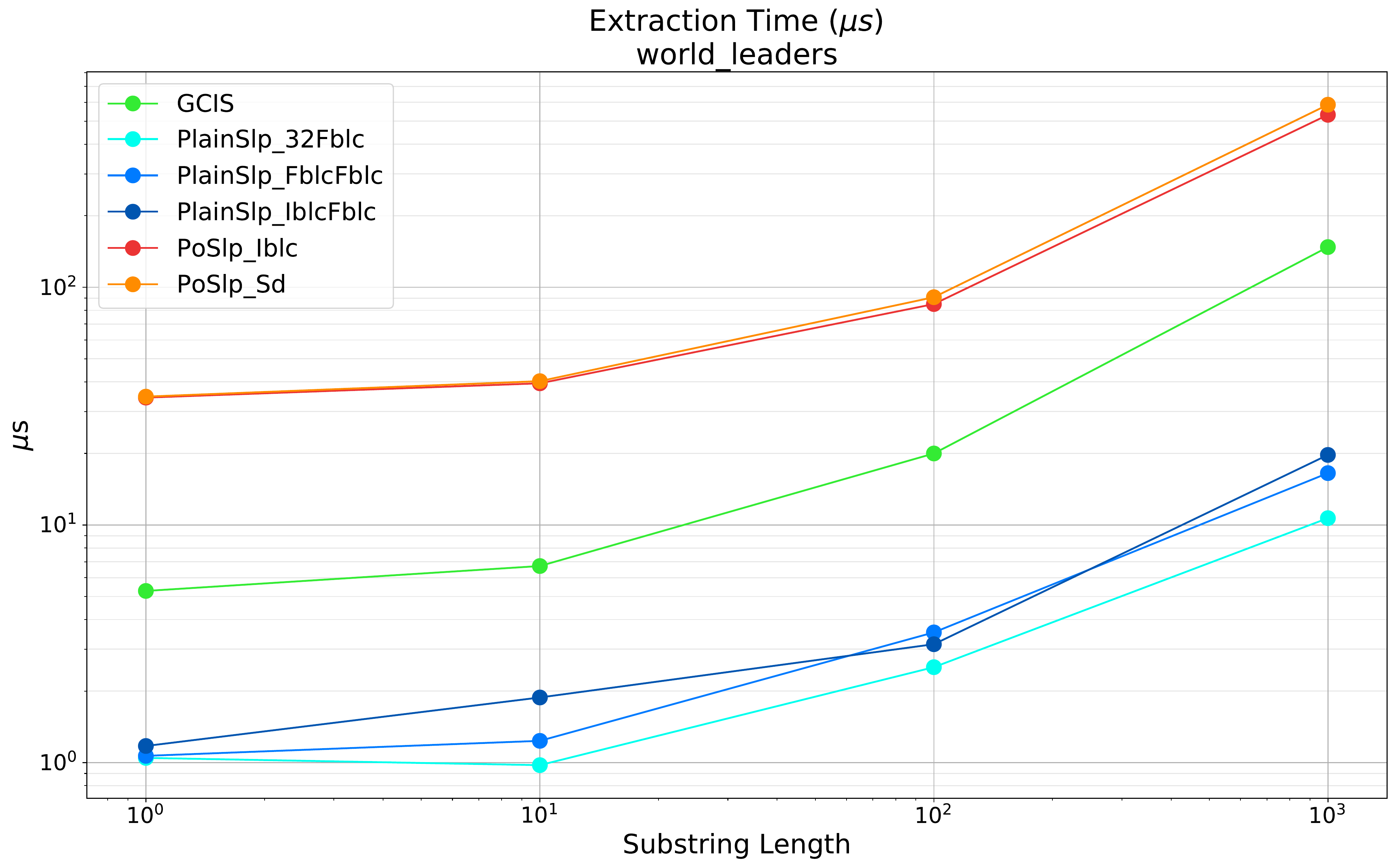}
		\label{fig:extract-world-leaders}
	\end{subfigure}
	\caption{Substring length \emph{vs.} extraction time (microseconds) on repetitive texts.}
	\label{fig:extract-repetitive-texts}
\end{figure}

\begin{figure}[t!]
	\centering
	\includegraphics[width=0.95\textwidth]{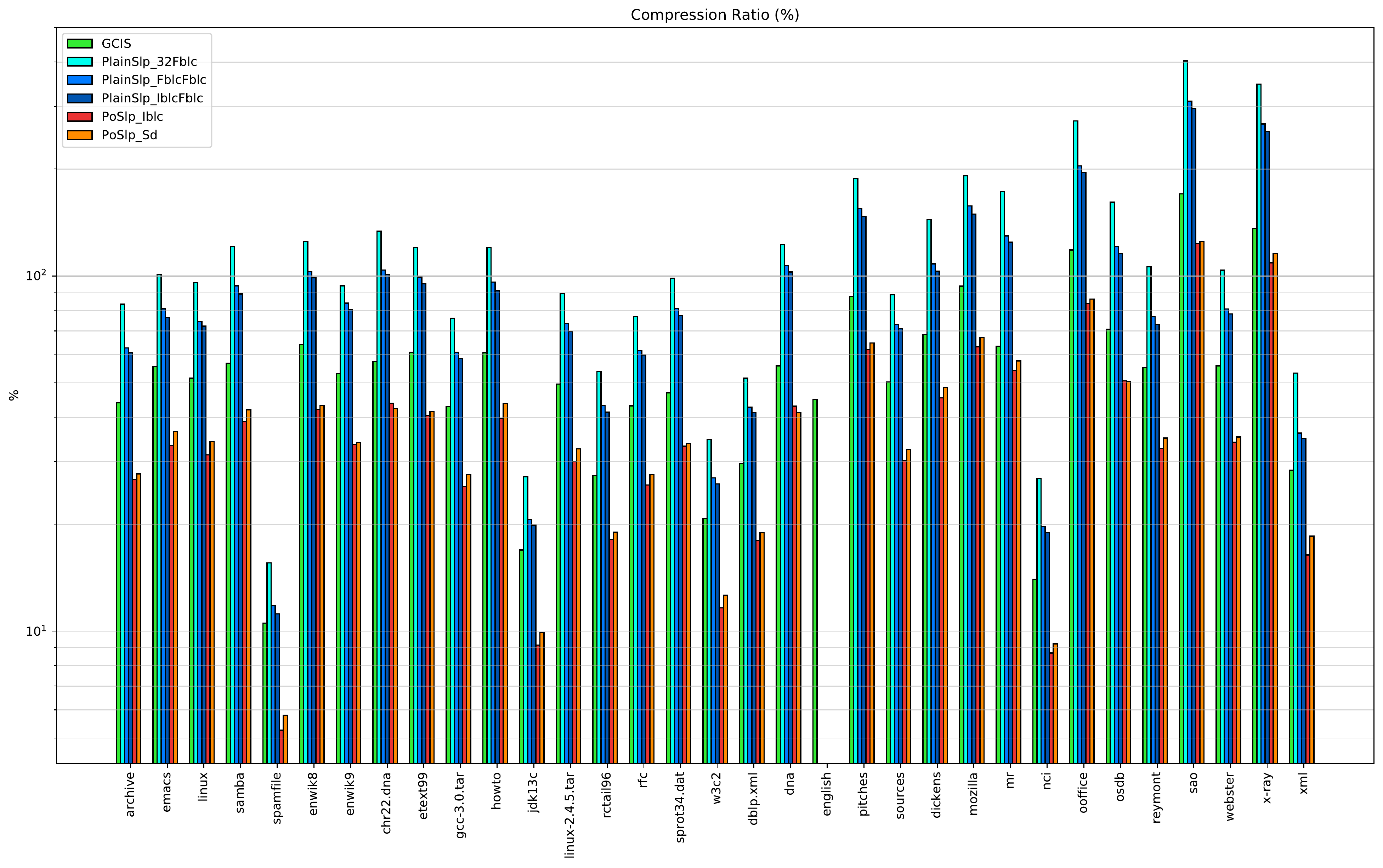}

	\vspace*{-5mm}
	\caption{Compression ratio of the extractors on regular texts.}
	\label{fig:compression-ratio-extractors-regular-text}
\end{figure}

\begin{figure}[t!]
	\centering
	\includegraphics[width=0.95\textwidth]{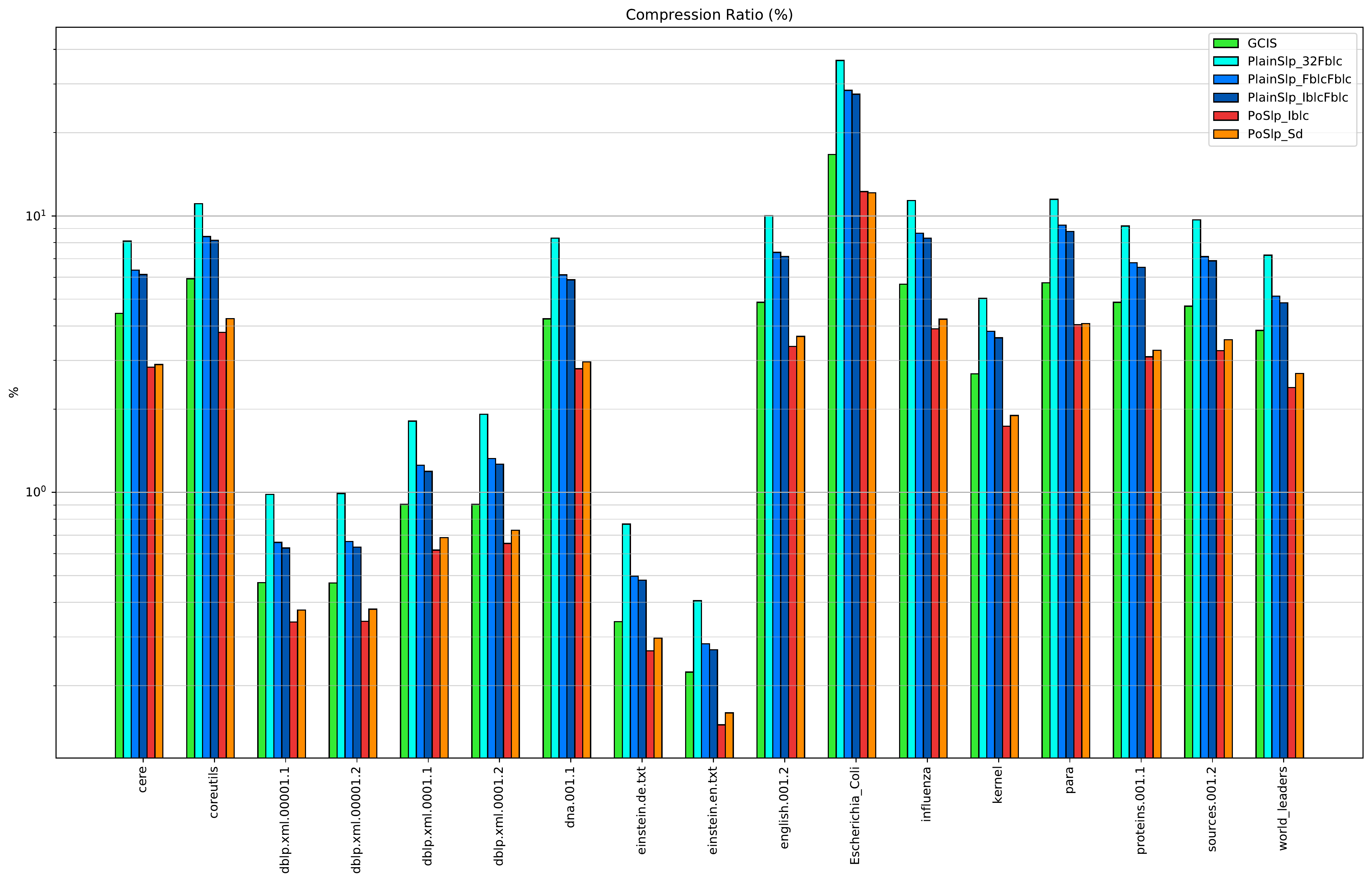}

	\vspace*{-5mm}
	\caption{Compression ratio of the extractors on repetitive texts.}
	\label{fig:compression-ratio-extractors-repetitive-text}
\end{figure}

Results depicted by Figures \ref{fig:extract-regular-texts} and \ref{fig:extract-repetitive-texts} show that \our is faster than the extractors on succinct encodings of \repair but slower than those running on the more straightfoward representation using integer arrays.
In turn, regarding space on regular and repetitive texts, \our is more space-efficient than the straightfoward encodings but less space-efficient than the POSLP alternatives, as shown in Figures \ref{fig:compression-ratio-extractors-regular-text} and \ref{fig:compression-ratio-extractors-repetitive-text}. \our is then a competitive alternative regarding the space-time trade-off.


\subsection{Suffix Array and LCP Construction}
\label{subsec:saca}


Considering the computation of \SA and \LCP arrays during decompression, we measured the total time to decompress the files with \our without generating a plain-text file, but instead inducing the \SA and the \LCP arrays.
We compared these results with {\tt SAIS} \cite{sais-lite-lcp} and {\tt divsufsort} \cite{divsufsort-lcp,DBLP:conf/stringology/0001K17} implementations based on those of Yuta Mori, which are known as the fastest suffix array construction algorithms in practice.

\begin{figure}[ht]
	\begin{subfigure}[t]{.45\textwidth}
		\centering
		\includegraphics[scale=.375]{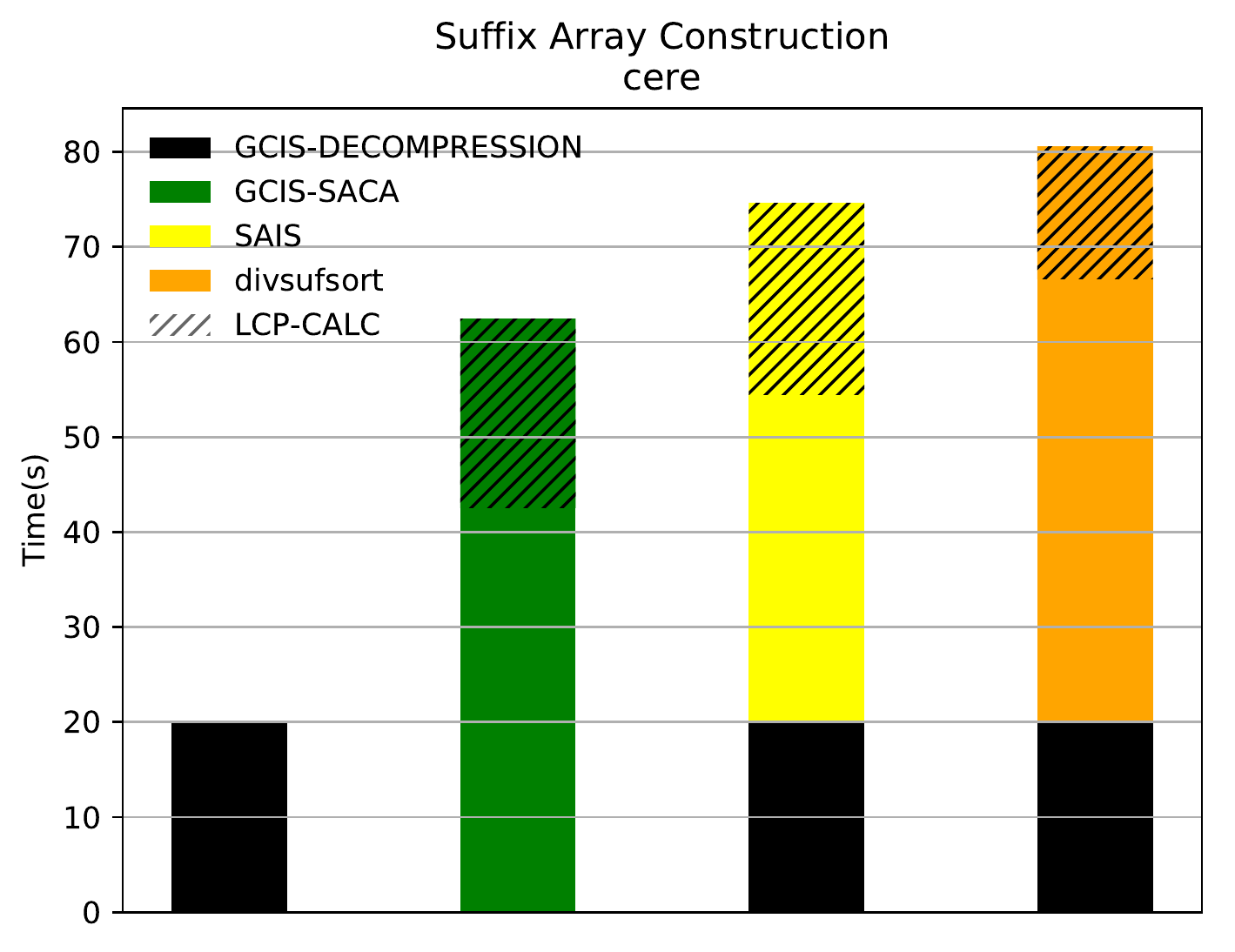}
	\end{subfigure}
	\begin{subfigure}[t]{.45\textwidth}
		\centering
		\includegraphics[scale=.375]{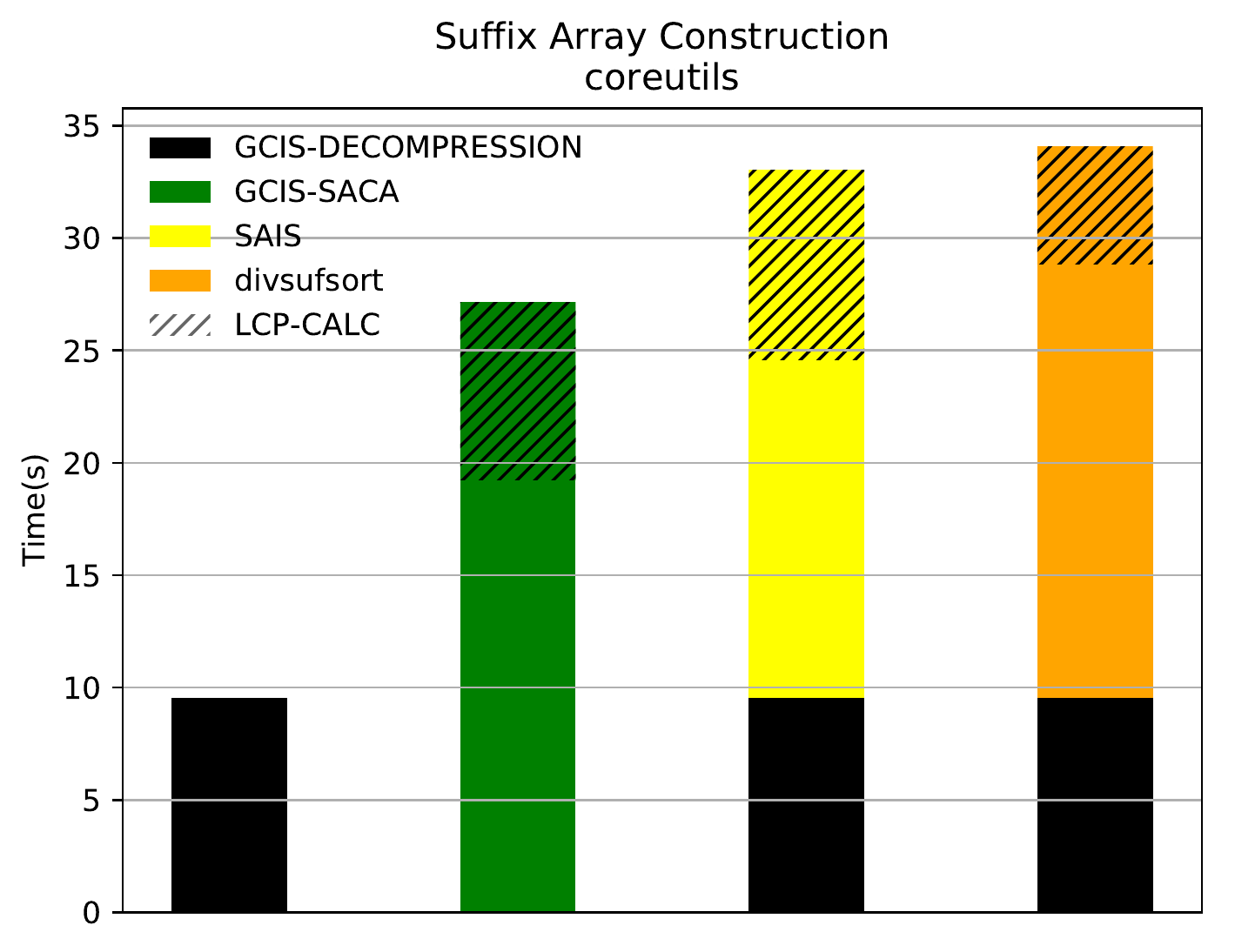}
	\end{subfigure}
	\begin{subfigure}[t]{.45\textwidth}
		\centering
		\includegraphics[scale=.375]{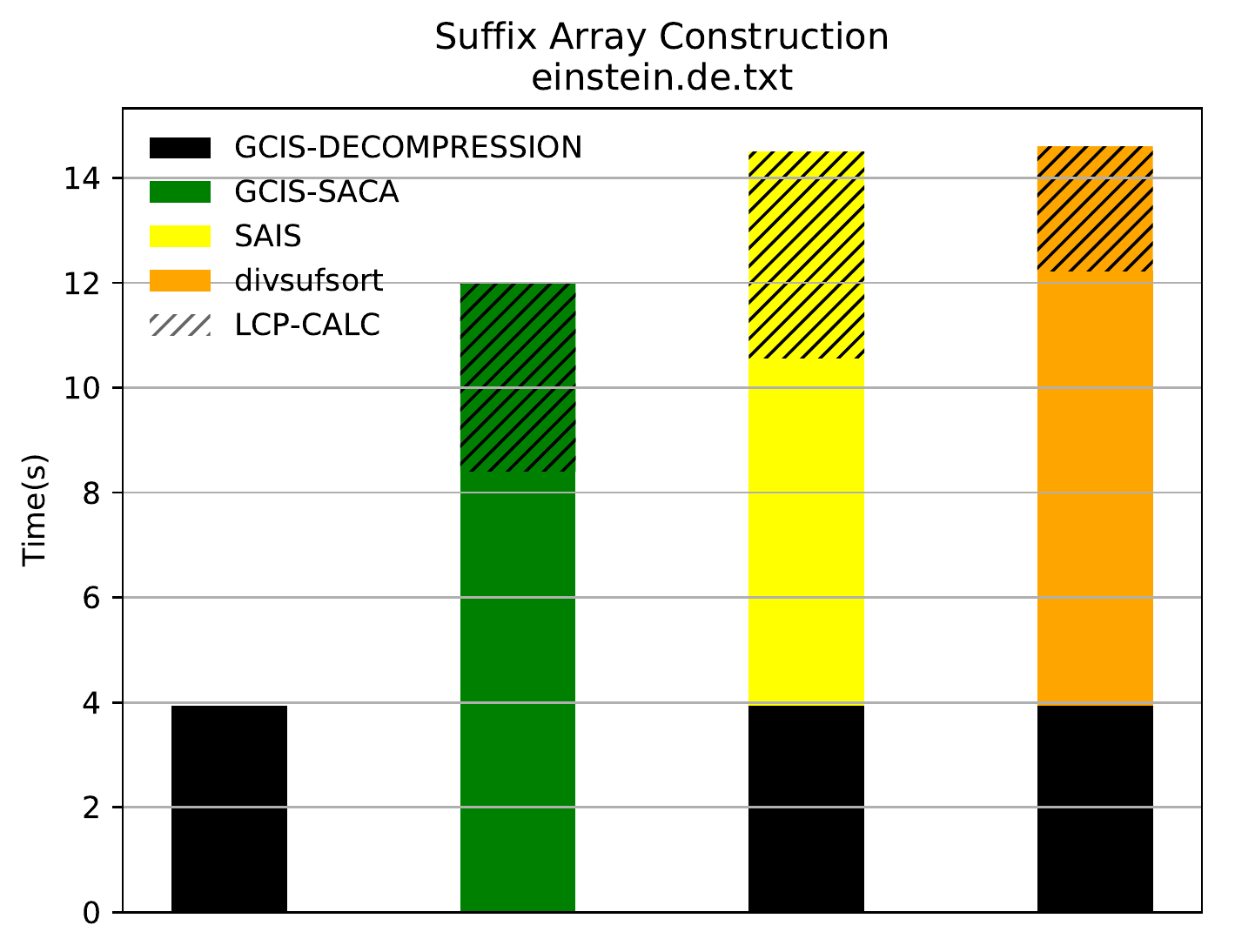}
	\end{subfigure}
	\begin{subfigure}[t]{.45\textwidth}
		\centering
		\includegraphics[scale=.375,ext=.en.txt.pdf]{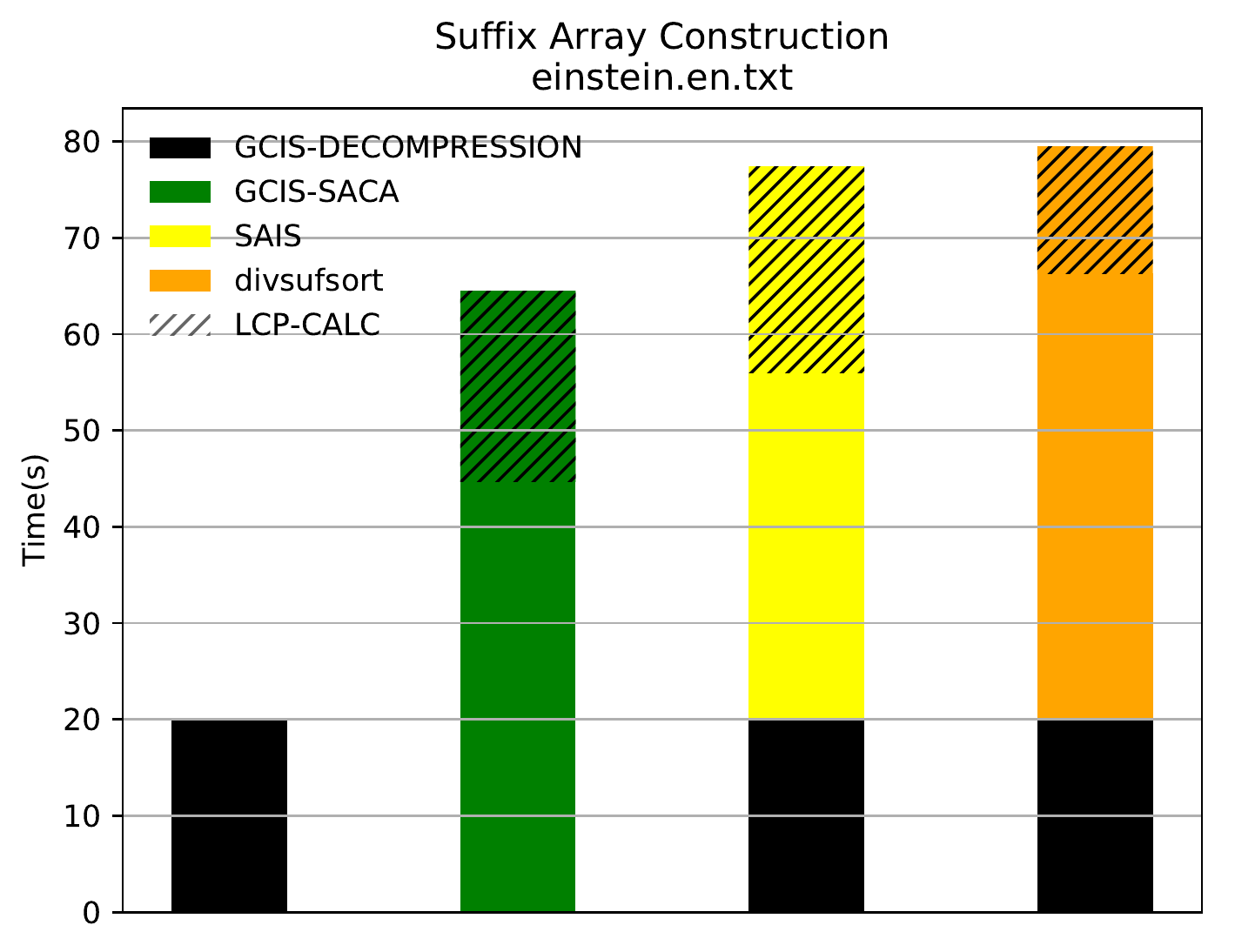}
	\end{subfigure}
	\begin{subfigure}[t]{.45\textwidth}
		\centering
		\includegraphics[scale=.375]{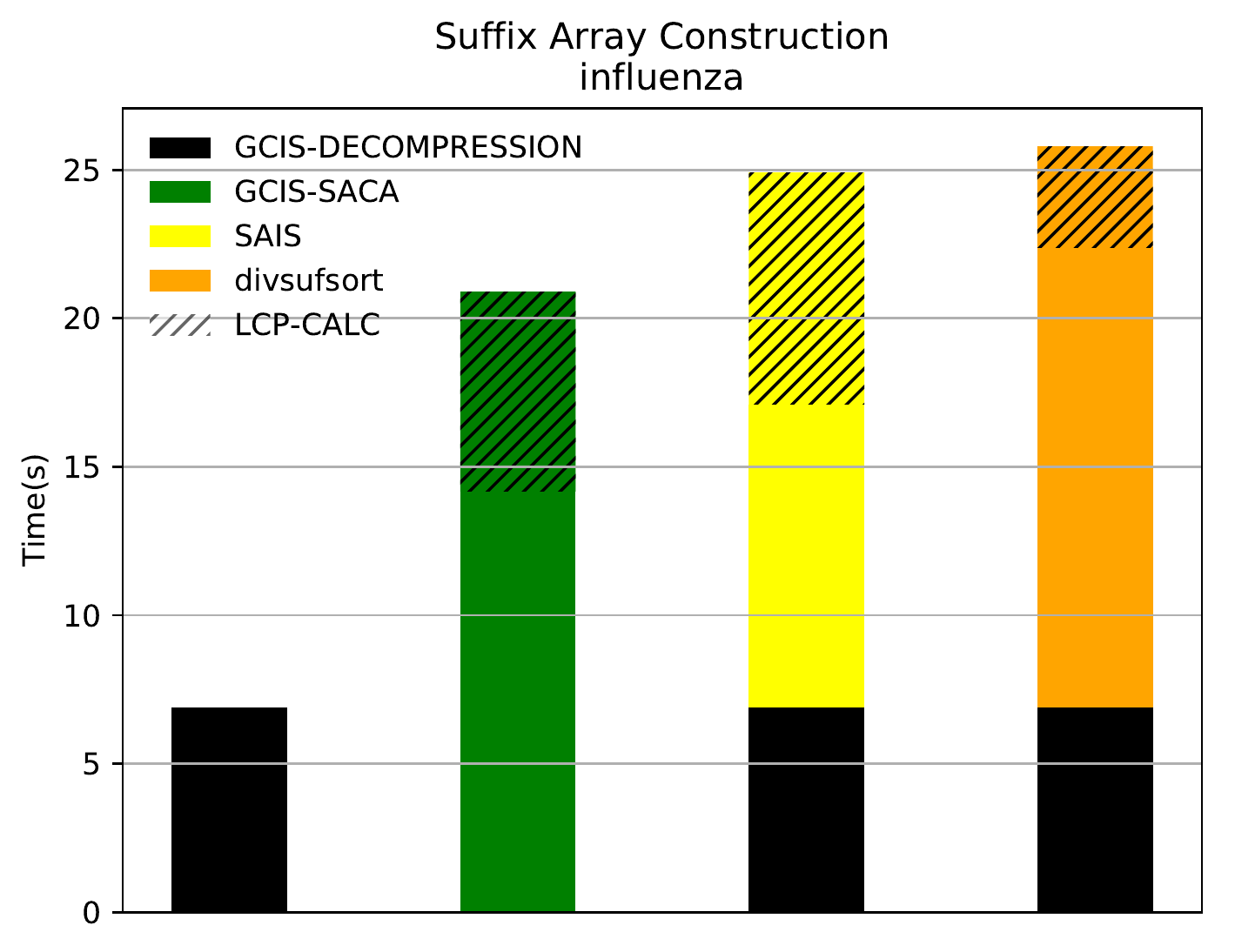}
	\end{subfigure}
	\begin{subfigure}[t]{.45\textwidth}
		\centering
		\includegraphics[scale=.375]{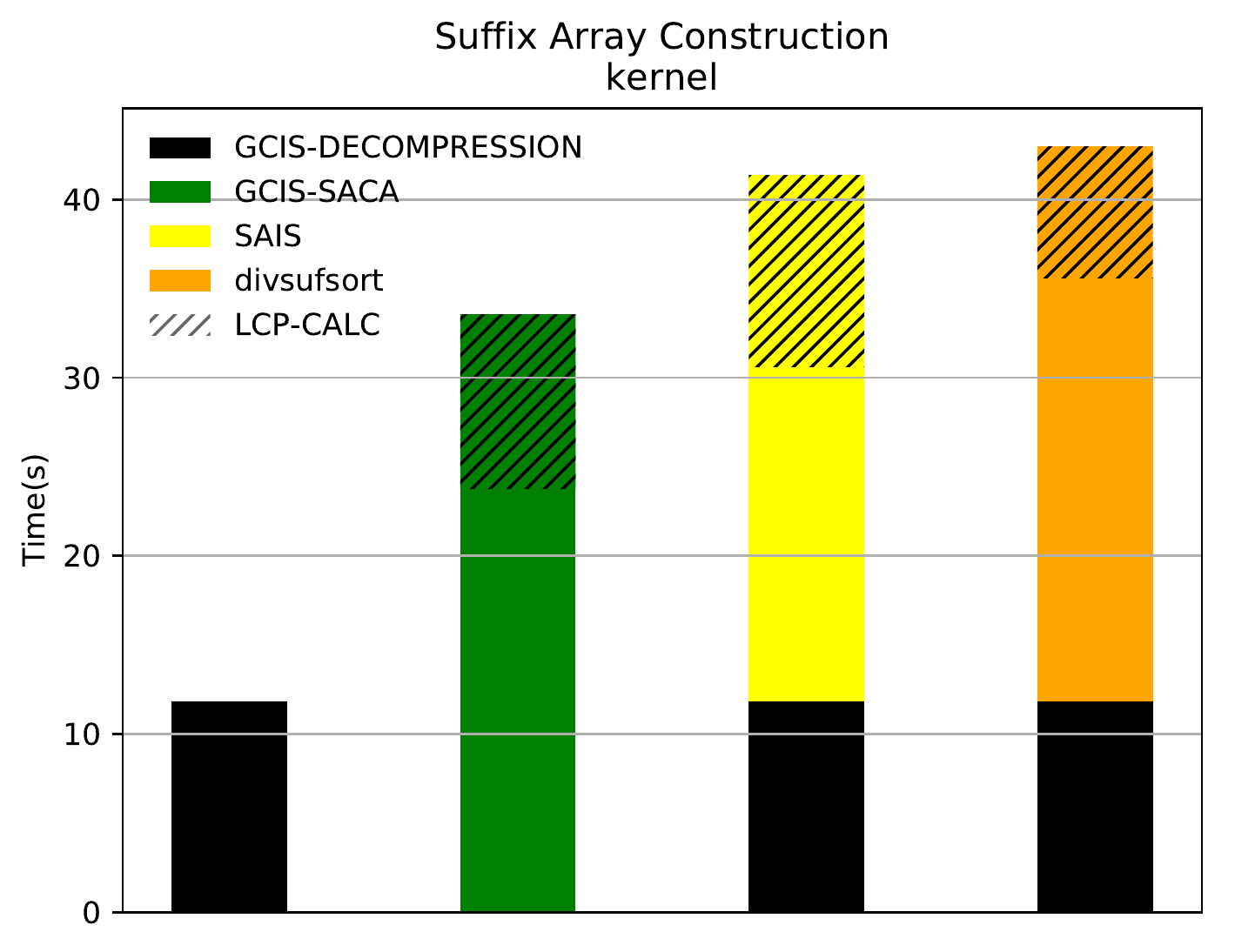}
	\end{subfigure}
	\begin{subfigure}[t]{.45\textwidth}
		\centering
		\includegraphics[scale=.375]{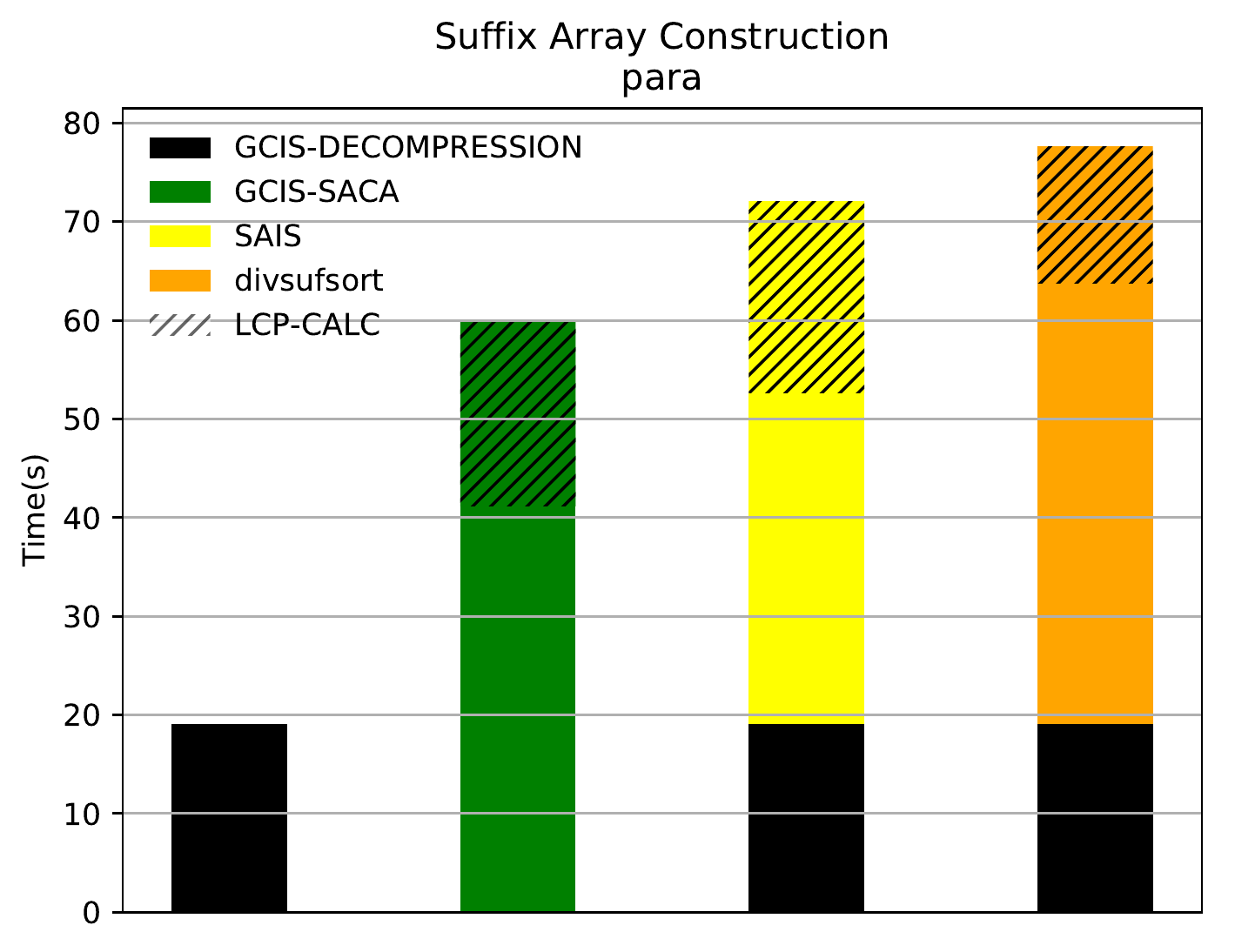}
	\end{subfigure}
	\begin{subfigure}[t]{.45\textwidth}
		\centering
		\includegraphics[scale=.375]{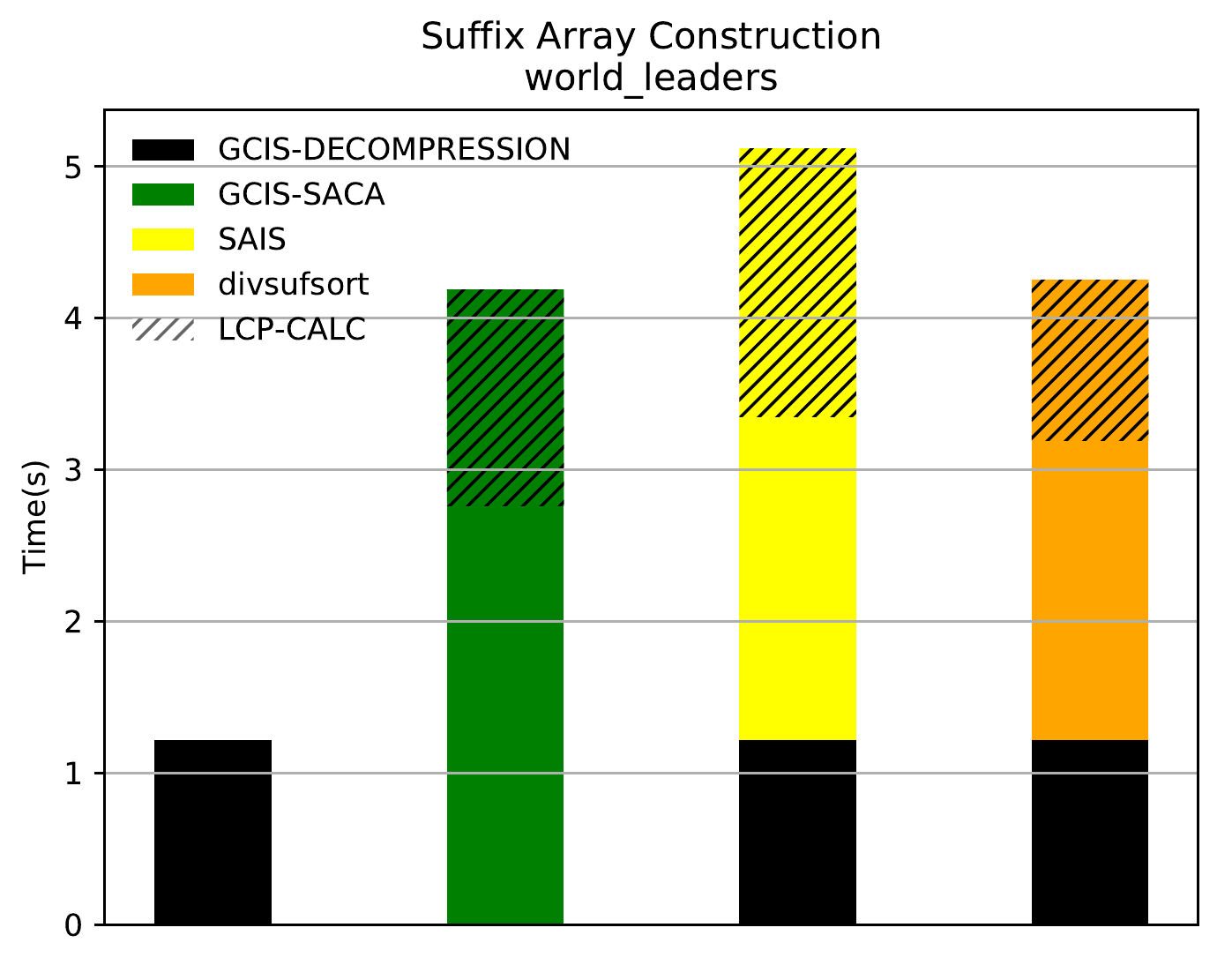}
	\end{subfigure}
	\caption{Time consumed during Suffix Array and LCP construction.}
	\label{fig:saca-lcp-1}
\end{figure}

Figure \ref{fig:saca-lcp-1} shows the \SA and \LCP construction on the $8$ most repetitive real texts when only the \our compressed texts are available. Very large texts were not considered because the implementations of \cite{sais-lite-lcp} and \cite{divsufsort-lcp} only deal with 32-bit integers.
\our builds the \SA and \LCP arrays faster than decompressing and then using the suffix array construction algorithms over the plain text.

The hatched part corresponds to the \LCP computation and the black bar corresponds to the time spent in decompressing the text with \our  to calculate \SA and \LCP values using the {\tt SAIS} and {\tt divsufsort} implementations.

%

\section{Conclusions}\label{s:conclusion}

We have introduced \our, a new grammar-based compression algorithm based on the induced suffix sorting framework of SAIS \cite{Nong2009a}. \our uses the meta-symbols introduced by SAIS to generate non-terminals  of a balanced grammar that reproduces the original text. Our experiments on repetitive texts show that \our compresses $3$--$7$ times faster than \repair and \szip. Compared to \repair, the grammar compressor that compresses the most, \our
compresses using $3$--$5$ times less memory, yet it obtains a compressed file twice as large (yet the absolute compression
is still attractive, below $5\%$ in most cases). \our decompresses $2$--$8$ times slower than \repair and \szip, though.

Grammar-based compression is attractive because, unlike Lempel-Ziv, it can be enriched to support fast extraction of  arbitrary text substrings. From this perspective and regarding the space-time relation, our experiments show that \our is a competitive option when compared to \repair-based extractors, being faster and less space-efficient than the succinct encoding of \repair extractors and slower, but more space-efficient, than the more straightforwardly encoded \repair extractors.

Finally, as a by-product of \our, the suffix array of the text can be obtained during the decompression algorithm, faster than decompressing and running on the original text.

All previously discussed features make \our especially attractive in scenarios where it is required to support random access on the compressed text. Grammar compression of very large files is challenging with \repair because of its large main memory footprint, for which \our offers an interesting alternative.  Given its slowness at decompression, the \our grammar is best suited as a compressed data structure to be repeatedly accessed without decompressing it completely. A possible further improvement would be to replace the variant of \citet{Nong2009a} used during compression by the more space-efficient SACA-K \cite{Nong2013} algorithm. This could decrease the working space used during compression. Grammar-based compressed indexes are of particular interest, all of which are based on \repair \cite{CNfi10, CNspire12.1}.  Our next goal is to build those compressed indexes on \our instead. This would yield indexes that might be built much faster, using much less memory, for moderate-sized repetitive texts, and that may lead to much more efficient search times.



\begin{acks}

	Part of the research of DSNN was performed during visits at University of Chile and supported by  FAP-DF, grants 13619.54.33621.1708/2016 and 0193.001380/2017. FAL was supported by the grant 2017/09105-0 from the S\~ao Paulo Research Foundation (FAPESP). MAR was supported by FAP-DF and CNPq, grants DE 193.001.369/2016 and PQ 307672/2017-4. GN was funded by Basal Funds FB0001 and Fondecyt Grant 1-200038, Conicyt, Chile.


\end{acks}

\clearpage
\bibliographystyle{ACM-Reference-Format}
\bibliography{refs}

\end{document}